\documentclass[journal,compsoc,screen]{IEEEtran}
\usepackage{color}
\usepackage{etoolbox}
\usepackage{listings}
\usepackage[T1]{fontenc}

\usepackage[scaled=0.8]{beramono}

\newcommand{\lstfontfamily}{\ttfamily}

\definecolor{darkviolet}{rgb}{0.5,0,0.4}
\definecolor{darkgreen}{rgb}{0,0.4,0.2} 
\definecolor{darkblue}{rgb}{0.1,0.1,0.9}
\definecolor{darkgrey}{rgb}{0.5,0.5,0.5}
\definecolor{lightblue}{rgb}{0.4,0.4,1}

\definecolor{stringColor}{rgb}{0.16,0.00,1.00}
\definecolor{annotationColor}{rgb}{0.39,0.39,0.39}
\definecolor{keywordColor}{rgb}{0.50,0.00,0.33}
\definecolor{commentColor}{rgb}{0.25,0.50,0.37}
\definecolor{javadocColor}{rgb}{0.25,0.37,0.75}
\definecolor{jTagColor}{rgb}{0.50,0.62,0.75}
\definecolor{eTagColor}{rgb}{0.50,0.62,0.75}
\definecolor{lineNumberColor}{rgb}{0.47,0.47,0.47}

\def\jTags{@author, @deprecated, @exception, @param, @return, @see, @serial, @serialData, @serialField, @since, @throws, @version}

\def\jAnnotations{
    classoffset=1,
    morekeywords={@Override, @Deperecated, @SuppressWarnings, @Retention, @Documented, @Target, @Inherited},
    keywordstyle=\color{annotationColor},
    classoffset=0
}

\def\eTags{FIXME, TODO, XXX}

\newrobustcmd{\markupJavadocs}[1]{%
\edef\mytok{\the\lst@token}%
\renewcommand*{\do}[1]{%
\ifdefstring{\mytok}{##1}%
{\color{jTagColor}\bfseries\listbreak}%
{}%
}%
{\color{javadocColor}%
\expandafter\docsvlist\expandafter{\jTags}%
\renewcommand*{\do}[1]{%
\ifdefstring{\mytok}{##1}%
{\color{eTagColor}\bfseries\listbreak}%
{}%
}%
\expandafter\docsvlist\expandafter{\eTags}%
#1}%
}%

\newrobustcmd{\markupComments}[1]{%
\edef\mytok{\the\lst@token}%
\renewcommand*{\do}[1]{%
\ifdefstring{\mytok}{##1}%
{\color{eTagColor}\bfseries\listbreak}%
{}%
}%
{\color{commentColor}%
\expandafter\docsvlist\expandafter{\eTags}#1}%
}%

\lstdefinestyle{eclipse}{
  basicstyle={\lstfontfamily},
  emphstyle=\bfseries,
  keywordstyle=\color{keywordColor}\bfseries,
  commentstyle=\markupComments,
  stringstyle=\color{stringColor},
  numberstyle=\color{lineNumberColor}\lstfontfamily,
  morecomment=[s][\markupJavadocs]{/**}{*/}, 
  showstringspaces=false,
  numbers=left,
}

\lstdefinestyle{black}{
  basicstyle=\small\lstfontfamily,
  numbers=left,
  columns=fullflexible,
  breaklines=true,
  mathescape=true,
  escapechar=\#,
  tabsize=4,
  frame=lines,
  showstringspaces=false
}

\lstdefinestyle{seminar}{
  basicstyle=\small\ttfamily,
  numbers=left,
  breaklines=true,
  mathescape=true,
  escapechar=\#,
  tabsize=4,
  showstringspaces=false
}

\expandafter\lstset\expandafter{\jAnnotations}
\usepackage{textcomp}
\usepackage{hyperref}
\usepackage{color}
\usepackage{tcolorbox}
\usepackage{balance}
\usepackage{makecell}
\usepackage{verbatim}
\usepackage{ulem}
\usepackage{mdframed}
\usepackage{lipsum}
\usepackage{amsmath}
\usepackage{amssymb}
\usepackage{amsfonts}
\usepackage[font=small]{caption}
\usepackage{ifpdf}
\usepackage{graphicx}
\usepackage{colortbl}
\usepackage{listings}

\usepackage{pifont}

\usepackage[inkscapelatex=false]{svg}

\usepackage{algorithmic}
\usepackage{array}

\usepackage{caption}
\usepackage{multirow}
\usepackage{float}
\usepackage{mwe}
\newcommand{\code}[1]{\texttt{#1}}

\usepackage{url}
\usepackage{todonotes}
\usepackage{subfigure}
\usepackage{tikz}
\usepackage{xcolor}

\usepackage{booktabs,caption,fixltx2e}
\usepackage[flushleft]{threeparttable}

 \usepackage{enumitem}

\hypersetup{
 colorlinks=true,
 linkcolor=blue,
 citecolor=blue,
 filecolor=black,
 urlcolor=black,
 pdfauthor = {},
 pdftitle = {},
 pdfkeywords = {}
}

\definecolor{bg}{HTML}{F8F9FB}  
\usepackage{algorithm2e}
\usepackage{xcolor}

\SetAlFnt{\footnotesize}

\SetAlCapSty{xAlCapSty}

\SetCommentSty{xCommentSty}

\SetNlSty{mynlfont}{}{} 

\LinesNumbered

\SetSideCommentRight

\DontPrintSemicolon

\RestyleAlgo{algoruled}

\usepackage[autolanguage]{numprint}

\newcommand*\np[2][z]{
\ifx z#1%
$\numprint{#2}$%
\else%
$\numprint[#1]{#2}$%
\fi\xspace%
}

\usepackage{fp}
\newcommand{\ShowAbsoluteNumber}[1]{%
\ifnum #1<10%
{\hspace*{0pt}#1}%
\else%
\ifnum #1<100%
{\hspace*{0pt}#1}%
\else%
\ifnum #1<1000%
{\hspace*{0pt}#1}%
\else%
{\numprint{#1}}%
\fi%
\fi%
\fi%
}

\newcommand{\ShowPercentage}[2]{%
\FPeval\percentage{round(#1/#2*100,0)}%
\FPeval\percentageOneDecimal{round(#1/#2*100,1)}%
\ifnum \percentage=0%
{\np[\%]{\FPprint{percentageOneDecimal}}}%
\else%
\ifnum \percentage<10%
{\np[\%]{\FPprint{percentageOneDecimal}}}%
\else%
{\np[\%]{\FPprint{percentageOneDecimal}}}%
\fi%
\fi%
\xspace
}

\newcommand{\ShowPercentageTwo}[2]{%
\FPeval\percentagetwo{round(#1/#2*100,0)}%
\FPeval\percentageTwoDecimal{round(#1/#2*100,2)}%
\ifnum \percentagetwo=0%
{\np[\%]{\FPprint{percentageTwoDecimal}}}%
\else%
\ifnum \percentagetwo<10%
{\np[\%]{\FPprint{percentageTwoDecimal}}}%
\else%
{\np[\%]{\FPprint{percentageTwoDecimal}}}%
\fi%
\fi%
\xspace
}

\newlength\BARSIZE  
\newcommand{\inlinechart}[2]{%
\FPeval{\BLACKBARSIZE}{#1/#2}\textcolor{black!80}{\rule{\BLACKBARSIZE\BARSIZE}{1.6ex}}%
\FPeval{\BLACKBARSIZE}{1 - (#1/#2)}\textcolor{black!10}{\rule{\BLACKBARSIZE\BARSIZE}{1.6ex}}%
}

\newcommand*\percent[3][v]{%
\ifx q#1%
    \np{#2}/\np{#3}(\ShowPercentage{#2}{#3})\else%
\ifx p#1%
    \np{#2}(\ShowPercentage{#2}{#3})\else%
\ifx m#1%
    \np{#2}%
    \inlinechart{#2}{#3}\else%
\ifx t#1%
    \ShowPercentageTwo{#2}{#3}%
    \inlinechart{#2}{#3}\else%
\ifx g#1%
    \ShowPercentageTwo{#2}{#3}\else%
\ifx c#1%
    \inlinechart{#2}{#3}%
\else%
    \np{#2}%
    \ifx r#1%
        /\np{#3}%
    \fi%
    \hspace*{0.5ex}(\ShowPercentage{#2}{#3}) %
    \inlinechart{#2}{#3}%
    \xspace
\fi\fi\fi\fi\fi\fi%
}

\newcommand{\ChatGPT}{\textit{ChatGPT }}

\newcommand{\revision}[1]{\textcolor{black}{#1}}

\newcommand{\minorrevision}[1]{\textcolor{black}{#1}}

\ifCLASSINFOpdf
\else
\fi
\begin{document}

\title{No Need to Lift a Finger Anymore? Assessing the Quality of Code Generation by ChatGPT}

\author{Zhijie~Liu,
Yutian~Tang,
Xiapu~Luo, 
Yuming~Zhou, and
Liang~Feng~Zhang   
\thanks{Zhijie Liu is with ShanghaiTech University, Shanghai 201210, China. E-mail: liuzhj2022@shanghaitech.edu.cn.}
\thanks{Yutian Tang is with University of Glasgow, United Kingdom. E-mail: yutian.tang@glasgow.ac.uk.}
\thanks{Xiapu Luo is with the Department of Computing, Hong Kong Polytechnic University, Hong Kong SAR, China. E-mail: csxluo@comp.polyu.edu.hk.}
\thanks{Yuming Zhou is with Nanjing University, China. E-mail: zhouyuming@nju.edu.cn.}
\thanks{Liang Feng Zhang is with ShanghaiTech University, Shanghai 201210, China. E-mail: zhanglf@shanghaitech.edu.cn.}
\thanks{Yutian Tang (yutian.tang@glasgow.ac.uk) is the corresponding author.}}

\markboth{Journal of \LaTeX\ Class Files,~Vol.~14, No.~8, August~2021}%
{Shell \MakeLowercase{\textit{et al.}}: A Sample Article Using IEEEtran.cls for IEEE Journals}



\IEEEtitleabstractindextext{%
\begin{abstract}
Large language models (LLMs) have demonstrated impressive capabilities across various natural language processing (NLP) tasks, such as machine translation, question answering, summarization, and so on. Additionally, LLMs are also highly valuable in supporting software engineering tasks, particularly in the field of code generation. Automatic code generation is a process of automatically generating source code or executable code based on given specifications or requirements, improving developer productivity. In this study, we perform a systematic empirical assessment \revision{to the quality} of code generation using \textit{ChatGPT}, a recent \revision{state-of-the-art product} LLM. \revision{We leverage 728 algorithm problems in five languages (i.e., C, C++, Java, Python, and JavaScript) and 18 CWEs with 54 code scenarios for the code generation task.} Our evaluation encompasses a comprehensive analysis of code snippets generated by \textit{ChatGPT}, focusing on three critical aspects: correctness, \revision{complexity}, and security. We also specifically investigate \textit{ChatGPT}'s ability to engage in multi-round \revision{fixing} process (i.e., \textit{ChatGPT}'s dialog ability\revision{, chatting between users and \textit{ChatGPT} for fixing generated buggy code}) of facilitating code generation. By delving into the generated code and examining the experimental results, this work provides valuable insights into the performance of \textit{ChatGPT} in tackling code generation tasks \revision{over the three critical aspects}. \revision{The experimental results demonstrate that (1) \textit{ChatGPT} is better at generating functionally correct code for problems before 2021 in different languages than problems after 2021 with $48.14\%$ advantage in \textit{Accepted} rate on judgment platform, but \textit{ChatGPT}'s ability to directly fix erroneous code with multi-round fixing process to achieve correct functionality is relatively weak; (2) the distribution of cyclomatic and cognitive complexity levels for code snippets in different languages varies. Furthermore, the multi-round fixing process with \ChatGPT generally preserves or increases the complexity levels of code snippets; (3) in \minorrevision{algorithm} scenarios \minorrevision{with languages of C, C++, and Java, and CWE scenarios with languages of C and Python3}, the code generated by \ChatGPT has relevant vulnerabilities. However, the multi-round fixing process for vulnerable code snippets demonstrates promising results, with more than $89\%$ of vulnerabilities successfully addressed; and (4) code generation may be affected by \textit{ChatGPT}'s non-determinism factor, resulting in variations of code snippets in functional correctness, complexity, and security.} Overall, our findings uncover potential issues and limitations that arise in the \textit{ChatGPT}-based code generation and lay the groundwork for improving AI and LLM-based code generation techniques.
\end{abstract}

}

\maketitle

\IEEEdisplaynontitleabstractindextext

%
\IEEEpeerreviewmaketitle

\newcommand{\blackding}[1]{\ding{\numexpr181+#1\relax}}
\newcommand{\whiteding}[1]{\ding{\numexpr171+#1\relax}}

\section{Introduction}\label{sec:introduction}

Automatic code generation is a process of automatically generating source code or executable code based on given specifications or requirements. It supports a range of capabilities that benefit software development greatly. By using automatic code generation, developers are able to enhance productivity, reduce development time, and assign more focus to higher-level tasks and core logic. Lots of studies on code generation leverage AI-based approaches~\cite{rabinovich2017abstract, ye2020leveraging, alon2019code2vec, bui2021infercode, TufanoDrainSvyatkovskiyDengSundaresan2020}, especially for using large language models (LLMs)~\cite{vaswani2017attention, svyatkovskiy2020intellicode, wang2021codet5, chen2021evaluating, feng2020codebert, copilot} such as the recent \textit{ChatGPT}~\cite{ChatGPT}. 

\noindent\textbf{AI-based Code Generation.} The emergence of AI-based code generation is driven by the increasing complexity of software systems and the desire for a more efficient development process~\cite{ai-based-code-generation}. Traditional code generation approaches~\cite{gulwani2017program} rely on predefined templates or rules (e.g., context-free grammar) and input-output specifications, which limits their flexibility and requires manual effort. AI-based approaches~\cite{allamanis2018survey, chen2021evaluating, copilot, ChatGPT} leverage the power of machine learning (deep learning) and natural language processing (NLP) to overcome these limitations and can offer more intelligent and adaptable code-generation capabilities. These approaches analyze directly input specifications or requirements expressed in natural language and generate corresponding code snippets or complete programs based on the provided input.

\noindent\textbf{Large Language Model and \textit{ChatGPT}.} Recently, large language models (LLMs) demonstrate remarkable capabilities in a wide range of NLP tasks, such as machine translation, question answering, summarization, text generation, grammar checking, and so on~\cite{Carlini:21, Brants:07, Raffel:22, nagata2021exploring}. These models possess a capacity for understanding and generating human-like text, approaching the level of humans. LLMs are primarily built on the Transformer architecture~\cite{vaswani2017attention}, with OpenAI's \textit{GPT-3} (Generative Pretrained Transformer 3)~\cite{Brown:20} being a prominent example. \textit{GPT-3} is trained on extensive amounts of textual data, resulting in exceptional performance. \textit{ChatGPT}~\cite{ChatGPT} is an implementation with dialog ability that is built upon the foundation of \revision{\textit{GPT-3.5}~\cite{gpt-3.5} (or \textit{GPT-4}~\cite{GPT-4})}. It exhibits outstanding performance in areas such as machine translation, question answering, summarization, and so on, and is found in widespread usage in various daily activities. Importantly, \textit{ChatGPT} also possesses the capability of code-related tasks, which can further expand its potential applications. \textit{ChatGPT} now has become an essential tool for individuals, academia, and industry, significantly enhancing productivity in various domains.

\noindent\textbf{Motivation.} While AI-based code generation, using LLMs, provides promising advantages in enhancing productivity and automating software development tasks, it is still essential to assess the generated code for showing better insights \revision{and understanding}. Code generation by LLMs is facing challenges. For example, whether the code generated by LLMs is functionally correct, \revision{complex}, and secure. \revision{The training datasets for LLMs come from the internet, but the quality of the data is uncertain. Subsequently, the quality of the code generated by LLMs also cannot be guaranteed~\cite{pearce2022asleep, nguyen2022empirical, liu2023your, fu2023security}. A deep analysis of these aspects can provide a more comprehensive understanding of AI and LLM-based code generation.} In this paper, we are interested in deeply and systematically evaluating the code generated by LLMs in terms of its correctness, \revision{complexity}, and security. Specifically, we leverage the state-of-the-art \textit{ChatGPT}\footnote{The version used in \textit{ChatGPT} is \textit{GPT-3.5} instead of \textit{GPT-4}.}, a recent product, as the representative of LLMs for evaluation\revision{, due to its advanced capabilities and widespread recognition~\cite{ChatGPT}}. We also assess \textit{ChatGPT}'s dialog ability (i.e., the multi-round \revision{fixing} process in one single conversation\revision{, chatting between users and \textit{ChatGPT} for fixing generated buggy code}) \revision{in the code generation task over} correctness, \revision{complexity}, and security. By conducting a comprehensive analysis, we seek to uncover potential issues and limitations that arise in the \textit{ChatGPT}-based code generation for improving AI and LLM-based code generation techniques. 

\noindent \textbf{Our Study.} To cope with the aforementioned challenges and explore the ability of \textit{ChatGPT}~\cite{ChatGPT} to generate code, we \revision{collect and leverage 728 algorithm problems in five languages (i.e., C, C++, Java, Python, and JavaScript) and 18 CWEs with 54 code scenarios from \textit{LeetCode} platform~\cite{LeetCode} and \cite{pearce2022asleep}, respectively, for the code generation task and} intend to answer the following research questions (RQs):

\noindent $\bullet$  \textbf{RQ1 (Functionally Correct Code Generation):} Is the code generated by \textit{ChatGPT} functionally correct?

\noindent $\bullet$  \textbf{RQ2 (Multi-round Fixing for Code Generation):} How effective is the multi-round fixing process in improving code generation \revision{for functional correctness}?

\noindent $\bullet$  \textbf{RQ3 (Code \revision{Complexity}):} How \revision{complex} is the code generated by \textit{ChatGPT}?

\noindent $\bullet$  \textbf{RQ4 (Security Code Generation):} Is the code generated by \textit{ChatGPT} secure?

\noindent $\bullet$  \revision{\textbf{RQ5 (Non-determinism of \textit{ChatGPT}):}} \revision{How does the non-deterministic output of \textit{ChatGPT} affect code generation?}

\revision{Our experimental results demonstrate that (1) \textit{ChatGPT} is better at generating functionally correct code for problems before 2021 in different languages than problems after 2021 with $48.14\%$ advantage in \textit{Accepted} rate on judgment platform, but \textit{ChatGPT}'s ability to directly fix erroneous code with multi-round fixing process to achieve correct functionality is relatively weak; (2) the distribution of cyclomatic and cognitive complexity levels for code snippets in different languages varies. Furthermore, the multi-round fixing process with \ChatGPT generally preserves or increases the complexity levels of code snippets; (3) in \minorrevision{algorithm} scenarios \minorrevision{with languages of C, C++, and Java, and CWE scenarios with languages of C and Python3}, the code generated by \ChatGPT has relevant vulnerabilities. However, the multi-round fixing process for vulnerable code snippets demonstrates promising results, with more than $89\%$ of vulnerabilities successfully addressed; and (4) code generation may be affected by \textit{ChatGPT}'s non-determinism factor, resulting in variations of code snippets in functional correctness, complexity, and security.}

\noindent \textbf{Contributions.} In summary, we make the following contributions to this paper:

\noindent $\bullet$ In this paper, we conduct a comprehensive empirical assessment \revision{to the quality} of \textit{ChatGPT}-based code generation;

\noindent $\bullet$ We systematically evaluate the \textit{ChatGPT}-based code generation, including multi-round process, from three aspects: correctness, \revision{complexity}, and security. \revision{The evaluated results reveal potential issues and limitations in \textit{ChatGPT}-based code generation over the three aspects}; and

\noindent $\bullet$ Our research contributes to advancing the potential knowledge and understanding of the capabilities of LLMs in enhancing software engineering practices, with a particular focus on code generation.

\noindent \textbf{Online Artifact.} The experimental scripts, results, and raw data are available at: \cite{artifact}.

\section{Background}\label{sec:background}

In this section, we briefly introduce LLMs, \textit{ChatGPT}, and the use of \textit{ChatGPT}.

\noindent\textbf{LLMs and \textit{ChatGPT}.} LLMs (large language models)~\cite{vaswani2017attention, Devlin:18, Raffel:20, Ouyang:22, radford:18, radford:19, Brown:20, ChatGPT} refer to a class of AI models that use an enormous amount of parameters and are designed to process and generate human-like text based on large-scale language datasets. These models utilize deep learning techniques, typically employing Transformer architectures\revision{~\cite{vaswani2017attention}, consisting of stacked encoders and decoders}, to learn patterns, relationships, and structures in languages. \revision{Transformer utilizes self-attention mechanism to weigh the importance of words in the input text, capturing long-range dependencies and relationships between words.} LLMs are trained on massive amounts of text data from various sources and show a strong ability in many NLP tasks, such as machine translation, question answering, summarization, and so on. \revision{\textit{GPT}~\cite{radford:18} and \textit{BERT}~\cite{Devlin:18} are based on the decoder (unidirectional) and encoder (bidirectional) components of the Transformer, respectively. They utilize pre-training and fine-tuning techniques. \textit{GPT-2}~\cite{radford:19} and \textit{GPT-3}~\cite{Brown:20} are the successors of \textit{GPT}, with \textit{GPT-2} having a larger model size in parameters than \textit{GPT}, and \textit{GPT-3} being even larger than \textit{GPT-2} with using 175 billion parameters. Additionally, with larger corpus, \textit{GPT-2} and \textit{GPT-3} introduce zero-shot and few-shot learning to enable adaptation to multitask scenarios. Moreover, \textit{GPT-3} has demonstrated performance comparable to state-of-the-art fine-tuned systems across various tasks. \textit{Codex}~\cite{chen2021evaluating} is obtained by training \textit{GPT-3} on GitHub code data. It serves as the underlying model for GitHub \textit{Copilot}~\cite{copilot}, a tool that can automatically generate and complete code automatically. To enhance the alignment between LLMs and users (humans), \textit{InstructGPT}~\cite{Ouyang:22} incorporates additional supervised learning and reinforcement learning from human feedback (RLHF) to fine-tune \textit{GPT-3}.} \textit{ChatGPT}~\cite{ChatGPT, Ouyang:22}, implemented atop \textit{GPT-3.5}~\cite{gpt-3.5} \revision{(or \textit{GPT-4}~\cite{GPT-4})}, is now the most ideal product LLM that adapts to human expression by using Instruct\revision{~\cite{Ouyang:22}}. \textit{ChatGPT} utilizes the same methods as \textit{InstructGPT} and provides the ability to answer follow-up questions (i.e., dialog ability) \revision{through RLHF}. \revision{The dialog ability~\cite{ChatGPT} enables \textit{ChatGPT} to communicate with users conversationally, continuously generating information or correcting previously incorrect ones. This property makes \textit{ChatGPT} even more powerful and versatile than previous LLMs. Thus, in this study, we take the state-of-the-art \textit{ChatGPT} (the default version of \textit{GPT-3.5}), the recent popular product, as the representative of LLMs for evaluation.}

\noindent\textbf{Use of \textit{ChatGPT}.} To use \textit{ChatGPT}, developers send a text message as input. The message is called prompt, used to guide \textit{ChatGPT}'s text generation. The prompt serves as a cue for the model to understand the desired output or the user's intent. \textit{ChatGPT} responds based on the input prompt and the knowledge it learns from its massive amounts of training data. \textit{ChatGPT} also supports answering follow-up questions (i.e., dialog ability), which allows users to engage in back-and-forth conversations. This capability enables users to send multiple text messages consecutively to \textit{ChatGPT} and receive responses that maintain context and continuity. 

When a user submits a series of messages to \textit{ChatGPT}, each message within the conversation context is considered by the model when generating a response. The messages can include user prompts, system instructions, and previous responses from \textit{ChatGPT} itself. By incorporating the conversation history, \textit{ChatGPT} can produce responses that are more coherent and informed, taking into account the ongoing dialogue.

\textit{ChatGPT} can be used in many software engineering tasks, including code generation. Additionally, by leveraging \textit{ChatGPT}'s dialog ability, users can provide a series of prompts (i.e., multi-round process) for code generation (e.g., fix functionally incorrect code).

\noindent\textbf{Example.} For example, the prompt, \textit{''Write a bubble sort program in Python to rank 1-10''}, can be used for \textit{ChatGPT} to generate a bubble sort program in Python. As shown in Fig. \ref{fig:chatgpt-bubble-sort}, \textit{ChatGPT} offers a sample code for the bubble sort program in Python, which fulfills the requirement given in the prompt. We can also append a new prompt \textit{''Output the sorted result in nonincreasing order''} in the same conversation that \textit{ChatGPT} is guided to replace comparison operator $>$ with $<$ in line 7 of the code snippet.

\begin{figure}[t]
    \centering
  \begin{lstlisting}
def bubble_sort(arr):
    n = len(arr)
    for i in range(n):
        # Last i elements are already sorted
        for j in range(0, n-i-1):
            # Swap if the element found is greater than the next element
            if arr[j] > arr[j+1]:
                arr[j], arr[j+1] = arr[j+1], arr[j]
# The input list containing numbers from 1 to 10 in random order
numbers = [3, 7, 2, 8, 10, 1, 5, 9, 4, 6]
# Sort the numbers using bubble sort
bubble_sort(numbers)
# Print the sorted list
print("Sorted list of numbers from 1 to 10:")
print(numbers)
  \end{lstlisting}
  \caption{\textit{ChatGPT}-generated Bubble Sort Algorithm in Python.}
  \label{fig:chatgpt-bubble-sort}
\end{figure}

\begin{figure*}[t]
    \centering
    \includegraphics[width=0.95\textwidth]{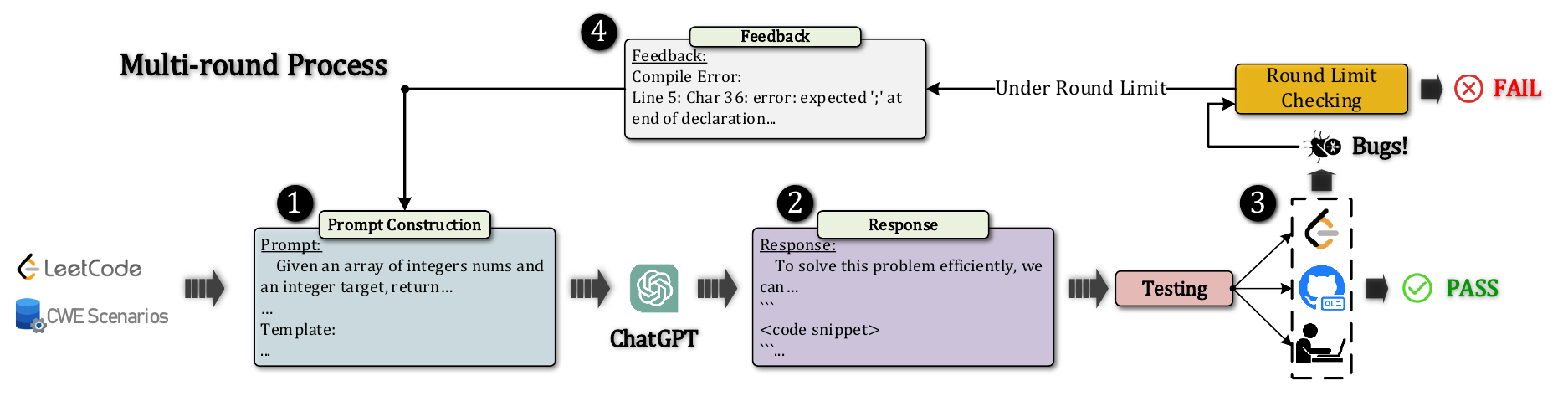}
    \caption{The workflow of interacting with \textit{ChatGPT} to generate code snippets.}
    \label{fig:workflow}
\end{figure*}

\section{Empirical Study Setup}\label{sec:studysetup}

In this section, we introduce the workflow of interacting with \textit{ChatGPT} to generate code and other study setups.

\subsection{Data Collection}

The assessment of our study involves the utilization of two datasets: \textit{LeetCode} problems~\cite{LeetCode} and CWE (Common Weakness Enumeration) scenarios (CWE's code scenarios) as provided in \cite{pearce2022asleep}. \revision{For \textit{LeetCode} problem dataset, we randomly collect 728 algorithm problems where 354 and 374 of them are published after 2021 and before 2021, respectively. The reason for splitting them is because \textit{ChatGPT} is trained on text data before 2021. \revision{For each problem, the problem description, input-output examples, and the method signature template in the specified language are used for code generation.} As for the CWE scenario dataset, it contains 18 CWEs with 54 scenarios in \textit{MITRE Top 25 CWEs}~\cite{mitre} (3 of them drop in rank to below 25 in 2022 \textit{MITRE Top 25 CWEs}). For each CWE, three different code scenarios (context) are provided for code generation.} The detailed introduction \revision{and preprocessing} of these datasets is presented in the corresponding subsections in Sec. \ref{sec:evaluation} \revision{(i.e., Sec. \ref{sec:FunctionallyCorrectCodeGeneration} and Sec. \ref{sec:securitycodegeneration})}.

\subsection{Methodology}\label{sec:methodology}
\noindent \textbf{Workflow.} The overall workflow of our study \revision{framework} is shown in Fig. \ref{fig:workflow}. \blackding{1} We construct a suitable prompt for the given \textit{LeetCode} problem or CWE scenario (i.e., one CWE's code scenario) and send the constructed prompt to \textit{ChatGPT}. \blackding{2} \textit{ChatGPT} generates a response based on the \revision{current round} provided prompt and the \revision{previous round} conversation context \revision{(first round has no previous round conversation context)}. We extract the code snippet by \textit{ChatGPT} between two triple backticks from the response. \blackding{3} For the generated code, we leverage \textit{LeetCode} online judgment to test its functional correctness, or we utilize \textit{CodeQL}~\cite{CodeQL} (with manual analysis) to detect CWE vulnerabilities. Here, we refer to them collectively as testing in Fig. \ref{fig:workflow}. If the testing result passes (e.g., pass all test cases or no vulnerability detected), the code generation process ends. \blackding{4} Otherwise, there are bugs (e.g., compile error) in the generated code snippet. If the (round) number in the conversation (i.e., dialog) with \textit{ChatGPT} does not exceed the round limit (e.g., the maximum round number of 5), we utilize the feedback provided from \textit{LeetCode} and \textit{CodeQL} to reconstruct a new prompt and input it to \ChatGPT for a new round of code generation (i.e., \revision{go back to \blackding{1} for} fixing). If the testing is consistently unpassed and the round number in the conversation exceeds the round limit, the code generation is considered failed. The entire process including multiple rounds in the conversation is called the multi-round \revision{(fixing)} process \revision{(one-round process with the maximum round number of 1 has no fixing property)}. The details of prompt construction, testing, and multi-round \revision{fixing} process are explained in \revision{the subsections} of Sec. \ref{sec:evaluation}.

\noindent \textbf{Principle of Prompt Design.} The goal of our prompt design is not to find the optimal prompt that maximizes \textit{ChatGPT}'s performance. Instead, our goal is to provide a reasonable prompt that simulates real-world usage scenarios, especially for code generation\revision{, which can also avoid overfitting to the specific prompts and datasets}. In developing the prompt template, we refer to online prompt templates (e.g., OpenAI Cookbook~\cite{openai-cookbook} and PromptBase~\cite{promptbase}) for code generation tasks and finally establish the following principle for prompt design: \textit{offering sufficient information to ChatGPT while leveraging its dialog ability}. 

\noindent \revision{\textbf{Subject LLM.} The default language model provided by OpenAI~\cite{OpenAI} for \textit{ChatGPT} is \textit{GPT-3.5}. This model contains 175 billion parameters, making it a highly capable and complex model. \textit{GPT-3.5} is engineered to handle a diverse range of natural language processing tasks, such as text generation, text completion, and other related tasks. In this study, we utilize the model version \textit{gpt-3.5-turbo-0301} of \textit{ChatGPT} for performing evaluation. We query \textit{ChatGPT} through a simple wrapper~\cite{chatgptwrapper} of OpenAI API~\cite{OpenAI} to easily control the dialog ability of \textit{ChatGPT}. The temperature of \textit{ChatGPT} is set to the default value of 0.7~\cite{ChatGPT} to simulate real-world usage scenarios. Furthermore, the token limitation of \textit{ChatGPT}~\cite{ChatGPT} is 4,096, which may influence the output from \textit{ChatGPT}. If the total length of the input prompt and the generated response exceeds this limitation, then the excess part is discarded and possibly produces incomplete code snippets, causing errors in the generated code. In our experiments, we impose strict length limitations on both the input prompt and the generated response. For each round in the multi-round process, we find that the current round prompt\footnote{Disregard the prompts and responses of previous rounds.} lengths and response lengths are all under 2,400 tokens and 800 tokens, respectively, which does not exceed \textit{ChatGPT}'s token limitation. Thus, for the one-round process (e.g., Sec. \ref{sec:FunctionallyCorrectCodeGeneration}), the outputs of \textit{ChatGPT} are not influenced by the token limitation problem. However, in the complete multi-round process, especially when performing code generation for \textit{LeetCode} problems, there are some cases where the token lengths used (include previous prompts, responses, and the current round prompt and response) can exceed the token limitation. To mitigate this issue when encountering the cases, we take a \textit{token-limitation} strategy of adding necessary information (e.g., \textit{LeetCode} problem descriptions) at the beginning of the current round prompt and remove as little of the beginning dialog content (in block granularity, i.e., one prompt or response) from the conversation as possible to keep the remaining token space for the response from \textit{ChatGPT} having at least 1000\footnote{In the first experiment in Sec. \ref{sec:FunctionallyCorrectCodeGeneration}, all lengths of responses are under 770. Thus, We slightly amplify 770 to 1000 as the length of remaining token space that should be guaranteed when generating responses.} in length. This strategy avoids missing the necessary details in tasks for \textit{ChatGPT}. Moreover, in our observation, the strategy guarantees that the generated code snippets are complete and at least ensures that the immediate previous round's response remains throughout the conversation such that \textit{ChatGPT} does not lose the most recent code generation-related information. The detailed introduction of this strategy is presented in Sec. \ref{sec:multiroundfixing}\footnote{The tasks in Sec. \ref{sec:FunctionallyCorrectCodeGeneration} and Sec. \ref{sec:securitycodegeneration} have no this issue. All token lengths used are lower than the token limitation.}.}

\subsection{Experiment Environment}
All experiments are conducted on a server with an Intel(R) Core(TM) i9-10900X CPU @ 3.70GHz (10 cores) and 128GB RAM. Its operating system is Ubuntu 20.04. The framework designed and scripts used in the experiments are developed in Python 3.10.9. \textit{CodeQL}~\cite{CodeQL} used is in version 2.12.2.

\section{Experiment and Evaluation}\label{sec:evaluation}

\subsection{Functionally Correct Code Generation}\label{sec:FunctionallyCorrectCodeGeneration}
\noindent \textbf{RQ1: Is the code generated by \textit{ChatGPT} functionally correct?}

\noindent \textbf{Motivation.}  Given an appropriate prompt, \textit{ChatGPT}~\cite{ChatGPT} is able to generate text consistent with the prompt based on knowledge learned. This ability may improve developer productivity~\cite{nguyen2022empirical, sobania2022choose, vaithilingam2022expectation, siddiq2022empirical}. In the first step, we focus on evaluating the ability of \textit{ChatGPT} to generate functionally correct code automatically \revision{in one-round process}.

\noindent \textbf{Approach.} We let \textit{ChatGPT} read the natural language description of the given problem to generate the corresponding code snippet \revision{in one-round process (i.e., the maximum round number is set to 1)}, and utilize the problems on \textit{LeetCode}~\cite{LeetCode} as our dataset. \textit{LeetCode} is an online platform that provides challenging coding problems and automatic judgment. At the time of writing, there are over 2,500 problems on \textit{LeetCode} with easy, medium, and hard levels, starting from the 2014 year. We collect all problems on \textit{LeetCode}, and divide them into two categories, problems before 2021 (Bef. problems) and problems after 2021 (Aft. problems), by using the time divider of 2022-01-01. Since \textit{ChatGPT}~\cite{ChatGPT} is trained on text data before 2021, Bef. problems and corresponding solutions may have a high probability to appear in its training set. This case may degenerate the code generation task for Bef. problems into querying code in the database (i.e., code reuse~\cite{haefliger2008code, bui2021infercode, zhang2019novel}). Code reuse is a commonly used software development practice that avoids creating new code from scratch (e.g., copy-paste). Therefore, we take both problems into account.

Specifically, we focus on Algorithm problems\footnote{https://leetcode.com/problemset/algorithms/} on \textit{LeetCode} \revision{since Algorithm problems are the most significant, numerous, and diverse problems on the platform}. The total numbers of Bef. problems and Aft. problems are 1,624 and 354, respectively. \revision{Furthermore, the difficulty level distribution to both of them is in the ratio of $1:2:1$ for hard, medium, and easy problems.} Among all the Bef. problems, we sample 374 of them randomly, having similar quantities to the Aft. problems and following the same difficulty level distribution as Aft. problems. \revision{The ratio of the numbers of hard, medium, and easy problems is also $1:2:1$ for both 354 Aft. problems and 374 Bef. problems, consistent with the difficulty level distribution of all problems on the \textit{LeetCode} platform.} \revision{Additionally, we also check if there are significant differences between Bef. problems and Aft. problems. If Aft. problems are just reformulations of Bef. problems, \textit{ChatGPT} may likely be able to easily solve them, which can affect the reliability of the experiment results in distinguishing between time periods. Specifically, we first use the "similar questions" provided for each problem on the \textit{LeetCode} platform to find similar problem pairs of Bef. problems and Aft. problems. The "similar questions"~\cite{LeetCode} represent two paired problems that have similar scenarios (e.g., processing string) or require using similar algorithms for solving (e.g., dynamic programming). In total, there are 142 pairs found. Then, we have two graduate students independently and manually check these problem pairs. Through a careful checking and discussion process, we find that these similar problems are either having similar scenarios but completely different solution goals, or different scenarios and conditions but can be solved using similar algorithms such as dynamic programming. After a careful manual analysis, we do not find any cases that Bef. problems can be easily reformulated to obtain Aft. problems. Thus, we consider Aft. problems and Bef. problems to be sufficiently different.} \revision{Moreover, }for each problem, we ask \textit{ChatGPT} to generate code in five different languages: C, C++, Java, Python3, and JavaScript. Moreover, we create a corresponding prompt using the same prompt template for each \textit{<problem, language>} pair. In total, there are 1,870 and 1,770 prompts for Bef. problems and Aft. problems, respectively. Due to the rate-limiting of queries to \textit{ChatGPT}, we input every prompt once into it to ask for generating code. Then, we submit parsed solutions to \textit{LeetCode} for functional correctness judgment and get submission statuses~\cite{LeetCode} including \textit{Accepted}, \textit{Wrong Answer}, \textit{Compile Error}, \textit{Time Limit Exceeded}, and \textit{Runtime Error}. They correspond to A., W.A., C.E., T.L.E., and R.E., respectively. One problem corresponds to one unique conversation to avoid triggering \textit{ChatGPT}'s reasoning from other problems. The status explanations are as follows:

\begin{itemize}
    \item \textit{Accepted}: The submitted code snippet passes all test cases.

    \item \textit{Wrong Answer}: The submitted code snippet has no compile errors but cannot pass all test cases.

    \item \textit{Compile Error}: The submitted code snippet cannot be compiled.

    \item \textit{Time Limit Exceeded}: The runtime of the submitted code snippet exceeds the permitted execution time.

    \item \textit{Runtime Error}: The execution of the submitted code snippet triggers a runtime error for at least one test case.
\end{itemize}

\noindent Note that code with W.A. does not necessarily mean it does not contain R.E. or T.L.E.. \revision{Several errors can occur at the same time.} Nevertheless, we study the functional correctness of code generation. Therefore, we take their priorities as \revision{C.E., R.E. $>$ W.A. $>$ T.L.E.} by default, meaning that we only focus on the judgment results returned by \textit{LeetCode} and this processing method has a negligible impact on the experimental conclusions. \revision{We prioritize C.E. and R.E. because these two errors lead to code running failures, which also implies W.A.. T.L.E. is set to the lowest priority because it mainly relates to non-functional requirements.} Moreover, \textit{LeetCode} online judgment platform terminates the testing process upon encountering the first failed test case. Thus, the test case pass rates \revision{(the percentage of predefined test cases that a submitted code snippet successfully passes)} provided by the platform may serve as a lower bound.

We evaluate \textit{ChatGPT}'s ability of code generation on the metric of status rate (SR) defined as follows:

\begin{equation}
    \label{equ:success-rate}
    SR = \frac{N_c}{N_i} \times 100\%
\end{equation}

Where, $N_c$ and $N_i$ are the number of code snippets generated belonging to the status and the number of prompts input, respectively. Status is either A., W.A., C.E., T.L.E., or R.E.. The deep analysis for code with W.A., C.E., T.L.E., or R.E. is presented in Sec. \ref{sec:multiroundfixing}.

We also conduct Wilcoxon rank-sum test~\cite{fay2010wilcoxon} and Cliff's Delta effect size measure~\cite{macbeth2011cliff} to compare two independent samples and determine whether there are significant differences between them and quantify the magnitude of the differences observed between the two independent samples. \revision{The null hypothesis for the Wilcoxon rank-sum test is that there is no significant difference between the two samples where the samples are the combinations of SR values in different conditions (e.g., A. rate values of five languages in different period problems).} If the obtained p-value from Wilcoxon rank-sum test is small (less than 0.05), it suggests that there is a statistically significant difference between the two independent samples. \revision{In cases of multiple comparisons, we apply Holm-Bonferroni correction~\cite{holm1979simple}, a commonly used technique, to adjust p-values to reduce the risk of Type I errors.} The absolute value of effect size (effect size value) obtained from Cliff's Delta measure ranges from 0 to 1. A value close to 0 indicates a small effect, meaning that there is minimal difference between the two independent samples, and a value close to 1 indicates a substantial effect size, meaning that there are significant differences between them. By combining the results from the Wilcoxon rank-sum test and Cliff's Delta, we can gain a comprehensive insight into the differences in code generation results, allowing us to draw more robust conclusions.


\begin{figure}[t]
    \centering
    \includegraphics[width=0.4\textwidth]{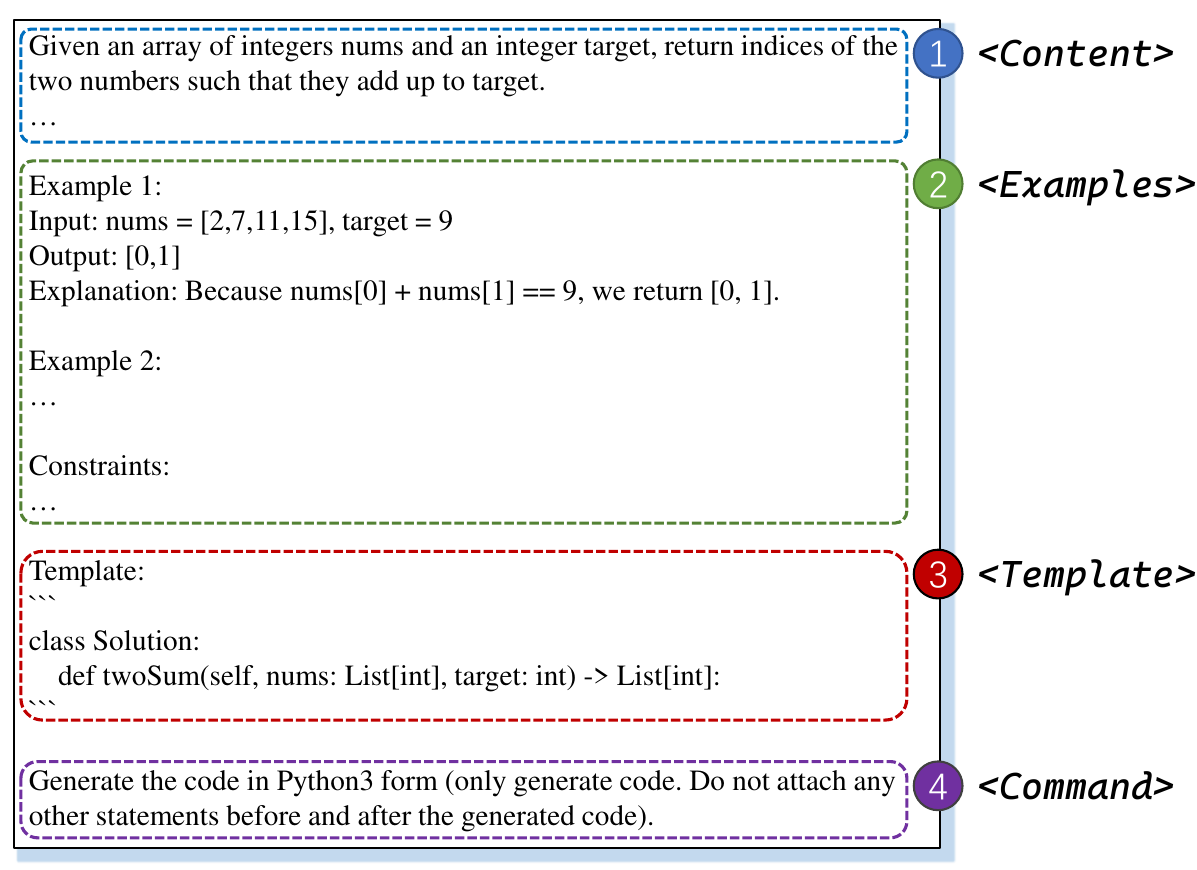}
    \vspace{-1em}
    \caption{An example of prompt for two sum problem in Python3.}
    \label{fig:chatgpt-Prompt-two-sum}
\end{figure}

\noindent \textbf{Prompt.} The prompt template designed consists of 4 components. They are \textit{<Content>}, \textit{<Examples>}, \textit{<Template>}, and \textit{<Command>}, aligning with the principle of prompt design (see Sec. \ref{sec:methodology}). Fig \ref{fig:chatgpt-Prompt-two-sum} shows an example of a prompt. \textit{<Content>} describes the problem in nature language, \textit{<Examples>} shows \textit{<input, output>} pairs of functionally correct code, \textit{<Template>} specifies the method signature of generated code, and \textit{<Command>} asks for generating code in a specific language.

\begin{table*}[ht]
\centering
\caption{Code-judged Result in C, C++, and Java Languages}
\renewcommand\arraystretch{1.3}
\scalebox{0.7}{\begin{tabular}{c|c|rrrrr|rrrrr|rrrrr} 
\toprule
\multirow{2}{*}{Period} & \multirow{2}{*}{Difficulty} & \multicolumn{5}{c|}{C}   & \multicolumn{5}{c|}{~C++~} & \multicolumn{5}{c}{Java}          \\ 
\cline{3-17}
                        &                             & \multicolumn{1}{c}{A.}  & \multicolumn{1}{c}{W.A.} & \multicolumn{1}{c}{C.E.} & \multicolumn{1}{c}{T.L.E.} & \multicolumn{1}{c|}{R.E.} & \multicolumn{1}{c}{A.} & \multicolumn{1}{c}{W.A.} & \multicolumn{1}{c}{C.E.} & \multicolumn{1}{c}{T.L.E.} & \multicolumn{1}{c|}{R.E.} & \multicolumn{1}{c}{A.} & \multicolumn{1}{c}{W.A.} & \multicolumn{1}{c}{C.E.} & \multicolumn{1}{c}{T.L.E.} & \multicolumn{1}{c}{R.E.} \\ 
\midrule
\multirow{4}{*}{Aft.}   & Hard                        & \percent[g]{0}{91}   & \percent[g]{57}{91}   & \percent[g]{25}{91}   & \percent[g]{2}{91}      & \percent[g]{7}{91}    & \percent[g]{0}{91}                    &   \percent[g]{57}{91}                &      \percent[g]{24}{91}                  &       \percent[g]{7}{91}               &         \percent[g]{3}{91}                 &  \percent[g]{0}{90}                     &    \percent[g]{64}{90}                    &     \percent[g]{16}{90}                   &   \percent[g]{5}{90}                        & \percent[g]{5}{90}                   \\

                        & Medium                      & \percent[g]{10}{170}  & \percent[g]{77}{170}   & \percent[g]{49}{170}   & \percent[g]{12}{170}     & \percent[g]{22}{170}     &   \percent[g]{23}{172}                   &  \percent[g]{99}{172}                      &  \percent[g]{23}{172}                      & \percent[g]{12}{172}                         & \percent[g]{15}{172}                        &   \percent[g]{26}{172}                   &  \percent[g]{119}{172}                     &  \percent[g]{13}{172}                      &  \percent[g]{10}{172}                        &  \percent[g]{4}{172}                   \\

                        & Easy                        & \percent[g]{44}{90}  & \percent[g]{33}{90}   & \percent[g]{9}{90}   & \percent[g]{0}{90}      & \percent[g]{4}{90}   &   \percent[g]{45}{88}                   &  \percent[g]{40}{88}                      &  \percent[g]{1}{88}                       &   \percent[g]{1}{88}                        &   \percent[g]{1}{88}                       & \percent[g]{45}{90}                     & \percent[g]{37}{90}                       & \percent[g]{4}{90}                        & \percent[g]{2}{90}                         & \percent[g]{2}{90}               \\ 
\cline{2-17}

                        & Total                       & \percent[g]{54}{351}  & \percent[g]{167}{351}  & \percent[g]{83}{351}   & \percent[g]{14}{351}     & \percent[g]{33}{351}    & \percent[g]{68}{351}                     & \percent[g]{196}{351}                      & \percent[g]{48}{351}                       & \percent[g]{20}{351}                         & \percent[g]{19}{351}                        & \percent[g]{71}{352}                     & \percent[g]{220}{352}                      & \percent[g]{33}{352}                       & \percent[g]{17}{352}                         & \percent[g]{11}{352}            \\

\midrule

\multirow{4}{*}{Bef.}   & Hard                        & \percent[g]{17}{93}  & \percent[g]{36}{93}   & \percent[g]{31}{93}   & \percent[g]{3}{93}      & \percent[g]{6}{93}     & \percent[g]{40}{94}                     & \percent[g]{18}{94}                       & \percent[g]{30}{94}                       & \percent[g]{2}{94}                          & \percent[g]{4}{94}                         & \percent[g]{47}{89}                     & \percent[g]{24}{89}                       & \percent[g]{12}{89}                       & \percent[g]{2}{89}                          & \percent[g]{4}{89}                    \\

                        & Medium                      & \percent[g]{80}{172}  & \percent[g]{35}{172}   & \percent[g]{36}{172}   & \percent[g]{5}{172}      & \percent[g]{16}{172}     & \percent[g]{125}{177}                    & \percent[g]{25}{177}                       & \percent[g]{21}{177}                       & \percent[g]{0}{177}                          & \percent[g]{6}{177}                         & \percent[g]{136}{174}                    & \percent[g]{21}{174}                       & \percent[g]{8}{174}                        & \percent[g]{6}{174}                          & \percent[g]{3}{174}                     \\

                        & Easy                        & \percent[g]{74}{97}  & \percent[g]{9}{97}    & \percent[g]{6}{97}    & \percent[g]{0}{97}      & \percent[g]{8}{97}   & \percent[g]{91}{102}                     & \percent[g]{7}{102}                        & \percent[g]{2}{102}                        & \percent[g]{1}{102}                          & \percent[g]{1}{102}                         & \percent[g]{95}{101}                     & \percent[g]{4}{101}                        & \percent[g]{2}{101}                        & \percent[g]{0}{101}                          & \percent[g]{0}{101}         \\ 
\cline{2-17}

                        & Total                       & \percent[g]{171}{362} & \percent[g]{80}{362}   & \percent[g]{73}{362}   & \percent[g]{8}{362}      & \percent[g]{30}{362}     & \percent[g]{256}{373}                    & \percent[g]{50}{373}                       & \percent[g]{53}{373}                       & \percent[g]{3}{373}                          & \percent[g]{11}{373}                        & \percent[g]{278}{364}                    & \percent[g]{49}{364}                       & \percent[g]{22}{364}                       & \percent[g]{8}{364}                          & \percent[g]{7}{364}                        \\

\midrule
-                       & Total                       & \percent[g]{223}{713} & \percent[g]{247}{713}  & \percent[g]{156}{713}  & \percent[g]{22}{713}     & \percent[g]{63}{713}  & \percent[g]{324}{724}                    & \percent[g]{246}{724}                      & \percent[g]{101}{724}                      & \percent[g]{23}{724}                         & \percent[g]{30}{724}                        & \percent[g]{349}{716}                    & \percent[g]{269}{716}                      & \percent[g]{55}{716}                       & \percent[g]{25}{716}                         & \percent[g]{18}{716}         \\
\bottomrule
\end{tabular}}
\label{tab:generation-1}
\end{table*}

\begin{table*}[ht]
\centering
\caption{Code-judged Result in Python3 and JavaScript Languages}
\renewcommand\arraystretch{1.3}
\scalebox{0.7}{\begin{tabular}{c|c|rrrr|rrrr} 
\toprule
\multirow{2}{*}{Period} & \multirow{2}{*}{Difficulty}                                                                                                              & \multicolumn{4}{c|}{Python3}                                                                               & \multicolumn{4}{c}{JavaScript}                                                                             \\ 
\cline{3-10}
                        &                              & \multicolumn{1}{c}{A.} & \multicolumn{1}{c}{W.A.} & \multicolumn{1}{c}{T.L.E.} & \multicolumn{1}{c|}{R.E.} & \multicolumn{1}{c}{A.} & \multicolumn{1}{c}{W.A.} & \multicolumn{1}{c}{T.L.E.} & \multicolumn{1}{c}{R.E.}  \\ 
\midrule
\multirow{4}{*}{Aft.}   & Hard                 & \percent[g]{2}{91}                      & \percent[g]{73}{91}                       & \percent[g]{11}{91}                         & \percent[g]{5}{91}                         & \percent[g]{1}{91}                      & \percent[g]{76}{91}                       & \percent[g]{5}{91}                          & \percent[g]{9}{91}                        \\
                        & Medium                     &   \percent[g]{31}{171}                   &     \percent[g]{105}{171}                   &     \percent[g]{19}{171}                      &   \percent[g]{16}{171}                       &  \percent[g]{29}{171}                    &  \percent[g]{116}{171}                     &  \percent[g]{17}{171}                        &  \percent[g]{9}{171}                       \\
                        & Easy                      & \percent[g]{51}{89}                     & \percent[g]{31}{89}                       & \percent[g]{2}{89}                          & \percent[g]{5}{89}                         & \percent[g]{49}{89}                     &  \percent[g]{34}{89}                      & \percent[g]{1}{89}                          & \percent[g]{5}{89}                         \\ 
\cline{2-10}
                        & Total                                   &  \percent[g]{84}{351}                    & \percent[g]{209}{351}                      & \percent[g]{32}{351}                         & \percent[g]{26}{351}                        & \percent[g]{79}{351}                     & \percent[g]{226}{351}                      & \percent[g]{23}{351}                         & \percent[g]{23}{351}                        \\ 
\midrule

\multirow{4}{*}{Bef.}   & Hard                             & \percent[g]{41}{90}                     & \percent[g]{30}{90}                       & \percent[g]{8}{90}                          & \percent[g]{11}{90}                        & \percent[g]{38}{90}                     & \percent[g]{42}{90}                       & \percent[g]{2}{90}                          & \percent[g]{8}{90}                         \\
                        & Medium                         &  \percent[g]{136}{172}                   & \percent[g]{28}{172}                       & \percent[g]{2}{172}                          & \percent[g]{6}{172}                         & \percent[g]{136}{169}                    & \percent[g]{22}{169}                       & \percent[g]{3}{169}                          & \percent[g]{8}{169}                        \\
                        & Easy                                          & \percent[g]{95}{99}                     & \percent[g]{4}{99}                        & \percent[g]{0}{99}                          & \percent[g]{0}{99}                         & \percent[g]{94}{101}                     & \percent[g]{6}{101}                        & \percent[g]{0}{101}                          & \percent[g]{1}{101}                         \\ 
\cline{2-10}
                        & Total                       &  \percent[g]{272}{361}                   & \percent[g]{62}{361}                       & \percent[g]{10}{361}                          & \percent[g]{17}{361}                        & \percent[g]{268}{360}                    & \percent[g]{70}{360}                       & \percent[g]{5}{360}                          & \percent[g]{17}{360}                        \\ 
\midrule
-                       & Total                                        & \percent[g]{356}{712}                    & \percent[g]{271}{712}                      & \percent[g]{42}{712}                         & \percent[g]{43}{712}                        & \percent[g]{347}{711}                    &  \percent[g]{296}{711}                     &  \percent[g]{28}{711}                        &   \percent[g]{40}{711}                      \\
\bottomrule
\end{tabular}}
\label{tab:generation-2}
\end{table*}

\noindent \textbf{Result.} Table \ref{tab:generation-1} and \ref{tab:generation-2} show code generation results judged by \textit{LeetCode} for five languages in two periods and in two forms, SR, and corresponding relative frequency bar chart. Columns of Python3 and JavaScript contain no C.E. since both of them are dynamic programming languages. From the overall results, \textit{ChatGPT} generates functionally correct code for Bef. problems at a significantly higher A. rate than Aft. problems. Specifically, the average A. rate ($68.41\%$) in five languages of Bef. problems exceeds Aft. problems' ($20.27\%$) by $48.14\%$. The performance in five languages of code generation in different periods is significantly different with a p-value of 0.008 and an effect size value of 1.

\begin{figure}[t]
    \centering
  \begin{lstlisting}[language=C]
int maximumGroups(int* grades, int gradesSize){
    ...
    qsort(grades, gradesSize, sizeof(int), cmpfunc); // sort grades in ascending order
    ...
}

int cmpfunc (const void * a, const void * b) {...}
  \end{lstlisting}
  \caption{Function in C code generated by \textit{ChatGPT} is not declared before invocation.}
  \label{fig:function-without-declared}
\end{figure}

\begin{figure}[t]
    \centering
    \includegraphics[width=0.4\textwidth]{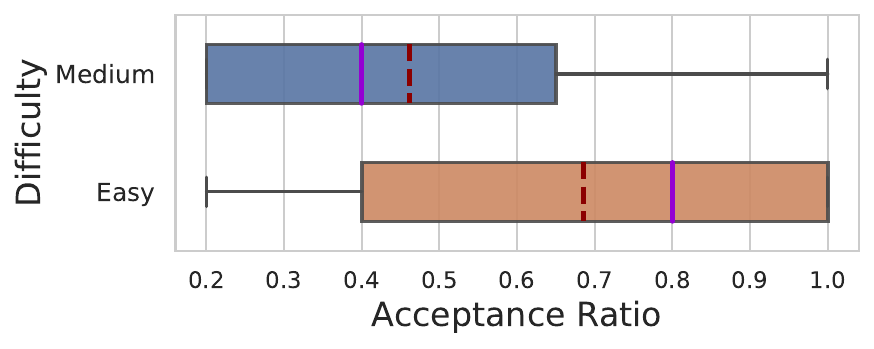}
    \vspace{-1em}
    \caption{Distribution of the ratios of languages accepted to corresponding Aft. problems. Where dark violet and dark red lines represent the median and mean, respectively.}
    \label{fig:af-l-p-distribution}
\end{figure}

\noindent $\bullet$ \textbf{Aft. Problems.} For Aft. problems, the overall A. rate is lower than $25\%$, where the A. rates of hard, medium, and easy problems are $0.66\%$, $13.90\%$, and $52.47\%$, respectively. The p-values \revision{adjusted using Holm-Bonferroni correction procedure} and effect size values between different difficulties in five languages are all less than 0.05 and equal to 1, respectively. The result indicates that \textit{ChatGPT}'s ability to functionally correct code generation decreases significantly as the difficulty of the problem increases in the face of Aft. problems. Additionally, even for easy problems, it is only able to answer half of them correctly. Out of these five/four metrics, the W.A. rate is the highest one reaching $58\%$ for all languages. Moreover, each W.A. \revision{code snippet} has an average of 109 test cases, however, the code generated by \textit{ChatGPT} can pass only $25\%$ of them. Hard, medium, and easy problems achieve $20.90\%$, $21.03\%$, and $38.41\%$ test case pass rates, respectively. Thus, regardless of the difficulty, the semantics of the code generated differs significantly from the logic of the corresponding problem descriptions. In addition, the C.E. rate and R.E. rate also reach $16\%$, and hard and medium problems' rates are significantly higher than easy problems. The code generated by \textit{ChatGPT} for hard and medium problems is more likely to contain both compile and runtime errors. The compile errors include undeclared variable, function declaration error, uninitialized variable, constant function (i.e., generate an empty body), and so on. For example, Fig. \ref{fig:function-without-declared} shows that the generated function \texttt{cmpfunc} is not declared before invocation. The syntax errors account for only a small fraction ($3.7\%$) of these errors. For runtime errors, there are null pointer dereference, out-of-bound, heap-buffer-overflow, type error, and so on, which are common in human-written code. As for T.L.E. rate, it does not dominate a high value ($6\%$), but the average pass rate of test cases is $51\%$ which is higher than W.A. \revision{code snippets'}. The average \revision{test case} pass rates of three difficulty \revision{levels in hard, medium, and easy} of T.L.E. \revision{problems} are $68\%$, $50\%$, and $1\%$ (easy problems can be neglected due to their T.L.E. rate close to $0\%$), respectively. Since T.L.E. \revision{code snippets'} \revision{test case} pass rate is partial, it is the lower bound for these problems, and at most, an additional $6\%$ of the generated code can be functionally correct, even though their time complexity may not be ideal. 

Breaking down to each language, language C, C++, Java, Python3, and JavaScript have A. rates of $15.38\%$, $19.37\%$, $20.17\%$, $23.93\%$, and $22.51\%$, respectively. Moreover, the A. rate distributions (acceptance ratio distributions) of combining five different languages to each problem (only consider problems have at least one correct solution) are shown in Fig. \ref{fig:af-l-p-distribution}. From the figure, we can see that both medium's mean and median lines are $\leq 0.5$, and easy's ones are all $\geq 0.6$. \textit{ChatGPT} is easier to generalize code generated to different languages for easy problems. The differences between easy and medium's median and mean are 0.4 and 0.22, respectively. Moreover, the average acceptance rate in humans for the problems accepted by \textit{ChatGPT} is $66\%$, and the one for the problems unaccepted by \textit{ChatGPT} is $48\%$. \textit{ChatGPT} has similarities with human performance. In addition, functionally correct code's runtime and memory overheads are excellent in the human ranking, averaging over $68\%$ and $51\%$ solutions, respectively. 


\begin{figure}[t]
    \centering
  \begin{lstlisting}[language=C]
int** matrixReshape(int** mat, int matSize, int* matColSize, int r, int c, int* returnSize, int** returnColumnSizes){
    // calculate the total number of elements...
    int totalElements = matSize * matColSize[0];
    // check the number of elements...
    if (r * c != totalElements) {...}
    // allocate memory for the new matrix
    int **newMat = (int **)malloc(sizeof(int *) * r);
    for (int i = 0; i < r; i++) {
        newMat[i] = (int *)malloc(sizeof(int) * c);
    }
    // copy the elements from the original matrix...
    ...
    // set the return values
    *returnSize = r;
    *returnColumnSizes = (int *)malloc(sizeof(int));
    (*returnColumnSizes)[0] = c;
    return newMat;
}
  \end{lstlisting}
  \caption{Code generated by \textit{ChatGPT} triggers heap-buffer-overflow runtime error.}
  \label{fig:code-heap-buffer-overflow}
\end{figure}

\begin{figure}[t]
    \centering
    \includegraphics[width=0.4\textwidth]{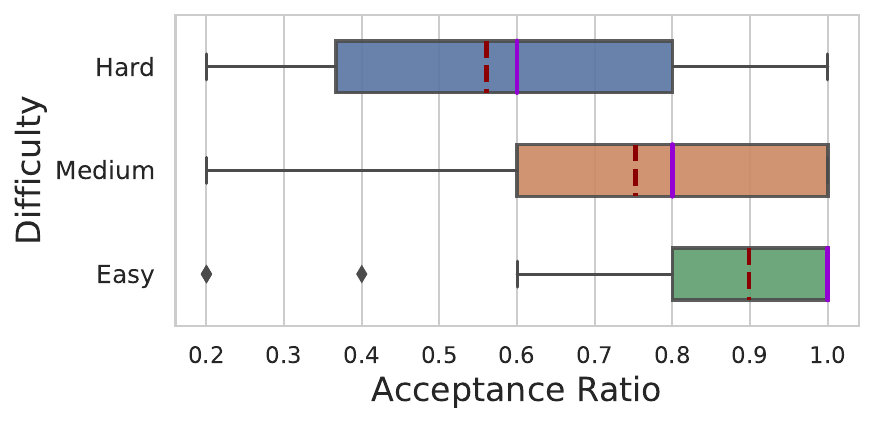}
    \caption{Distribution of the ratios of languages accepted to corresponding Bef. problems (the meaning of two lines is presented in Fig. \ref{fig:af-l-p-distribution}).}
    \vspace{-1em}
    \label{fig:be-l-p-distribution}
\end{figure}

\noindent $\bullet$ \textbf{Bef. Problems.} As for Bef. problems, the A. rates of hard, medium, and easy problems are $40.13\%$, $70.95\%$, and $89.80\%$, respectively, which are much higher than Aft. problems' though there still exist significant differences among different difficulties. The p-values \revision{adjusted using Holm-Bonferroni correction procedure} and effect size values between hard and medium and hard and easy difficulties in five languages are all less than 0.05 and greater than 0.9, respectively. The \revision{adjusted} p-value and effect size value between medium and easy difficulties in five languages are 0.056 and 0.76, respectively. \textit{ChatGPT} performs better against problems that may appear in the training set before 2021, especially for medium and easy problems. The A. rate of solving hard problems has increased by $40\%$ but is still below $50\%$, which indicates \textit{ChatGPT}'s ability to generate code for logically complex problems still has a big room for improvement. The overall W.A. rate decreases to $17.03\%$, and hard, medium, and easy problems' W.A. rates are $32.89\%$, $15.05\%$, and $6\%$, respectively. The code generated can still pass $25\%$ of average 112 test cases. Hard, medium, and easy problems achieve $19.19\%$, $31.12\%$, and $47.32\%$ test case pass rates, respectively. Both the latter two have a $10\%$ improvement, which indicates that \textit{ChatGPT} has a better understanding of Bef. problems. However, the C.E. rate and R.E. rate still reach $13\%$ close to Aft. problems' $16\%$ with a p-value and effect size value, between two periods, of 0.328 and 0.3125, respectively, and hard problems have the highest rate, followed by the medium ones. The compile errors and runtime errors are similar to Aft. problems' including undeclared variable, uninitialized variable, null pointer dereference, out-of-bounds, heap-buffer-overflow, type error, and so on. For example, the code shown in Fig. \ref{fig:code-heap-buffer-overflow} is used to reshape a given 2-dimensional matrix but triggers runtime errors at line 15 that allocates a wrong size of memory to \texttt{*returnColumnSizes}. To T.L.E. rate, the value decreases to $1.87\%$ with an average of $74\%$ test case pass rate. 

Breaking down to each language, C, C++, Java, Python3, and JavaScript have A. rates of $47.24\%$, $68.63\%$, $76.37\%$, $75.35\%$, and $74.44\%$, respectively. The rate values of the last four languages are close to each other and substantially higher than the rate value of C, the lowest-level language, for at least $20\%$. Fig. \ref{fig:be-l-p-distribution} shows the same as Fig. \ref{fig:af-l-p-distribution} but for Bef. problems. From the figure, we can see that medium and easy's mean and median lines are $\geq 0.75$, and the differences between their median and mean are smaller than previous ones of Aft. problems by half. Moreover, hard's mean and median lines are both $\geq 0.55$. \textit{ChatGPT} is easier to generalize code to different languages for Bef. problems. The average acceptance rate in humans for the problems accepted by \ChatGPT is $55\%$, and the one for the problems unaccepted by \ChatGPT is $47\%$. Functionally correct code’s runtime and memory overheads are also excellent in the human ranking, averaging over $71\%$ and $54\%$ solutions, respectively. 


\revision{We also sample 50 problems from all problems (25 Aft. problems and 25 Bef. problems, and each problem has 5 solutions in 5 different languages) to investigate how many times \textit{ChatGPT} accurately generates the exact solutions (token-by-token) to ground truth solutions. We collect 5 distinct ground truth solutions to each \textit{ChatGPT}-generated solution and the ground truth solutions are obtained from the LeetCode platform and \cite{leetcodesolution}\footnote{It is notable that the analysis result is a lower bound since it is impossible to check all the ground truth for each solution.}. By our manual analysis, we find that none of the solutions are generated on a token-by-token basis. However, for Bef. problems we find that 6 solutions in easy and medium difficulties are Type-2 Clone (i.e., have some renamed unique identifiers)~\cite{hu2023code2img} to ground truth solutions. This result indicates that \textit{ChatGPT} may have a certain probability (>5\%) of reproducing similar solutions from the training set for Bef. problems, especially in easy and medium difficulties.} 

\begin{tcolorbox}[boxrule=1pt,boxsep=1pt,left=2pt,right=2pt,top=2pt,bottom=2pt,title=Answer to RQ1: Functionally Correct Code Generation]

\noindent \blackding{1} \textit{ChatGPT} is better at generating functionally correct code for Bef. problems in different languages than Aft. problems. Specifically, the average A. rate of the former exceeds the one of the latter by $48.14\%$. Additionally, different levels of difficulty also have an impact on \textit{ChatGPT}-based code generation; 

\noindent \blackding{2} For both problems, \textit{ChatGPT} is able to generate code with smaller runtime and memory overheads than at least $50\%$ human solutions;

\noindent \blackding{3} Regardless of the periods of problems, \textit{ChatGPT} has a similar probability of $14.23\%$ on average to generate code with compile or runtime errors; and

\revision{\noindent \blackding{4} The A. rate values of C++, Java, Python3, and JavaScript with $44.75\%$, $48.74\%$, $50.00\%$, and $48.80\%$ are close to each other and substantially higher than the rate value of C with $31.28\%$ for all problems.}
\end{tcolorbox} 


\subsection{Multi-round Fixing for Code Generation}\label{sec:multiroundfixing}
\noindent \textbf{RQ2: How effective is the multi-round fixing process in improving code generation \revision{for functional correctness}?}

\noindent \textbf{Motivation.} \ChatGPT supports multiple rounds of conversations (i.e., dialog ability), and users can use this feature to continuously generate code to get functionally correct code in one conversation. In this RQ, we study the multi-round \revision{fixing} process for fixing code snippets with W.A., C.E., R.E., or T.L.E. errors.

\noindent \textbf{Approach.} We ask \textit{ChatGPT} in multiple rounds of conversation to fix the error code snippets. Moreover, before analyzing the results of the multi-round fixing process, we also further \revision{manually analyze} the causes of each category of errors including W.A., C.E., T.L.E., and R.E. for the code generated by \textit{ChatGPT} for a better understanding. \revision{The exact procedures for analyzing different categories of errors are shown in the \textit{\textbf{Analyzing}} part in each following subsection.}

\noindent \textbf{Result.} We first analyze the corresponding errors and then apply multi-round fixing for code generation (multi-round code generation).

\begin{table}[t]
\centering
\caption{The Statistics of 157 \textit{<problem, language>} Pairs}
\begin{tabular}{lcccc} 
\toprule

   & \textbf{Hard} & \textbf{Medium} & \textbf{Easy}  & \textbf{Total} \\

\midrule
   
 C & 5 & 12 & 5  & 22 \\
 C++ & 5 & 17 & 7 & 29 \\  
 Java &  6 & 22 & 5 & 33 \\
 Python3 &  8 & 16 & 8 & 32 \\
JavaScript &  9 & 22 & 10 & 41 \\

\midrule

Total & 33  & 89 & 35 & 157 \\

\midrule

Problem & 12  & 25 & 13 & 50 \\

\bottomrule
\end{tabular}
\label{tab:code-wa-statistics}
\end{table}

\begin{table*}[t]
\centering
\caption{Defect Classification of the 157 \textit{<problem, language>} Pairs}
\scalebox{0.85}{\begin{tabular}{l|l|l|c|c|c|c|c}

\toprule

\multicolumn{1}{c|}{\textbf{Defect Class}} & \multicolumn{1}{c|}{\textbf{Subclass}} & \multicolumn{1}{c|}{\textbf{Definition}} & \textbf{Logic} & \textbf{Hard} &  \textbf{Medium} & \textbf{Easy} & \textbf{Total} \\

\midrule

\multirow{4}{*}[-6ex]{Multi-hunk} & {Similar (M-S)} &\multicolumn{1}{m{7cm}|}{{Similar single-hunk defects are at multiple discontinuous locations of code snippets.}} & WD, MCC &{0} & {2}& {4}& {6} \\ 

\cline{2-8}

& {Unique (M-U)} &\multicolumn{1}{m{7cm}|}{{Diverse single-hunk defects are at multiple discontinuous locations of code snippets, and the total lines of fixes are no more than five lines.}} &WD, MCC &{0} & {3}& {6}&{9}  \\

\cline{2-8}

& {Need Large Fix (M-L)} &\multicolumn{1}{m{7cm}|}{{The defect is neither M-S nor M-U and needs to edit more than five lines at multiple locations of code snippets.}} &MCC &{9} & {32}& {10}&{51}  \\

\cline{2-8}

& {Code Block (M-B)} &\multicolumn{1}{m{7cm}|}{{The fix of defect needs moving or exchanging code blocks with few (<5) additional statements inserted, updated, or deleted.}} & WD, MCC &{0}& {2}& {0} & {2} \\

\midrule

\multirow{6}{*}[-5ex]{Single-hunk} & {Operator Mutation (S-O)} &\multicolumn{1}{m{7cm}|}{{Replace arithmetic/logical/relational/bitwise operator with another operator or insert/delete operators and relevant operands or modify operator precedence.}} & WD &{0} & {2} & {6} & {8} \\ 
\cline{2-8}

& {Array Mutation (S-A)} &\multicolumn{1}{m{7cm}|}{{Replace the array access with other constant/variable, operands with arithmetic operators, or replace an array with another array.}} &WD &{0} & {1}& {0}&{1}  \\
\cline{2-8}
& {Function Call Mutation (S-F)} &\multicolumn{1}{m{7cm}|}{{Replace function call with another function call or change function arguments.}} &WD &{0} & {0}& {1}&{1}  \\
\cline{2-8}
& {Add Statements (S-AS)} &\multicolumn{1}{m{7cm}|}{{Insert a continuous chunk of statements.}} &WD &{0} & {0}& {1}&{1}  \\
\cline{2-8}
& {Delete Statements (S-DS)} &\multicolumn{1}{m{7cm}|}{{Delete a continuous chunk of statements.}} &WD &{0} & {1}& {0}&{1}  \\
\cline{2-8}
& {Higher Order (S-HO)} &\multicolumn{1}{m{7cm}|}{{A single-hunk patch that combines multiple single-hunk bugs.}} &WD &{0} & {13}& {2}&{15}  \\

\midrule

\multirow{1}{*}{Algorithm-related} & {Misaligned Algorithm} &\multicolumn{1}{m{7cm}|}{{The algorithm used is misaligned with the requirement given in the problem description.}} & MP &{24} & {33}& {5}& {62} \\

\midrule

Total & \multicolumn{1}{c|}{-} & \multicolumn{1}{c|}{-} & - &{33} & {89}&{35} & {157}  \\ 

\bottomrule

\end{tabular}}
\label{tab:code-wa-analysis}
\end{table*}

\subsubsection{Code with \textit{Wrong Answer}}\label{sec:code-with-wrong-answer}
 \noindent \textcolor{red}{$\blacktriangleright$} \textit{\textbf{Analyzing:}} We randomly choose 50 problems with W.A. code snippets to analyze the causes of \textit{Wrong Answer}. In total, there are 157 \textit{<problem, language>} pairs and among these 50 problems, there is an average of 3.14 pairs per problem that are in W.A.. The basic statics of these pairs are shown in Table \ref{tab:code-wa-statistics}. To analyze the causes, two graduate students with experience in algorithm analysis manually check each W.A. code snippet and assign each of them to one defect class. During this process, if there is any disagreement, the two students further discuss it with a senior software analyst to resolve the disagreement. Out of these 157 pairs, 115 pairs receive consistent classifications and the consistency ratio is 0.7325. The remaining 42 pairs reach a unanimous classification through discussion. 

We utilize the defect categories from the fixing perspective used in \cite{tan2017codeflaws} (CodeFlaws) and \cite{fan2022automated}. \cite{tan2017codeflaws} analyzes the programs submitted in Codeforces~\cite{codeforces} and classifies the defects in these programs into multiple classes, and \cite{fan2022automated} follows the defect classification in \cite{tan2017codeflaws} to construct the code defects generated by \textit{Codex}~\cite{chen2021evaluating}. The defect classes, definitions, and classified results are shown in Table \ref{tab:code-wa-analysis}. As we can see, most of the defects fall under the M-L subclass of Multi-hunk and Misaligned Algorithm subclass of Algorithm-related with 51 and 62, respectively. The other subclasses of defects are relatively much less, especially for S-A, S-F, S-AS, and S-DS which are all 1. From the perspective of difficulty, hard problems require more fixing work than medium and easy ones since hard problems have only M-L and Misaligned Algorithm while medium and easy ones have other subclasses. Moreover, during our manual analysis of these code snippets, we also find that the causes of errors from logic perspective could be divided into three categories as follows (also see Table \ref{tab:code-wa-analysis}):

\begin{figure}[t]
    \centering
  \begin{lstlisting}[language=Python]
class Solution:
    def smallestNumber(self, pattern: str) -> str:
        n = len(pattern)
        result = [0] * (n + 1)
        curr_min, curr_max = 0, n
        for i in range(n):
            if pattern[i] == 'I':
                result[i] = curr_min
                curr_min += 1
            else:
                result[i] = curr_max
                curr_max -= 1
        result[n] = curr_min
        ans = ''
        for i in range(n + 1):
            ans += str(result[i] + 1)
        return ans
  \end{lstlisting}
  \vspace{-1em}
  \caption{An example code snippet with WD error stemming from a misunderstanding of the meaning of lexicographically smallest in the given problem description.}
  \label{fig:wrong-detail-word}
\end{figure}

\begin{figure}[t]
    \centering
  \begin{lstlisting}[language=C]
char *categorizeBox(...) {
    ...
    if (...) {
        category = "Bulky";
    }
    else if (...) {
        category = "Heavy";
    }
    else {
        category = "Neither";
    }
    if (strcmp(category, "Bulky") == 0 && strcmp(category, "Heavy") == 0) {
        category = "Both";
    }
    else if (strcmp(category, "Bulky") == 0 && strcmp(category, "Heavy") != 0) {
        category = "Bulky";
    }
    else if (strcmp(category, "Heavy") == 0 && strcmp(category, "Bulky") != 0) {
        category = "Heavy";
    }
    return category;
}
  \end{lstlisting}
  \caption{An example code snippet with WD error stemming from the generated code that is not consistent with the understanding of the problem.}
  \label{fig:wrong-detail-code}
\end{figure}

\whiteding{1} \textbf{Wrong Detail (WD)}: The code generated by \ChatGPT has errors in some details. These detail errors stem from a little misunderstanding (e.g., a word) of the given problem or the generated code that is not consistent with the understanding of the problem. Fig. \ref{fig:wrong-detail-word} and \ref{fig:wrong-detail-code} show two code examples corresponding to the two detail errors, respectively. Fig. \ref{fig:wrong-detail-word} is an example of having a little misunderstanding of the given problem. The given problem 2375\footnote{https://leetcode.com/problems/construct-smallest-number-from-di-string/} is a medium one asks to generate the lexicographically smallest possible string that meets conditions. However, the generated code does not completely hold the meaning of lexicographically smallest due to line 11 decreasing number from the maximum one. As for the code snippet in Fig. \ref{fig:wrong-detail-code}, it is an example of the generated code that is not consistent with the understanding of the problem. The given problem 2525\footnote{https://leetcode.com/problems/categorize-box-according-to-criteria/} is an easy one that asks to categorize a given box into one of the categories according to its properties. \textit{ChatGPT} understands the meaning of the problem but fails to transfer the meaning of the problem description in natural language to code semantics. This is reflected in conditional expressions of lines 12, 15, and 18. We take line 12 as an example. The condition in natural language is "If the box is both "\texttt{Bulky}" and "\texttt{Heavy}", then its category is \texttt{Both}", and \ChatGPT utilizes \texttt{strcmp} twice to compare if the \texttt{category} string is both "\texttt{Bulky}" and "\texttt{Heavy}". The meaning of the natural language description is not equivalent to the code semantics. 

WD errors are also easy to be fixed by humans since the generated code logic is roughly correct. The defect subclasses corresponding to WD errors are mainly the subclasses other than M-L and Misaligned Algorithm, based on our manual analysis.

\begin{figure}[t]
    \centering
  \begin{lstlisting}[language=C]
int* findArray(...){...
    arr[0] = pref[0];
    for (int i = 1; i < prefSize; i++) {
        arr[i] = arr[i-1] ^ pref[i];
    }
    *returnSize = prefSize;
    return arr;
}
  \end{lstlisting}
  \vspace{-1em}
  \caption{An example code snippet with MCC error that reasons a wrong recurrence formula.}
  \label{fig:mcc-recurrence}
\end{figure}

\whiteding{2} \textbf{Misunderstanding Certain Content (MCC)}: The code generated by \ChatGPT does not hold the main condition of the given problem. However, the algorithm used by the generated code is suitable. Fig. \ref{fig:mcc-recurrence} shows an example code snippet. The corresponding problem 2433\footnote{https://leetcode.com/problems/find-the-original-array-of-prefix-xor/description/} is a medium one and asks to find the solution satisfying one \texttt{xor}-based condition. The key to this problem is to solve the correct recurrence formula according to this \texttt{xor}-based condition. \ChatGPT reasons out a recursive formula in lines 2-5, but this recursive formula does not satisfy the condition required by the problem. Other typical examples are the problems using dynamic programming (DP) that the generated code uses wrong DP equations. 

MCC errors are more difficult to fix than WD errors by humans since the core of the code needs to be modified to meet conditions provided by problems. The defect class corresponding to MCC errors is Multi-hunk based on our manual analysis.

\begin{figure}[t]
    \centering
  \begin{lstlisting}[language=C++]
public class Solution {
    public int totalSteps(int[] nums) {
        int steps = 0;
        for (int i = 1; i < nums.length; i++) {
            if (nums[i - 1] > nums[i]) {
                steps++;
                nums[i] = nums[i - 1];}}
        return steps;}}
  \end{lstlisting}
  \vspace{-1em}
  \caption{An example code snippet with MP error.}
  \label{fig:mp-total-wrong}
\end{figure}

\whiteding{3} \textbf{Misunderstanding Problem (MP)}: \ChatGPT misunderstands or does not understand the problem description given. The generated code does not hold all conditions and uses wrong (misaligned) algorithms. Fig. \ref{fig:mp-total-wrong} shows an example for the problem 2289\footnote{https://leetcode.com/problems/steps-to-make-array-non-decreasing/description/}.

MP errors are the most difficult to fix among these three kinds of errors by humans since the code needs rewriting completely. The defect subclass corresponding to MP errors is Misaligned Algorithm based on our manual analysis.

\noindent \textcolor{red}{$\blacktriangleright$} \textit{\textbf{Multi-round Fixing:}} We take each code snippet of \textit{<problem, language>} pair to \ChatGPT to continuously generate code in one unique conversation with multiple rounds. The round limit number is set to 5, providing a reasonable maximum number of fixes to 5 times~\cite{dong2023self}. For each pair, we create an initial prompt by leveraging the corresponding problem (i.e., \textit{<Content>} and \textit{<Examples>} in Fig. \ref{fig:chatgpt-Prompt-two-sum}), the code snippet, and error message. The error message is returned by \textit{LeetCode} online judgment, which is suitable to be taken as feedback provided to \textit{ChatGPT}. One example is shown below (where bolded words are filled in according to each pair's information):

\begin{lstlisting}[
    basicstyle=\ttfamily\footnotesize,
    xleftmargin=0.5ex,
    backgroundcolor=\color{bg},
    breaklines=true,
    numbers=none,
    escapeinside=||
]
|\underline{\textbf{Prompt}}:|
|\textbf{Given four integers length, width, height, and...}|

The code in |\textbf{C}| below cannot pass all test cases:
```
|\textbf{char *categorizeBox(...) \{...\}}|
```

Error Message:
|\textbf{Last test case: 2909 3968 3272 727}|
|\textbf{Code output: "Bulky"}|
|\textbf{Expected output: "Both"}|

Fix the code and generate the fixed code.
\end{lstlisting}

\noindent If the newly generated code snippet is still not accepted (i.e., W.A, C.E., R.E., and T.L.E.), the corresponding error message is taken directly as a new prompt provided to \textit{ChatGPT} to fix and generate a new code snippet, in the same conversation. It is appropriate to use the error message directly as the new prompt since \ChatGPT has the ability to dialog. The whole process lasts for a maximum of five rounds (one round corresponds to one newly generated code snippet) if the generated code is never accepted. \revision{However, there are cases that the cumulative token length of previous prompts, responses, and the current round prompt and response exceeds the token limitation of \textit{ChatGPT}. We mitigate this problem with \textit{token-limitation} strategy (see Sec. \ref{sec:methodology}) through reusing the initial prompt template with the current round's error code and message, which avoids missing necessary information of problem description to \textit{ChatGPT}. Moreover, it guarantees completely generated code snippets and also at least remains the immediate previous round's response in practice (Sec. \ref{sec:methodology}).}

\begin{table}[t]
\centering
\caption{Result of Multi-round Code Generation for W.A. Code Snippets}
\scalebox{0.8}{\begin{tabular}{l|r|r|r|r} 
\toprule

   & \multicolumn{1}{c|}{\textbf{Hard}} & \multicolumn{1}{c|}{\textbf{Medium}} & \multicolumn{1}{c|}{\textbf{Easy}}  & \multicolumn{1}{c}{\textbf{Total}} \\

\midrule
   
 C & \percent[q]{0}{5} & \percent[q]{0}{12} & \percent[q]{1}{5}  & \percent[q]{1}{22} \\
 C++ & \percent[q]{0}{5} & \percent[q]{1}{17} & \percent[q]{3}{7} & \percent[q]{4}{29} \\  
 Java &  \percent[q]{0}{6} & \percent[q]{0}{22} & \percent[q]{3}{5} & \percent[q]{3}{33} \\
 Python3 &  \percent[q]{0}{8} & \percent[q]{3}{16} & \percent[q]{4}{8} & \percent[q]{7}{32} \\
JavaScript &  \percent[q]{2}{9} & \percent[q]{3}{22} & \percent[q]{5}{10} & \percent[q]{10}{41} \\

\midrule

Total & \percent[q]{2}{33}  & \percent[q]{7}{89} & \percent[q]{16}{35} & \percent[q]{25}{157} \\

\midrule

Problem & \percent[q]{2}{12}  & \percent[q]{6}{25} & \percent[q]{12}{13} & \percent[q]{20}{50} \\

\bottomrule
\end{tabular}}
\label{tab:code-wa-multi}
\end{table}

The result of multi-round code generation is shown in Table \ref{tab:code-wa-multi}, where '/'s left hand and right hand represent the accepted (i.e., code snippets accepted in five rounds) number and the total number, respectively. From the result, we can see that the majority of these 157 \textit{<problem, language>} pairs cannot be fixed by automation. Only 25 pairs are fixed in 5 different languages, and 16 of them are problems at easy level. The pairs at medium level are fixed with only 7 pairs though its total number of pairs is more than twice as many as the pairs at easy level. Pairs at hard level are nearly impossible to be fixed. The percentage of fixes for pairs under all difficulties is less than half. However, judging from the fixes of the problems, 12 out of 13 easy problems are fixed. The percentage of fixes for hard and medium problems is still below 30\%. The average number of rounds per fixed pair is 1.32. 21 of the 25 can be fixed with just one round. The defect classes of the 25 pairs are mostly Multi-hunk, where M-S, M-U, M-L, and M-B account for 3, 3, 12, and 1, respectively. The redundant is in Single-hunk and Algorithm-relate, where  S-O, S-AS, S-HO, and Misaligned Algorithm account for 2, 1, 2, and 1, respectively. 

\begin{table}[t]
\centering
\caption{Result of 10-round Code Generation for 10 W.A. Code Snippets}
\scalebox{0.7}{\begin{tabular}{lccc} 
\toprule

 \textbf{Problem Title}  & \textbf{Language} & \textbf{Difficulty} & \textbf{Result}  \\

\midrule

\cellcolor{black!20}{Make Array Zero by Subtracting Equal Amounts}
 &  C++ & Easy & W.A.  \\   

\cellcolor{black!40}{Maximum Enemy Forts That Can Be Captured}
 &  JavaScript & Easy & W.A.  \\  

 \cellcolor{white!0}{Construct Smallest Number From DI String} & JavaScript & Medium & \cellcolor{white!0}{A.}   \\
 \cellcolor{black!40}{Construct Smallest Number From DI String} & Python3 & Medium & R.E.  \\  
 \cellcolor{black!40}{Frog Jump II} &  JavaScript & Medium & W.A.  \\   
 \cellcolor{black!40}{Longest Nice Subarray} &  Java & Medium & W.A.  \\  
  \cellcolor{white!0}{Minimum Number of Steps to Make Two Strings Anagram II} &  JavaScript & Medium & \cellcolor{white!0}{A.}  \\

 \cellcolor{black!40}{Make the XOR of All Segments Equal to Zero} &  Java & Hard & R.E.  \\  
 
 \cellcolor{black!40}{Maximum Number of Points From Grid Queries} &  JavaScript & Hard & W.A.  \\  
 \cellcolor{black!40}{Minimum Difference in Sums After Removal of Elements} &  C & Hard & W.A.  \\ 

\bottomrule
\end{tabular}}
\label{tab:code-wa-10-rounds}
\end{table}

To further analyze why most of the code snippets cannot be fixed under the multi-round process, we randomly select 10 more pairs from these unfixed pairs and expand the round limit number to 10 for multi-round fixing. The results are shown in Table \ref{tab:code-wa-10-rounds}. In these 10 code snippets, 2 of them are successfully fixed under 10 rounds. The remaining 8 still fail to be fixed, including 2 that are eventually fixed as R.E.. We manually check these failed pairs’ final generated code snippets and find that 7 of them marked as \revision{dark grey} deviate significantly from the meaning of the corresponding problems (i.e., MCC and MP). The remaining one marked as \revision{light grey} is almost correct, with only a very small logical error (i.e., WD and single-hunk), \revision{where this error persists throughout the multi-round process}. Moreover, there are only 5 W.A. code snippets with single-hunk fixed under the previous 5-round fixing.


Therefore, we conclude that there are 2 reasons why \textit{ChatGPT} cannot automatically fix W.A. code snippets through multi-round fixing. On one hand, \ChatGPT lacks the ability to grasp logical details, even though these details may be straightforward for humans. \ChatGPT struggles to notice and make corresponding fixes to them. Thus, \ChatGPT needs improving for its implementation ability for logical details. On the other hand, \ChatGPT lacks in dealing with problems that require complex reasoning (for W.A. code snippets), resulting in the code newly generated still deviating from the actual meaning of the problems. As a result, these kinds of W.A. code snippets are difficult to be fixed directly and automatically, but it is not always the case (e.g., \textit{<Construct Smallest Number From DI String, Python3>} and \textit{<Construct Smallest Number From DI String, JavaScript>} in Table \ref{tab:code-wa-10-rounds} where the latter one is fixed at the 10-th round).

\noindent \textcolor{magenta}{$\bigstar$} \textbf{Summary 1.} Most of the defects of code with W.A. fall under the M-L subclass and Misaligned Algorithm subclass with 51 and 62, respectively. The other subclasses of defects are relatively much less. 

\noindent \textcolor{magenta}{$\bigstar$} \textbf{Summary 2.} After our manual analysis, we conclude that W.A. code snippets can be divided into three categories of WD, MCC, and MP, from logic perspective.

\noindent \textcolor{magenta}{$\bigstar$} \textbf{Summary 3.} By applying multi-round fixing, \ChatGPT has difficulty fixing W.A. code snippets. We conclude for two reasons: (1) \ChatGPT lacks the ability to grasp logical details, and (2) \ChatGPT lacks in dealing with problems that require complex reasoning.

\begin{table*}[t]
\centering
\caption{Compile Error Classification of All C.E. Code Snippets}
\begin{tabular}{l|l|c|c|c|c}

\toprule

 \multicolumn{1}{c|}{\textbf{Class}} & \multicolumn{1}{c|}{\textbf{Explanation}} & \textbf{Hard} &  \textbf{Medium} & \textbf{Easy} & \textbf{Total} \\

\midrule

 {Label error} &\multicolumn{1}{m{7cm}|}{{A label can only be part of a statement and a declaration is not a statement.}} & {0} & {0}& {1}& {1} \\ 

\midrule

 {Redefinition} &\multicolumn{1}{m{7cm}|}{{A symbol has been defined in multiple places.}} &{0} & {4}& {1}&{5}  \\

\midrule

 {Function declaration error} &\multicolumn{1}{m{7cm}|}{{Function is invoked before declaration.}} &{4} & {19}& {1}&{24}  \\

\midrule

 {Undeclared variable} &\multicolumn{1}{m{7cm}|}{{Variable is referenced or used in a program without being previously declared or defined.}} & {7}& {5}& {2} & {14} \\

\midrule

 {Wrong method name} &\multicolumn{1}{m{7cm}|}{{Use a different method (or function) name than the template provided by \textit{LeetCode} to define a method (or function).}} & {8} & {2} & {2} & {12} \\ 
\midrule

 {Constant function} &\multicolumn{1}{m{7cm}|}{{Function (or method) is generated with empty body.}} &{81} & {71}& {7}&{159}  \\

\midrule

 {Redefinition of \texttt{main}} &\multicolumn{1}{m{7cm}|}{{Generated code snippet contains \texttt{main} function already provided by \textit{LeetCode}.}} &{10} & {12}& {7}&{29}  \\
\midrule
 {Use undeclared function} &\multicolumn{1}{m{7cm}|}{{Generated code snippet uses undeclared and undefined functions (mainly caused by EBL).}} &{18} & {25}& {3}&{46}  \\

\midrule

 {Incompatible parameter types} &\multicolumn{1}{m{7cm}|}{{Mismatch between the expected parameter type and the actual argument type passed to a method or function.}} &{2} & {0}& {1}&{3}  \\
\midrule
 {Uninitialized variable} &\multicolumn{1}{m{7cm}|}{{Variable is being declared and initialized but not given a valid initial value before using.}} &{2} & {2}& {0}&{4}  \\

\midrule

 {Invalid operators} &\multicolumn{1}{m{7cm}|}{{Invalid operands to binary operators (e.g., \texttt{\%}, \texttt{+}, \texttt{/}, and so on).}} & {2} & {1}& {0}& {3} \\
\midrule
  {Error of \texttt{\#include}} &\multicolumn{1}{m{7cm}|}{{Indicated file cannot be compiled.}} &{1} & {4}& {0}&{5}  \\

\midrule

 {Syntax error} &\multicolumn{1}{m{7cm}|}{{Generated code snippet has syntax errors.}} & {2} & {4}& {0}& {6} \\
\midrule
  {No attribute} &\multicolumn{1}{m{7cm}|}{{The accessed member does not exist in the corresponding data structure.}} &{1} & {2}& {0}&{3}  \\

\midrule

Total &  \multicolumn{1}{c|}{-} & {138} & {151}&{25} & {314}  \\ 

\bottomrule

\end{tabular}
\label{tab:code-ce-analysis}
\end{table*}

\subsubsection{Code with \textit{Compile Error}}

 \noindent \textcolor{red}{$\blacktriangleright$} \textit{\textbf{Analyzing:}} We analyze all C.E. code snippets and classify them manually based on the compile error messages returned by \textit{LeetCode}. There are 312 code snippets with C.E. in three different languages, C, C++, and Java. 

The compile error classes, explanations, and classified results are shown in Table \ref{tab:code-ce-analysis}. From the table, we can see that the majority of compile errors are in the class of constant function, accounting for half (159/314) of all compile errors. The code snippets having constant function compile error means that the functions or methods in code snippets have empty body (i.e., the generated codes are the same as corresponding code templates provided). Thus, this type of compile error is not a real compile error for generated code since it is a case of failure of code generation by \textit{ChatGPT} (it is not a failure of response). For other three special compile errors of classes of wrong method name, redefinition of \texttt{main}, and incompatible parameter types, they are related to \textit{LeetCode} online judgment platform, inconsistent with the settings in \textit{LeetCode} but not real compile errors. For example, a compile error-free code snippet generated by \textit{ChatGPT} may contain \textit{main} function but \textit{LeetCode} has set another internal \textit{main} for running test cases, which causes compile error of redefinition of \texttt{main}. Nevertheless, for wrong method name and incompatible parameter types, though they do not indicate real compile errors, it shows that \textit{ChatGPT} may have a certain chance to generate code regardless of the requirement (i.e., code template) given in the prompt. More interestingly, we also find that for code snippets with compile error of wrong method name, a few method names used for Aft. problems are method names of Bef. problems. For example, problem 2449\footnote{https://leetcode.com/problems/minimum-number-of-operations-to-make-arrays-similar/} requires using \texttt{makeSimilar} as method name but \textit{ChatGPT} generates method name of \texttt{minOperations} which is used in problem 1658\footnote{https://leetcode.com/problems/minimum-operations-to-reduce-x-to-zero/}, which may point to inference attack problem~\cite{rigaki2020survey, hu2022membership}. As for the remaining classes of compile errors, they are the real compile errors not triggered by \textit{LeetCode} platform. Table \ref{tab:code-ce-analysis} provides the explanations of corresponding classes of these compile errors, and examples of these classes can be found at our online artifact~\cite{artifact}.

\noindent \textcolor{red}{$\blacktriangleright$} \textit{\textbf{Multi-round Fixing:}} We follow the settings in W.A.'s multi-round fixing. The prompt used in C.E.'s multi-round fixing has a little bit different from the previous one. One example is shown below:

\begin{lstlisting}[
    basicstyle=\ttfamily\footnotesize,
    xleftmargin=0.5ex,
    backgroundcolor=\color{bg},
    breaklines=true,
    numbers=none,
    escapeinside=||
]
|\underline{\textbf{Prompt}}:|
|\textbf{Write a function to find the longest...}|

The code in |\textbf{C}| below has compile errors:
```
|\textbf{char *longestCommonPrefix(...) \{...\}}|
```

Error Message:
|\textbf{Compile Error}|
|\textbf{solution.c: In function ‘longestCommonPrefix’}|
|\textbf{Line 22: Char 5: error: a label can only be part of...}|
|\textbf{char *prefix = (char *)malloc((prefixLen + 1) * sizeof(char));}|
|\textbf{\^ ~~~}|

Fix the code and generate the fixed code.
\end{lstlisting}

\noindent Where the error message also comes from \textit{LeetCode} online judgment. The entire fixing process continues until the generated code snippet is accepted or the process reaches the maximum round number of 5. We take the final status (e.g., A.) in one conversation as the final generation result for the corresponding \textit{<problem, language>} pair. \revision{The strategy of mitigating token limitation follows the setting in W.A. multi-round fixing.} In addition, we do fixing for all classes except for the class of constant function, since fixing constant function is equivalent to regenerating the entire code snippets for \textit{<problem, language>} pairs.

\begin{table}[t]
\centering
\caption{Result of Multi-round Code Generation for C.E. Code Snippets}
\begin{tabular}{c|c|c|c}

\toprule

 \textbf{Class} & \textbf{Hard} &  \textbf{Medium} & \textbf{Easy}   \\ 
 
 \midrule
\midrule

Label error &   -   &    -  &   0:1:0 \\\hline

Redefinition &   -   &   3:0:1   &  1:0:0 \\\hline

Function declaration error &  3:1:0    &   14:4:1   & 1:0:0 \\\hline

Undeclared variable &   6:0:1   &   3:0:2   & 2:0:0 \\\hline

Wrong method name &   8:0:0   &   2:0:0   &  2:0:0   \\\hline

Redefinition of \texttt{main}  &     9:0:1   &   8:0:4   &  6:0:1   \\\hline

Use undeclared function &   12:3:3   &    16:6:3  &  3:0:0     \\\hline

Incompatible parameter types &    2:0:0  &   -   & 1:0:0 \\\hline

Uninitialized variable &   2:0:0   &     2:0:0 & - \\\hline

Invalid operators &   1:0:1   &   1:0:0   & - \\\hline

Error of \texttt{\#include} &    0:1:0  &  3:1:0    & - \\\hline

Syntax error &   1:1:1   &   1:1:2   & - \\\hline

No attribute & 1:0:0     &  1:0:1    & - \\ \hline

\midrule

Total & 45:6:7     &  54:12:14   &  16:1:1  \\

\bottomrule

\end{tabular}
\label{tab:code-ce-fixing}
\end{table}

The result is shown in Table \ref{tab:code-ce-fixing}. The \textit{x}:\textit{y}:\textit{z} in the table is the generation result under different conditions, where \textit{x}, \textit{y}, and \textit{z} represent the number of fixed code snippets (i.e., C.E. $\rightarrow$ A., W.A., or T.L.E.), the number of errors retained in code snippets (i.e., C.E. $\rightarrow$ the same C.E.), and the number of errors changed in code snippets (i.e., C.E. $\rightarrow$ other C.E. or R.E.), respectively. From the result, we can see that most of the code snippets can be fixed. 19 and 22 code snippets in C and C++ get retained errors and changed errors, respectively. For the 115 fixed code snippets, 40 of them are accepted, containing 30, 7, and 3 in C, C++, and Java, respectively. For the 40 code snippets, their classes of compile errors contain redefinition (1), function declaration error (8), undeclared variable (1), wrong method name (2), redefinition of \texttt{main} (8), use undeclared function (12), incompatible parameter types (1), uninitialized variable (3), invalid operators (1), and error of \texttt{\#include} (1). As for the code snippets with retained errors and changed errors, we manually analyze them and divide the causes of the two errors into two categories as follows:

\begin{figure}[t]
    \centering
  \begin{lstlisting}[language=C]
...
char* longestCommonPrefix(...) {
    ...
    goto exit_loop;
    ...
exit_loop:
    char* prefix = ...;
    ...}
  \end{lstlisting}
  \vspace{-1em}
  \caption{An example code snippet in C with retained error (label error) of EIL. The code snippet is the final generated one in the conversation.}
  \label{fig:code-retained-error-label-error}
\end{figure}

\begin{figure}[t]
    \centering
  \begin{lstlisting}[language=C]
int* smallestTrimmedNumbers(...){
    ...
    qsort(nums, numsSize, sizeof(char*), cmp);
    ...}
  \end{lstlisting}
  \vspace{-1em}
  \caption{An example code snippet in C with changed error (use undeclared function) of EIL. The code snippet is the final generated one in the conversation.}
  \label{fig:code-retained-error-use-undeclared-function}
\end{figure}

\whiteding{1} \textbf{Errors in Languages (EIL)}: EIL errors arise from the properties of the language used (i.e., C and C++) to implement the code. The errors include both retained errors and changed errors. Fig. \ref{fig:code-retained-error-label-error} shows an example code snippet of retained error (label error), where the code snippet is the final generated one in the conversation. In this particular case, the retained error is related to label error, where the code violates the rule of C by placing a label (line 6) before a declaration (line 7). \textit{ChatGPT} fails to fix the error in 5 rounds even though the error message contains the explanation of label error shown in Table \ref{tab:code-ce-analysis}. Regarding changed errors, all of them are runtime error, except one no attribute error and two use undeclared function errors, which means that almost all C.E. errors are fixed. \textit{ChatGPT} tries to generate functionally correct code, but R.E. errors are triggered in the implementation of algorithms. For instance, the final generated code may have an out-of-bound error. We further discuss R.E. subsequently. For the two classes of C.E. errors, we show an example of use undeclared function in Fig. \ref{fig:code-retained-error-use-undeclared-function} that \textit{ChatGPT} fixes its original syntax error on '\texttt{\}}' but introduces another C.E. error. The specific error is related to the comparison function \texttt{cmp} used as an argument for the \texttt{qsort} function. However, \texttt{cmp} is not defined in the code snippet, resulting in a compile error.

\begin{figure}[t]
    \centering
  \begin{lstlisting}[language=C]
#include <cstring>
#include <algorithm>
...
char *subStrHash(...) {...}
  \end{lstlisting}
  \vspace{-1em}
  \caption{An example code snippet in C with retained error (error of \texttt{\#include}) of EBL. The code snippet is the final generated one in the conversation.}
  \label{fig:code-retained-error-error-of-include}
\end{figure}

\whiteding{2} \textbf{Errors between Languages (EBL)}: Different from EIL, EBL errors arise from the similarity between languages of C and C++. The errors still include both retained errors and changed errors, and all changed errors are C.E. errors which are the same as the C.E. errors in retained errors of EBL. Thus, we only use the retained error as an example. Fig. \ref{fig:code-retained-error-error-of-include} shows an example code snippet of retained error (error of \texttt{\#include}). The language for this \textit{<problem, language>} pair is C but the generated code uses \texttt{<cstring>} and \texttt{<algorithm>} header files belonging to C++.

We observe the entire multi-round process for these unfixed \textit{<problem, language>} pairs of which the final errors are C.E. errors. We find that in most cases, although the error messages provided contain the causes of the C.E. errors and the corresponding locations, and \textit{ChatGPT} is aware of the error problem from its natural language, the newly generated code snippets still have the same errors. One example is Fig. \ref{fig:code-retained-error-label-error} that \textit{ChatGPT} notices the error in each round but fails to fix it. For these errors, a potentially appropriate approach is to add information to prompts from human knowledge that triggers \textit{ChatGPT} to truly fix errors. For instance, for the example of Fig. \ref{fig:code-retained-error-error-of-include}, we can supply extra information (e.g. "the code snippet is in language C, you cannot use C++ header files") to \textit{ChatGPT} to fix the error.

Additionally, for each \textit{<class, difficulty>} (e.g., \textit{<use undeclared function, medium>}) in Table \ref{tab:code-ce-fixing} except for label error, it has at least one code snippet can be fixed. Thus, \textit{ChatGPT}'s multi-round code fixing for errors (include R.E. errors. See Table \ref{tab:code-re-fixing}) may also be related to the randomness (i.e., Temperature) of \textit{ChatGPT} itself or the code snippets it receives.

\noindent \textcolor{magenta}{$\bigstar$} \textbf{Summary 1.} More than half of \textit{ChatGPT}'s C.E. errors (in static languages) are unreal compile errors, especially for constant function, wrong method name, and incompatible parameter types. This experimental result indicates that \textit{ChatGPT}'s code generation stability (avoid generating empty body) and alignment with human attention (meet user requirements such as method signature provided) are the potentially severe issues that need to be strengthened.

\noindent \textcolor{magenta}{$\bigstar$} \textbf{Summary 2.} By applying multi-round fixing, most ($70\%$) of C.E. code snippets can be fixed, including $26\%$ of them can be fixed to A.. After analyzing unfixed code snippets, it can be inferred that the unfixed reasons can be concluded to EIL and EBL. Additionally, a potential approach to help \textit{ChatGPT} fix the unfixed errors is to add human knowledge.


\begin{table*}[t]
\centering
\caption{Runtime Error Classification of All R.E. Code Snippets}
\begin{tabular}{l|l|c|c|c|c}

\toprule

 \multicolumn{1}{c|}{\textbf{Class}} & \multicolumn{1}{c|}{\textbf{Explanation}} & \textbf{Hard} &  \textbf{Medium} & \textbf{Easy} & \textbf{Total} \\

\midrule

 {Integer-overflow} &\multicolumn{1}{m{7cm}|}{{The result of an arithmetic operation exceeds the maximum value that can be represented by a given integer data type (e.g., \texttt{int}).}} & {4} & {17}& {2}& {23} \\ 

\midrule

 {Heap-buffer-overflow} &\multicolumn{1}{m{7cm}|}{{A program writes data beyond the allocated memory block in the heap, potentially overwriting adjacent data or causing a crash due to corrupted memory.}} &{9} & {22}& {8}&{39}  \\

\midrule

 {Undefined-behavior} &\multicolumn{1}{m{7cm}|}{{The executing of a certain code statement is not defined or unpredictable, potentially leading to unexpected runtime errors or crashes (e.g., left shift of negative value).}} &{1} & {1}& {1}&{3}  \\

\midrule

 {Out-of-bound} &\multicolumn{1}{m{7cm}|}{{Access an array, list, or other data structure by using an index or pointer that exceeds the valid range of elements or memory.}} & {18}& {21}& {4} & {43} \\

\midrule

 {Constant function} &\multicolumn{1}{m{7cm}|}{{Function (or method) is generated with empty body.}} & {3} & {6} & {1} & {10} \\ 
\midrule

 {Null pointer dereference} &\multicolumn{1}{m{7cm}|}{{A program attempts to access or manipulate data through a null pointer, which is a pointer that does not point to a valid memory location.}} &{4} & {8}& {2}&{14}  \\

\midrule

 {Wrong method name} &\multicolumn{1}{m{7cm}|}{{Use a different method (or function) name than the template provided by \textit{LeetCode} to define a method (or function).}} &{3} & {4}& {6}&{13}  \\
\midrule
 {Type error} &\multicolumn{1}{m{7cm}|}{{An operation is performed on an object of an incompatible type (e.g., try to concatenate a string with an integer in Python3).}} &{9} & {10}& {2}&{21}  \\

\midrule

 {Value error} &\multicolumn{1}{m{7cm}|}{{A method (function) or operation receives a valid type of input, but the value itself is not suitable or within the expected range for the specific operation (e.g., pass an empty \textit{List} to function \texttt{max} in Python3).}} &{0} & {4}& {1}&{5}  \\
\midrule
 {Heap-use-after-free} &\multicolumn{1}{m{7cm}|}{{Use a memory block in the heap after it has been deallocated (freed).}} &{1} & {1}& {0}&{2}  \\

\midrule

 {Recursion error} &\multicolumn{1}{m{7cm}|}{{A function or method calls itself recursively without reaching a base case or termination condition.}} & {4} & {3}& {0}& {7} \\
\midrule
  {Uninitialized variable} &\multicolumn{1}{m{7cm}|}{{Variable is being declared and initialized but not given a valid initial value before using.}} &{1} & {1}& {0}&{2}  \\

\midrule

 {Syntax error} &\multicolumn{1}{m{7cm}|}{{Generated code snippet has syntax errors.}} & {3} & {0}& {0}& {3} \\
\midrule
{Undeclared variable} &\multicolumn{1}{m{7cm}|}{{Variable is referenced or used in a program without being previously declared or defined.}} & {2} & {2}& {0}& {4} \\

\midrule

{No attribute} &\multicolumn{1}{m{7cm}|}{{The accessed member does not exist in the corresponding data structure.}} &{0} & {3}& {0}&{3}  \\
 \midrule
{Divided by zero} &\multicolumn{1}{m{7cm}|}{{Attempt to divide a number by zero.}} & {0} & {2}& {0}& {2} \\

\midrule

Total &  \multicolumn{1}{c|}{-} & {62} & {105}&{27} & {194}  \\ 

\bottomrule

\end{tabular}
\label{tab:code-re-analysis}
\end{table*}

\begin{figure}[t]
    \centering
  \begin{lstlisting}[language=Python]
class Solution:
    def largestInteger(self, num: int) -> int:
        num_str = str(num)
        ...
        while swapped:
            ...
            for i in range(n):
                for j in range(i+1, n):
                    if (num_str[i] % 2 == num_str[j] % 2) and (num_str[i] < num_str[j]):
                        ...
                ...
        return int(num_str)
  \end{lstlisting}
  \vspace{-1em}
  \caption{An example code snippet in Python3 with type error.}
  \label{fig:re-type-error}
\end{figure}

\subsubsection{Code with \textit{Runtime Error}}

\noindent \textcolor{red}{$\blacktriangleright$} \textit{\textbf{Analyzing:}} We also analyze all R.E. code snippets and classify them manually based on the runtime error messages returned by \textit{LeetCode}. There are 194 code snippets with R.E. in five different languages, C, C++, Java, Python3, and JavaScript. 

The runtime error classes, explanations, and classified results are shown in Table \ref{tab:code-re-analysis}. From the table, we can see that there are a small number of code snippets in the class of constant function for dynamic languages (i.e., Python3 and JavaScript), which is different from C.E.'s results. Moreover, like C.E. errors, wrong method name in R.E. errors is not a real runtime error and we also find examples that Aft. problems use method names of Bef. problems (e.g., problem 2164\footnote{https://leetcode.com/problems/sort-even-and-odd-indices-independently/} and problem 905\footnote{https://leetcode.com/problems/sort-array-by-parity/}). Additionally, the majority of runtime errors are overflow runtime errors (i.e., integer-overflow, heap-buffer-overflow, and out-of-bound) and the languages used in these errors are also mainly in static languages, C, C++, and Java, which is similar to humans making runtime errors. For dynamic languages (i.e., Python3 and JavaScript), the majority of runtime errors are in the class of type error. The error occurs when an operation is performed on an object of an incompatible type. One example is shown in Fig. \ref{fig:re-type-error} that in line 9, the code statement performs modulus operations (\texttt{\%}) on characters \texttt{num\_str[i]} and \texttt{num\_str[j]}, which is not valid. As for the remaining classes, their explanations are provided in Table \ref{tab:code-re-analysis} and their examples can be found at our online artifact~\cite{artifact}.

\begin{table}[t]
\centering
\caption{Result of Multi-round Code Generation for R.E. Code Snippets}
\begin{tabular}{c|c|c|c}

\toprule

 \textbf{Class} & \textbf{Hard} &  \textbf{Medium} & \textbf{Easy}   \\ 
 
 \midrule
\midrule

Integer-overflow &   2:2:0   &    16:1:0  &   2:0:0 \\\hline

Heap-buffer-overflow &   5:1:3   &   14:8:0   &  5:3:0 \\\hline

Undefined-behavior &  0:1:0    &   1:0:0   & 1:0:0 \\\hline

Out-of-bound &   14:3:1   &   15:6:0   & 4:0:0 \\\hline

Null pointer dereference &   3:1:0   &   6:2:0   &  1:1:0   \\\hline

Wrong method name  &     1:0:2   &   4:0:0   &  6:0:0   \\\hline

Type error &   8:1:0   &    8:1:1  &  2:0:0     \\\hline

Value error &    -  &   4:0:0   & 0:1:0 \\\hline

Heap-use-after-free &   0:0:1   &     1:0:0 & - \\\hline

Recursion error &   4:0:0   &   3:0:0   & - \\\hline

Uninitialized variable &    1:0:0  &  1:0:0    & - \\\hline

Syntax error &   3:0:0   &   -   & - \\\hline

Undeclared variable & 2:0:0     &  2:0:0    & - \\ \hline

No attribute & -     &  2:0:1    & - \\ \hline

Divided by zero & -     &  2:0:0    & - \\ \hline

\midrule

Total & 43:9:7     &  79:18:2   &  21:5:0  \\

\bottomrule

\end{tabular}
\label{tab:code-re-fixing}
\end{table}

\begin{figure}[t]
    \centering
  \begin{lstlisting}[language=Python]
class Solution:
    def minOperationsToFlip(...) -> int:
        ...
        def evaluate(mapping):
            stack = []
            for char in expression:
                if char == ')':
                    ...
                elif char in {'0', '1', '&', '|', '(', ')'}:
                    stack.append(mapping[char])
            ...
        ...
        # count the number of unique variables
        variables = set([char for char in expression if char.isalpha()])
        # generate all possible mappings ...
        mappings = []
        for i in range(2**n):
            mapping = {}
            for j, var in enumerate(variables):
                mapping[var] = (i >> (n-j-1)) & 1
            mappings.append(mapping)
        ...  # later invoke evaluate(mapping)
  \end{lstlisting}
  \caption{An example code snippet in Python3 with a retained error (value error).}
  \label{fig:re-re-type-error}
\end{figure}

\noindent \textcolor{red}{$\blacktriangleright$} \textit{\textbf{Multi-round Fixing:}} We follow the settings in C.E.'s multi-round fixing, and the prompt template used here is modified, turning "The code in \textit{<language>} below has compile errors:" to "The code in \textit{<language>} below has runtime errors:".

The result is shown in Table \ref{tab:code-re-fixing}. The \textit{x}:\textit{y}:\textit{z} in the table is the same as Table \ref{tab:code-ce-fixing}. From the result, we can see that most of the code snippets can be fixed, and there are 32 and 9 code snippets in five different languages that get retained errors and changed errors, respectively. For the 143 fixed code snippets, 52 of them are accepted, containing 23, 12, 2, 10, and 5 in C, C++, Java, Python3, and JavaScript, respectively. For the 52 code snippets, their classes of runtime errors contain integer-overflow (11), heap-buffer-overflow (12), undefined-behavior (1), out-of-bound (10), null pointer dereference (4), wrong method name (3), type error (5), value error (1), heap-use-after-free (1), recursion error (1), uninitialized variable (2), and divided by zero (1). As for the code snippets with retained errors and changed errors, they are few in number (41) and most of them belong to overflow error. Regarding retained errors, by manual analysis, we believe that the main reason why the runtime errors cannot be eliminated is the algorithm implementation by \textit{ChatGPT}. It is similar to those (e.g., WD) in the W.A. errors. One example of overflow is shown in Fig. \ref{fig:code-heap-buffer-overflow} that \ChatGPT fails to fix line 15 under the 5 rounds of dialogue. Fig. \ref{fig:re-re-type-error} shows an example of value error. The problem (or conflict) in the code snippet is between line 14 and line 9 (i.e., \texttt{char.isalpha()} and \texttt{char in {'0', '1', '\&', '|', '(', ')'}}). The \texttt{char.isalpha()} condition checks whether a character is an alphabetic character (a-z or A-Z), while the \texttt{char in {'0', '1', '\&', '|', '(', ')'}} condition checks for specific characters which are not alphabetic. Regarding the 9 changed errors, 1 is changed to compile error (use undeclared function) and the remaining errors are still runtime errors including heap-buffer-overflow (1), undefined-behavior (2), out-of-bound (2), type error (2), and recursion error (1). These new errors are introduced as the code snippets continue to be fixed and some parts of code snippets conflict.

We also observe the entire multi-round process for these unfixed \textit{<problem, language>} pairs of which the final errors are R.E. errors. Like C.E., in most cases, \textit{ChatGPT} notices errors based on the error messages provided, however, the
newly generated code snippets still have the same errors. Fig. \ref{fig:code-heap-buffer-overflow} and Fig. \ref{fig:re-re-type-error} are two examples. To fix runtime errors, also like C.E. errors, a potentially appropriate approach is to add information to prompts from
human knowledge. For instance, for fixing Fig. \ref{fig:code-heap-buffer-overflow}, we can supply extra information (e.g., "\texttt{*returnColumnSizes = (int *)malloc(sizeof(int));} allocates a wrong size of memory to \texttt{*returnColumnSizes}") to fix the error.

\noindent \textcolor{magenta}{$\bigstar$} \textbf{Summary 1.} The majority of runtime errors for static languages and dynamic languages are overflow (105) and type error (21), respectively. Additionally, for Python3 and JavaScript, 10 code snippets are constant function and 13 are wrong method name. The result indicates that \textit{ChatGPT}’s code generation stability and alignment with human attention are also potentially severe issues for dynamic languages.

\noindent \textcolor{magenta}{$\bigstar$} \textbf{Summary 2.} By applying multi-round fixing, like C.E., most ($78\%$) of R.E. code snippets can be fixed, including $28\%$ of them can be fixed to A.. After analyzing unfixed code snippets, it can be concluded that \ChatGPT is flawed in the details of the algorithm implementation and a potential approach to trigger \ChatGPT to fix the unfixed errors is to add human knowledge.

\subsubsection{Code with \textit{Time Limit Exceeded}}


\noindent \textcolor{red}{$\blacktriangleright$} \textit{\textbf{Analyzing:}} There are 140 code snippets with T.L.E. in five different languages, C, C++, Java, Python3, and JavaScript. \revision{Two graduate students together analyze each T.L.E. code snippet and categorize all these code snippets based on the analysis results.} \revision{After} manually analyzing these code snippets, we classify their timeout reasons into three categories:

\begin{figure}[t]
    \centering
  \begin{lstlisting}[language=C]
bool reachingPoints(int sx, int sy, int tx, int ty){
    while (tx >= sx && ty >= sy) {
        ...
        if (tx > ty) {
            tx -= ty;
        } else {
            ty -= tx;
        }
    }
    return false;
}
  \end{lstlisting}
  \caption{An example code snippet in C with AIAI.}
  \label{fig:tle-aiai}
\end{figure}

\whiteding{1} \textbf{Aligned but Inefficient Algorithm Implementation (AIAI)}: The algorithm used in code generated by \textit{ChatGPT} is aligned with the requirement
given in the problem description, but some parts are not efficient. One classic example of AIAI is \texttt{gcd} functions, \revision{which use the modulo operator \texttt{\%} or subtraction operator \texttt{-}}. Both two functions \revision{with either the modulo operator or subtraction operator} use the Euclidean algorithm, but their time complexities are $O(\log n)$ and $O(n)$, respectively. Some T.L.E. errors are caused by AIAI. Fig. \ref{fig:tle-aiai} shows an example\footnote{https://leetcode.com/problems/reaching-points} similar to \textit{gcd}. It uses subtraction operator \texttt{-} rather than modulo operator \texttt{\%}, which increases the time complexity and fails to pass all test cases in the limited time set by \textit{LeetCode} platform.

\begin{figure}[t]
    \centering
  \begin{lstlisting}[language=Python]
class Solution:
    def countSpecialNumbers(self, n: int) -> int:
        def is_special(x):
            digits = set()
            while x > 0:
                digit = x % 10
                ...
                x //= 10
            return True 
        count = 0
        for i in range(1, n+1):
            if is_special(i):
                count += 1
        return count
  \end{lstlisting}
  \caption{An example code snippet in Python3 with CMA. The time complexity of the code snippet is $O(n \log n)$.}
  \label{fig:tle-cia}
\end{figure}

\whiteding{2} \textbf{Functionally Correct but Misaligned Algorithm (CMA)}: The algorithm used by \textit{ChatGPT} is functionally correct to the corresponding problem but it is inefficient for the limited time set by \textit{LeetCode} platform. So, the algorithm is misaligned. In CMA, the most common examples of the generated code are to solve problems using the brute-force method. Although the algorithm (or code) using the brute-force method is functionally correct, the time complexity may also be very high (e.g., $O(2^n)$ time complexity), and thus the code is judged as timeout by \textit{LeetCode} online judgment. One example\footnote{https://leetcode.com/problems/count-special-integers/} is shown in Fig. \ref{fig:tle-cia} whose time complexity is exponential (the input size of the code is $\log n$, representing the input number $n$).

\begin{figure}[t]
    \centering
  \begin{lstlisting}[language=Java]
var maximumGroups = function(grades) {
    grades.sort((a, b) => b - a);
    let groups = 0;
    while (grades.length > 0) {
        let sum = 0;
        let count = 0;
        for (let i = 0; i < grades.length; i++) {
            if (sum + grades[i] <= count) {
                sum += grades[i];
                count++;
                grades.splice(i, 1);
                i--;
            }
        }
        groups++;
    }
    return groups;
};
  \end{lstlisting}
  \caption{An example code snippet in JavaScript with IA.}
  \label{fig:tle-iia}
\end{figure}

\whiteding{3} \textbf{Functionally Incorrect Algorithm (IA)}: The algorithm used by \textit{ChatGPT} is functionally incorrect (i.e., WD, MCC, and MP) to the corresponding problem, which is also slow (due to the incorrect function) to resolve given problem instances (i.e., test cases). The algorithm may be aligned or misaligned (e.g., brute-force method). One example\footnote{https://leetcode.com/problems/maximum-number-of-groups-entering-a-competition/} is shown in Fig. \ref{fig:tle-iia}, using greedy algorithm. The algorithm is aligned, but its function is incorrect. In line 8, the condition \texttt{sum + grades[i] <= count} can lead the outer while loop to be an infinite loop if the condition is \texttt{false}, causing T.L.E. error.

\begin{table}[t]
\centering
\caption{Result of Multi-round Code Generation for T.L.E. Code Snippets}
\scalebox{0.8}{\begin{tabular}{l|r|r|r|r}

\toprule

   & \multicolumn{1}{c|}{\textbf{Hard}} & \multicolumn{1}{c|}{\textbf{Medium}} & \multicolumn{1}{c|}{\textbf{Easy}}  & \multicolumn{1}{c}{\textbf{Total}} \\

\midrule
   
 C & \percent[q]{2}{3} & \percent[q]{7}{17} & 0/0 ($0.0 \%$)  & \percent[q]{9}{20} \\
 C++ & \percent[q]{2}{9} & \percent[q]{1}{12} & \percent[q]{2}{2} & \percent[q]{5}{23} \\  
 Java &  \percent[q]{2}{8} & \percent[q]{8}{16} & \percent[q]{0}{2} & \percent[q]{10}{26} \\
 Python3 &  \percent[q]{3}{19} & \percent[q]{7}{21} & \percent[q]{0}{2} & \percent[q]{10}{42} \\
JavaScript &  \percent[q]{2}{8} & \percent[q]{8}{20} & \percent[q]{0}{1} & \percent[q]{10}{29} \\

\midrule

Total & \percent[q]{11}{47}  & \percent[q]{31}{86} & \percent[q]{2}{7} & \percent[q]{44}{140} \\

\bottomrule

\end{tabular}}
\label{tab:code-tle-fixing}
\end{table}

\noindent \textcolor{red}{$\blacktriangleright$} \textit{\textbf{Multi-round Fixing:}} We follow the settings in W.A.'s multi-round fixing.

The result is shown in Table \ref{tab:code-tle-fixing}. There are 44/140 code snippets across different languages and difficulty levels in total that can be fixed (i.e., accepted) by \textit{ChatGPT}. By manual analysis, we find that these code snippets are in AIAI, CMA, and IA. For code snippets in AIAI and IA, \textit{ChatGPT} tends to generate patches or rewrite code in different algorithms for fixing. One example of generating patches is Fig. \ref{fig:tle-aiai} that \textit{ChatGPT} modifies subtraction operator \texttt{-} to modulo operator \texttt{\%}. The example of rewriting code is that for problem 1047\footnote{https://leetcode.com/problems/remove-all-adjacent-duplicates-in-string/} in C, \textit{ChatGPT} changes aligned stack-based algorithm to array-based algorithm without using stack. As for code snippets in CMA, \textit{ChatGPT} tends to change the algorithms used. As for the remaining 96 code snippets not fixed, their newly generated code snippets are judged as W.A. (64), R.E. (8), and T.L.E. (24) by \textit{LeetCode} online judgment. For the new ones judged as W.A., \textit{ChatGPT} is able to fix the conflict parts causing T.L.E. but the incorrect functions cannot be fixed (e.g., Fig. \ref{fig:tle-iia}), or it changes the algorithms used but the new algorithms are functionally incorrect (e.g., Fig. \ref{fig:tle-cia} in Java version). For the new ones judged as R.E., the runtime errors of them include integer-overflow, heap-buffer-overflow, out-of-bound, value error, and out-of-memory\footnote{A computer program tries to allocate memory from the heap, but there is not enough available memory to fulfill the request}. For the new ones judged as T.L.E., 16 of them can pass $> 75 \%$ test cases and only 4 of them pass $ < 50 \%$ test cases. By our manual analysis, we find that \textit{ChatGPT} tends to change algorithms used to avoid T.L.E., but the algorithms used have some inefficient parts (i.e., AIAI or IA). Thus, to fix these inefficient parts, a potentially appropriate approach is to tell \textit{ChatGPT} the inefficient locations and fixing suggestions by humans.

\noindent \textcolor{magenta}{$\bigstar$} \textbf{Summary 1.} By manually analyzing T.L.E. code snippets, it can be concluded that the timeout reasons are AIAI, CMA, and IA.

\noindent \textcolor{magenta}{$\bigstar$} \textbf{Summary 2.} By applying multi-round fixing, only $31\%$ code snippets can be fixed. For fixed code snippets in AIAI and IA, \textit{ChatGPT} tends to generate patches or rewrite code in different algorithms, and for code snippets in CMA, \textit{ChatGPT} tends to change the algorithm used. The unfixed ones have 24 in T.L.E., caused by AIAI and IA. To fix them, a potential approach is to provide inefficient locations and fixing suggestions by humans.

\begin{tcolorbox}[boxrule=1pt,boxsep=1pt,left=2pt,right=2pt,top=2pt,bottom=2pt,title=Answer to RQ2: Multi-round Fixing for Code Generation]

\noindent \blackding{1} Multi-round fixing process can only fix a small fraction ($< 32\%$) of code snippets with W.A., C.E., R.E., or T.L.E. to A.. For fixing \revision{code snippets with} C.E., R.E., \revision{or} T.L.E. to A., W.A., or T.L.E. (exclude T.L.E. $\rightarrow$ T.L.E.), most ($\geq 70\%$) of \revision{them} can be fixed by using multi-round fixing;

\noindent \blackding{2} Our analysis identifies several factors for errors in code snippets and unfixed cases under the multi-round fixing process. The findings contribute to the ongoing research focused on improving functionally correct code generation.

\end{tcolorbox}

\subsection{Code \revision{Complexity}}
\noindent \textbf{RQ3: How \revision{complex} is the code generated by \textit{ChatGPT}?}

\noindent \textbf{Motivation.} \revision{The complexity of code is a critical factor influencing code readability, maintainability, and overall quality~\cite{dantas2021readability, nguyen2022empirical, scalabrino2017automatically}.} In this RQ, we evaluate the \revision{complexity} of the code generated by \textit{ChatGPT}.

\noindent \textbf{Approach.} We utilize \textit{SonarQube}~\cite{sonarqube} and \textit{cccc}~\cite{cccc} to calculate two metrics for evaluating the \revision{complexity} of Bef. and Aft. generated code, including the code generated in multi-round fixing. The metrics are cyclomatic complexity and cognitive complexity~\cite{nguyen2022empirical, dantas2021readability, mccabe1976complexity}, and their specific meanings are as follows:

\begin{itemize}
    \item \textbf{Cyclomatic Complexity:} The complexity counts the number of linearly independent paths through a given source code. It determines how difficult the given code is to test. A high cyclomatic complexity can potentially lead to a high probability of errors and bugs. 

    \item \textbf{Cognitive Complexity:} The complexity refers to a measure of how difficult it is to understand and reason about a piece of code from the human perspective. It takes factors into account like control structures but slightly different from cyclomatic complexity. Its specific methodology can be found in~\cite{cognitive-complexity}. A high cognitive complexity can affect code maintainability and increase the risk of bugs or errors. 
\end{itemize}

\noindent Where cognitive complexity is measured for three languages (Java, Python3, and JavaScript) due to the limitation of \textit{SonarQube} and \textit{cccc}.

We also utilize \textit{LeetCode} solutions~\cite{leetcodesolution} in C++ and Python3 written by humans (lack solutions in other languages) to compare with \textit{ChatGPT}'s, for observing their different extent in code \revision{complexity}. 

Note that the complexity is measured in terms of problems as a unit (most solutions have only one method).

\noindent \textbf{Result.} We first examine code snippets not generated in multi-round fixing. The analysis for code snippets generated in multi-round fixing is discussed in \textbf{Multi-round Comparisons} in this section. Table \ref{tab:cyclomaticcomplexity} and \ref{tab:cognitivecomplexity} show the cyclomatic and cognitive complexity values of code generated by \textit{ChatGPT} in five languages. Table \ref{tab:humancyclomaticcomplexity} and \ref{tab:humancognitivecomplexity} show the cyclomatic and cognitive complexity values of code written by humans in two languages.

\begin{table}[t]
\centering
\caption{Cyclomatic Complexity Result of Generated Code}
\scalebox{0.9}{\begin{tabular}{clr} 
\toprule
\multicolumn{1}{l}{\textbf{Language}} & \multicolumn{1}{c}{\textbf{Cyclomatic Complexity Level}} & \multicolumn{1}{c}{\textbf{Number of Code}}  \\ 
\midrule
\multirow{4}{*}{\textbf{C}}           & Low complexity  (1-4)                                    &   \percent{148}{676}                   \\
                                      & Moderate complexity (5-7)                                &   \percent{223}{676}                   \\
                                      & High complexity (8-10)                                   &   \percent{140}{676}                   \\
                                      & Very High complexity (11+)                               &   \percent{165}{676}                   \\ 
\midrule
\multirow{4}{*}{\textbf{C++}}         & Low complexity (1-4)                                     &  \percent{218}{638}                    \\
                                      & Moderate complexity (5-7)                                &  \percent{241}{638}                    \\
                                      & High complexity (8-10)                                   &   \percent{90}{638}                    \\
                                      & Very High complexity (11+)                               &   \percent{89}{638}                    \\ 
\midrule
\multirow{4}{*}{\textbf{Java}}        & Low complexity (1-4)                                     &  \percent{225}{667}                     \\
                                      & Moderate complexity (5-7)                                & \percent{240}{667}                     \\
                                      & High complexity (8-10)                                   & \percent{100}{667}                     \\
                                      & Very High complexity (11+)                               & \percent{102}{667}                     \\ 
\midrule
\multirow{4}{*}{\textbf{Python3}}     & Low complexity (1-4)                                     & \percent{314}{705}                     \\
                                      & Moderate complexity (5-7)                                & \percent{226}{705}                     \\
                                      & High complexity (8-10)                                   & \percent{105}{705}                     \\
                                      & Very High complexity (11+)                               & \percent{60}{705}                      \\ 
\midrule
\multirow{4}{*}{\textbf{JavaScript}}  & Low complexity (1-4)                                     & \percent{221}{681}                     \\
                                      & Moderate complexity (5-7)                                & \percent{234}{681}                     \\
                                      & High complexity (8-10)                                   & \percent{115}{681}                     \\
                                      & Very High complexity (11+)                               & \percent{111}{681}                      \\
\bottomrule
\end{tabular}}
\label{tab:cyclomaticcomplexity}
\end{table}

\begin{table}[t]
\centering
\caption{Cognitive Complexity Result of Generated Code}
\scalebox{0.9}{\begin{tabular}{clr} 
\toprule
\multicolumn{1}{l}{\textbf{Language}} & \multicolumn{1}{c}{\textbf{Cognitive Complexity Level}} & \multicolumn{1}{c}{\textbf{Number of Code}}  \\ 
\midrule
\multirow{4}{*}{\textbf{Java}}        & Low complexity ($<$5)                                     &  \percent{244}{667}                     \\
                                      & Moderate complexity (6-10)                               & \percent{231}{667}                     \\
                                      & High complexity (11-20)                                   & \percent{147}{667}                     \\
                                      & Very High complexity (21+)                               & \percent{45}{667}                     \\ 
\midrule
\multirow{4}{*}{\textbf{Python3}}     & Low complexity ($<$5)                                     & \percent{301}{706}                     \\
                                      & Moderate complexity (6-10)                                & \percent{211}{706}                     \\
                                      & High complexity (11-20)                                 & \percent{147}{706}                     \\
                                      & Very High complexity (21+)                               & \percent{47}{706}                      \\ 
\midrule
\multirow{4}{*}{\textbf{JavaScript}}  & Low complexity ($<$5)                                      & \percent{250}{682}                     \\
                                      & Moderate complexity (6-10)                                & \percent{235}{682}                     \\
                                      & High complexity (11-20)                                  & \percent{146}{682}                     \\
                                      & Very High complexity (21+)                                & \percent{51}{682}                      \\
\bottomrule
\end{tabular}}
\label{tab:cognitivecomplexity}
\end{table}

\begin{table}[t]
\centering
\caption{Cyclomatic Complexity Result of Written Code}
\scalebox{0.9}{\begin{tabular}{clr} 
\toprule
\multicolumn{1}{l}{\textbf{Language}} & \multicolumn{1}{c}{\textbf{Cyclomatic Complexity Level}} & \multicolumn{1}{c}{\textbf{Number of Code}}  \\ 
\midrule
\multirow{4}{*}{\textbf{C++}}        & Low complexity (1-4)                                     &  \percent{243}{619}                     \\
                                      & Moderate complexity (5-7)                               & \percent{203}{619}                     \\
                                      & High complexity (8-10)                                   & \percent{84}{619}                     \\
                                      & Very High complexity (11+)                               & \percent{89}{619}                     \\ 
\midrule
\multirow{4}{*}{\textbf{Python3}}     & Low complexity (1-4)                                     & \percent{323}{705}                     \\
                                      & Moderate complexity (5-7)                                & \percent{205}{705}                     \\
                                      & High complexity (8-10)                                 & \percent{87}{705}                     \\
                                      & Very High complexity (11+)                               & \percent{90}{705}                      \\ 
\bottomrule
\end{tabular}}
\label{tab:humancyclomaticcomplexity}
\end{table}

\begin{table}[t]
\centering
\caption{Cognitive Complexity Result of Written Code}
\scalebox{0.9}{\begin{tabular}{clr} 
\toprule
\multicolumn{1}{l}{\textbf{Language}} & \multicolumn{1}{c}{\textbf{Cognitive Complexity Level}} & \multicolumn{1}{c}{\textbf{Number of Code}}  \\ 
\midrule

\multirow{4}{*}{\textbf{Python3}}  & Low complexity ($<$5)                                      & \percent{357}{705}                     \\
                                      & Moderate complexity (6-10)                                & \percent{176}{705}                     \\
                                      & High complexity (11-20)                                  & \percent{112}{705}                     \\
                                      & Very High complexity (21+)                                & \percent{60}{705}                      \\
\bottomrule
\end{tabular}}
\label{tab:humancognitivecomplexity}
\end{table}


\begin{figure*}[t]
    \centering
    \subfigure{
    \includegraphics[width=0.35\textwidth]{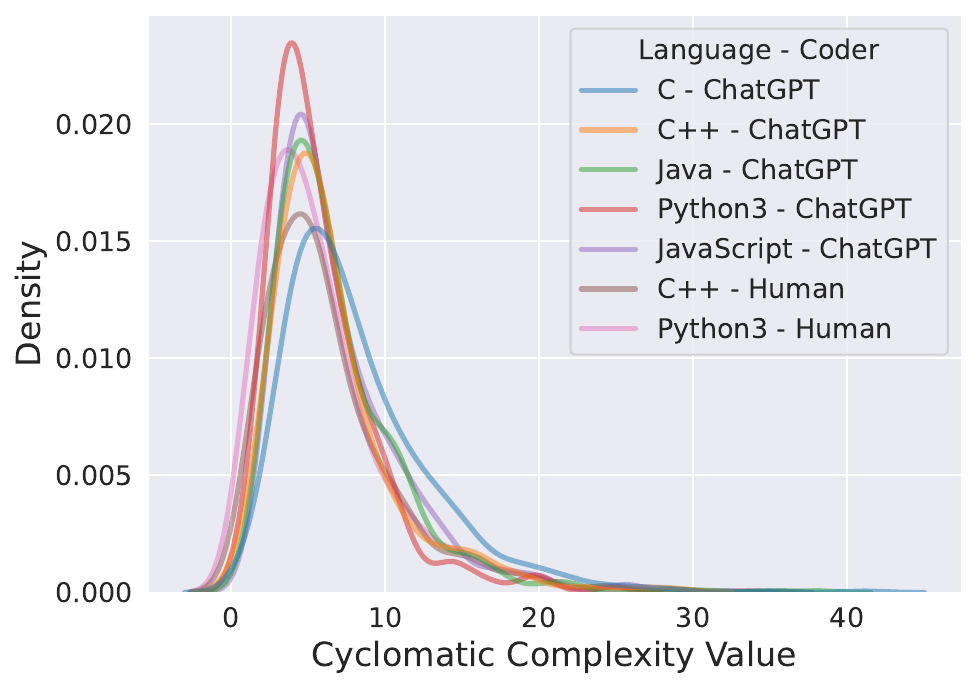}} \hspace{1.5cm}
    \subfigure{
    \includegraphics[width=0.357\textwidth]{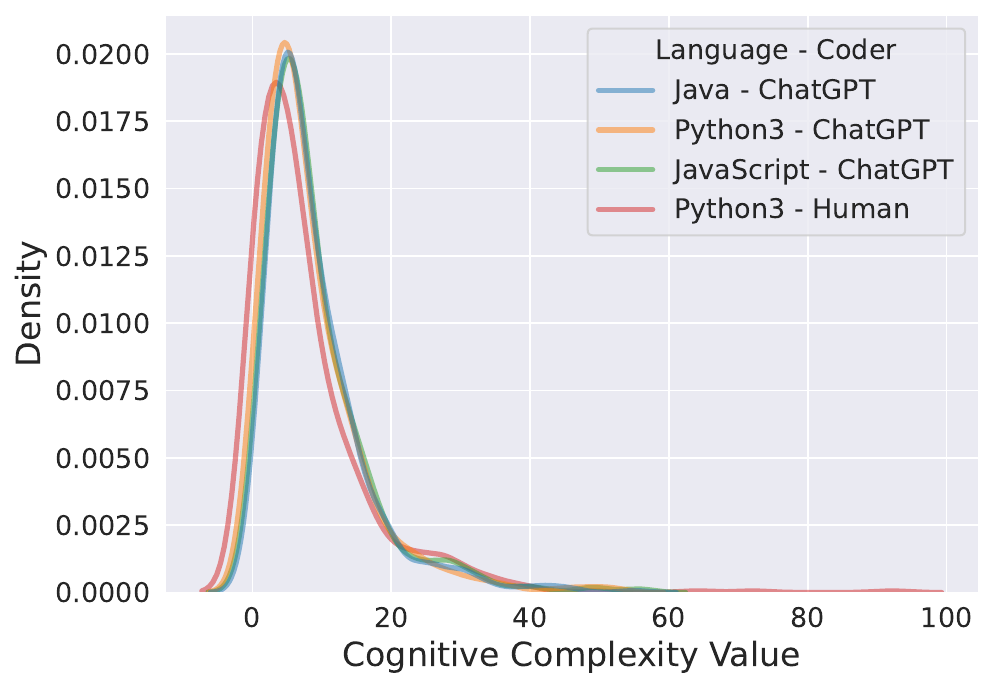}}
    \caption{Density graph of cyclomatic and cognitive complexity.}
    \label{fig:cyc-cog-distribution}
\end{figure*}

\noindent $\bullet$ \textbf{Cyclomatic Complexity.} Based on the official documentation of PMD~\cite{pmd}, cyclomatic complexity can be categorized into four classes low (1-4 cyclomatic complexity value), moderate (5-7), high (8-10), and very high complexity ($\geq$ 11). From Table \ref{tab:cyclomaticcomplexity}, we can find that both low and moderate complexities dominate more than $50\%$ in five languages for generated code, with C having the lowest percentage at $54.9\%$, while other languages exceed $66\%$. The difference is at least $11\%$. For the other four languages, their complexity distributions are similar. Specifically, the differences between the maximum and minimum of four complexity levels from low to high are $12\%$, $5.7\%$, $2.8\%$, and $7.8\%$. When excluding Python3, the differences decrease to $1.7\%$, $ 3.4\%$, $2.8\% $, and $2.4\%$. Notably, Python3 has a much higher percentage ($44.5\%$) of low complexity and a much lower percentage ($8.5\%$) of very high complexity, while C++, Java, and JavaScript have more similar distributions of the four complexity levels. 


Compared with human solutions in C++ and Python3 (see Table \ref{tab:humancyclomaticcomplexity}), we find that the complexity distributions of the generated code for both languages closely resemble those of the written code. For C++, the written code's percentage of low complexity is $5\%$ higher than the generated code's, and correspondingly, the one of moderate complexity is $5\%$ lower than the generated code's. They have similar percentages of high and very high complexities. As for Python3, both generated code and written code have similar percentages of low complexity, while the percentages of moderate and high complexities of generated code are higher than written code's by $3\%$ and $2.6\%$. Consequently, the former's percentage of very high complexity is lower than the latter's with $4.3\%$.

\noindent $\bullet$ \textbf{Cognitive Complexity.} According to \cite{cognitive-complexity}, cognitive complexity can also be categorized into four classes low ($<$5 cognitive complexity value), moderate (6-10), high (11-20), and very high complexity ($\geq$ 21). From Table \ref{tab:cognitivecomplexity}, we can see that both low and moderate complexities dominate more than $70\%$ in five languages. The generated code in Java and JavaScript has similar distributions of four complexity levels. For Python3, it has a $6\%$ higher percentage ($42.6 \%$) of low complexity, and correspondingly, its percentage of moderate complexity is lower than the other two languages' with $4.7\%$. Python3's percentages of high and very high complexities are also similar to those in Java and JavaScript. 

Compared with human solutions in Python3 (see Table \ref{tab:humancyclomaticcomplexity}), we find that the written code has a higher percentage of low complexity with $8\%$ than the generated code. Correspondingly, the former's percentages of moderate and high complexity are lower than the latter's with double $4.9\%$, respectively. However, the difference between the two percentages of very high complexity for the two kinds of code is only $2\%$.

\noindent \textcolor{magenta}{$\bigstar$} \textbf{Summary 1.} Fig. \ref{fig:cyc-cog-distribution} also further shows the density graphs of cyclomatic and cognitive complexities for each language and corresponding coder (i.e., \textit{ChatGPT} and \textit{Human}) pair. The horizontal coordinate is the complexity value and the vertical one is the corresponding density. By analyzing the two figures, we can gain a more intuitive insight that \textit{ChatGPT}'s Python3's distributions have comparatively smaller mean, while C's ones have larger mean. The distributions of C++, Java, and JavaScript are nearly overlapping. In addition, the distributions of code written by humans skew to the left compared with code generated in corresponding languages by \textit{ChatGPT}. Therefore, we can conclude that the level of \revision{complexity} of code generated by \textit{ChatGPT} varies among the five programming languages. The generated code in C is more complex than the other languages, while the complexity of code in C++, Java, and JavaScript is comparable. The code in Python3 is the least complex. Moreover, the complexity of code generated by \textit{ChatGPT} in C++ and Python3 is slightly higher but nearly equal to that of code written by humans.


\begin{figure}[t]
    \centering
    \includegraphics[width=0.45\textwidth]{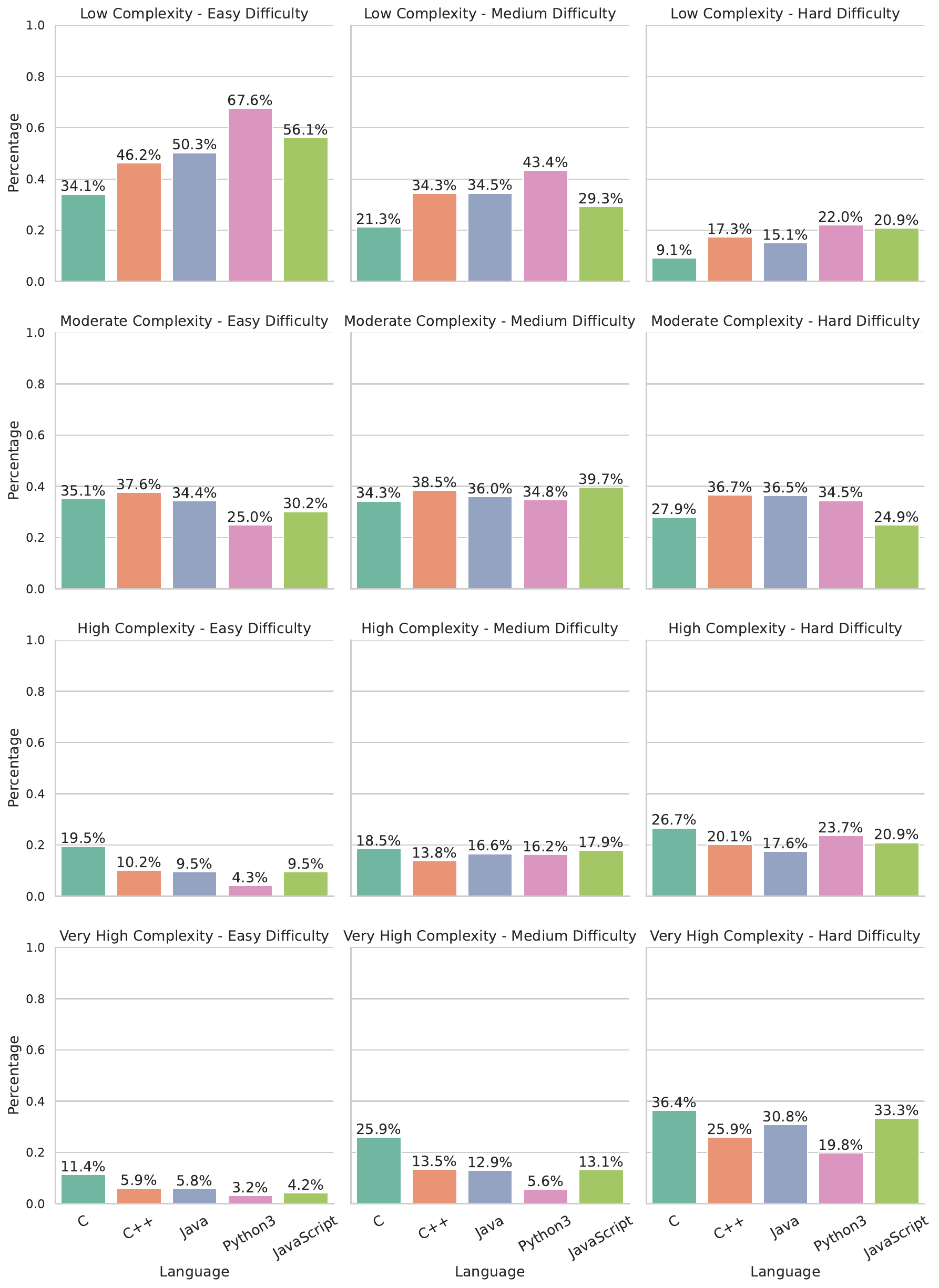}
    \caption{Distribution of cyclomatic complexity under three difficulty levels of problems.}
    \label{fig:cyc-d}
\end{figure}

\begin{figure}[t]
    \centering
    \includegraphics[width=0.45\textwidth]{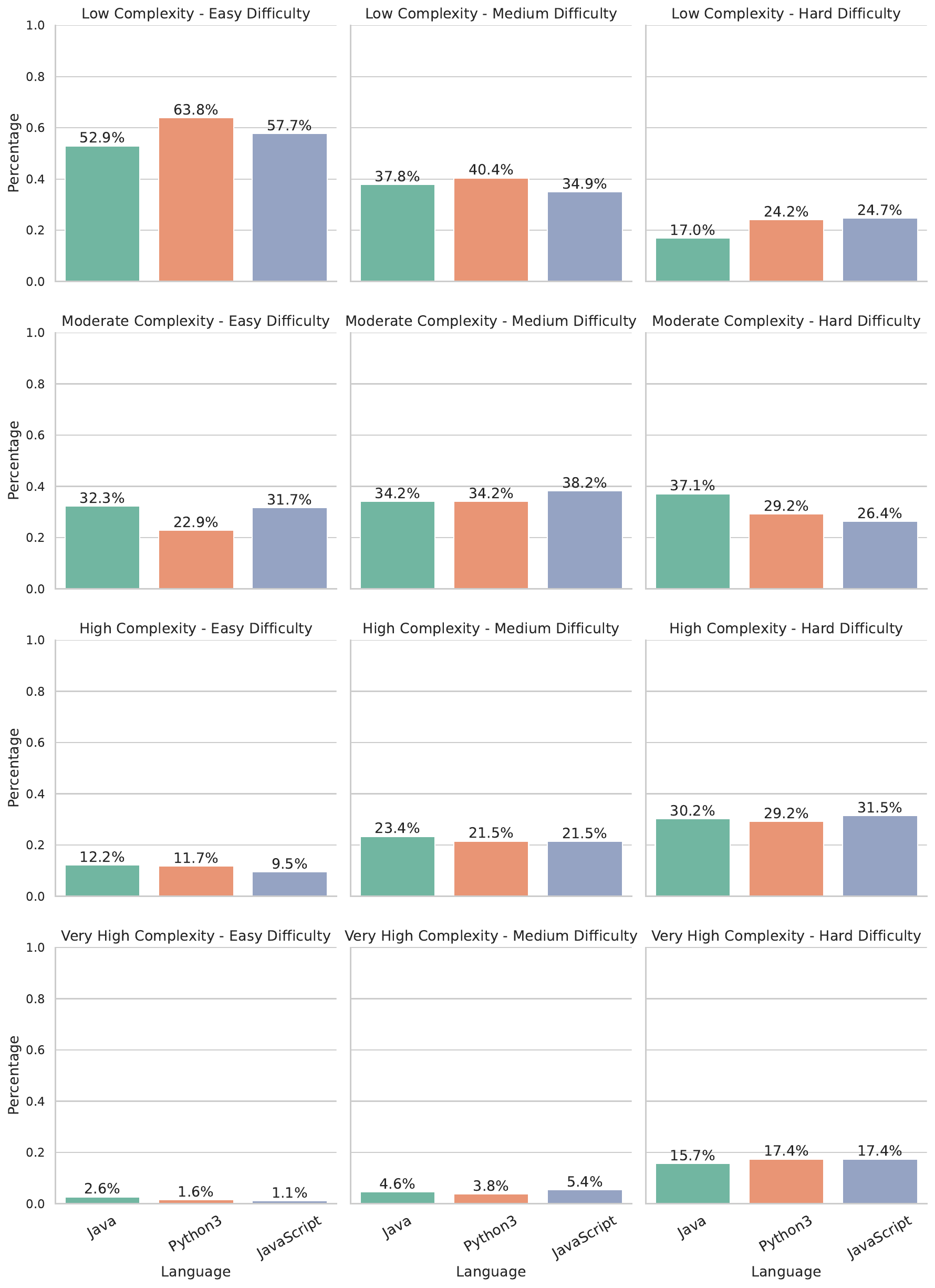}
    \caption{Distribution of cognitive complexity under three difficulty levels of problems.}
    \label{fig:cog-d}
\end{figure}

\begin{table*}
\centering
\caption{\revision{Comparision between Code Generated by \textit{ChatGPT} and Code Written by Humans to the Distributions of Cyclomatic and Cognitive Complexities under Three Difficulty Levels of Problems}}
\begin{tabular}{l|l|ll|ll|ll|ll|ll|ll}

\toprule

\multicolumn{1}{c|}{\multirow{2}{*}{Metric}} & \multicolumn{1}{c|}{\multirow{2}{*}{Level}} & \multicolumn{4}{c|}{Easy Difficulty}                    & \multicolumn{4}{c|}{Medium Difficulty}                 & \multicolumn{4}{c}{Hard Difficulty}                    \\

\cline{3-14}

                                   &                        & \multicolumn{1}{c}{C++}     & \multicolumn{1}{c|}{C++-H} & \multicolumn{1}{c}{Python} & \multicolumn{1}{c|}{Py-H} & \multicolumn{1}{c}{C++}    & \multicolumn{1}{c|}{C++-H} & \multicolumn{1}{c}{Python} & \multicolumn{1}{c|}{Py-H} & \multicolumn{1}{c}{C++}   & \multicolumn{1}{c|}{C++-H} & \multicolumn{1}{c}{Python} & \multicolumn{1}{c}{Py-H}  \\

\cline{1-14}

\multirow{4}{*}{Cyclomatic}        & Low                    & 46.2\% & 56.5\%      & 67.6\%   & 69.0\%             & 34.3\% & 39.0\%        & 43.4\%   & 46.0\%             & 17.3\% & 16.3\%      & 22.0\%     & 21.2\%            \\
                                   & Moderate               & 37.6\%  & 31.5\%      & 25.0\%     & 21.9\%           & 38.5\% & 34.3\%      & 34.8\%   & 33.0\%             & 36.7\% & 31.1\%      & 34.5\%   & 29.1\%            \\
                                   & HIgh                   & 10.2\%  & 9.2\%       & 4.3\%    & 8.0\%              & 13.8\% & 15.3\%      & 16.2\%   & 11.2\%           & 20.1\% & 15.6\%      & 23.7\%   & 19.0\%              \\
                                   & Very High              & 5.9\%   & 2.7\%       & 3.2\%    & 1.1\%            & 13.5\% & 11.3\%      & 5.6\%    & 9.7\%            & 25.9\% & 37.0\%        & 19.8\%   & 30.7\%            \\

\cline{1-14}
                                
\multirow{4}{*}{Cognitive}         & Low                    &    \multicolumn{1}{c}{-}   &      \multicolumn{1}{c|}{-}     & 63.8\%   & 73.3\%           &  \multicolumn{1}{c}{-}    &     \multicolumn{1}{c|}{-}      & 40.4\%   & 51.3\%           &   \multicolumn{1}{c}{-}   &       \multicolumn{1}{c|}{-}    & 24.2\%   & 25.7\%            \\
                                   & Moderate               &    \multicolumn{1}{c}{-}   &      \multicolumn{1}{c|}{-}     & 22.9\%   & 18.2\%           &   \multicolumn{1}{c}{-}   &     \multicolumn{1}{c|}{-}      & 34.2\%   & 27.4\%           &   \multicolumn{1}{c}{-}   &       \multicolumn{1}{c|}{-}    & 29.2\%   & 27.4\%            \\
                                   & HIgh                   &     \multicolumn{1}{c}{-}  &      \multicolumn{1}{c|}{-}     & 11.7\%   & 8.0\%              &   \multicolumn{1}{c}{-}   &  \multicolumn{1}{c|}{-}         & 21.5\%   & 16.5\%           &    \multicolumn{1}{c}{-}  &       \multicolumn{1}{c|}{-}    & 29.2\%   & 22.9\%            \\
                                   & Very High              &    \multicolumn{1}{c}{-}   &    \multicolumn{1}{c|}{-}       & 1.6\%    & 0.5\%            &   \multicolumn{1}{c}{-}   &    \multicolumn{1}{c|}{-}       & 3.8\%    & 4.7\%            &   \multicolumn{1}{c}{-}   &      \multicolumn{1}{c|}{-}     & 17.4\%   & 24.0\% \\    

                                \bottomrule
\end{tabular}
\label{tab:crossdifficulty-compared-to-human}
\end{table*}

\noindent $\bullet$ \textbf{Cross-difficulty Comparisons.} We further examine the distributions of cyclomatic complexity and cognitive complexity under different difficulty levels. The results are shown in Fig. \ref{fig:cyc-d} and Fig. \ref{fig:cog-d}. Each row in the figures corresponds to the percentages of the same complexity level at different difficulty levels, while each column represents to the percentages of different complexities at the same difficulty level. The percentages in the graph indicate the proportion of a certain complexity level within the same difficulty level. 

Regardless of cyclomatic complexity or cognitive complexity, the low complexity of each language decreases as the difficulty of the problem increases. On the other hand, high and very high complexity increase with the difficulty of the problem increasing. Moderate complexity shows no significant changes as the difficulty of the problem increases. The high and very high complexity percentages increase as the difficulty of the problem increases. \revision{Compared to code written by humans~\cite{leetcodesolution} which is shown in Table \ref{tab:crossdifficulty-compared-to-human} (suffix -H represents human-based results), the complexity trend in \textit{ChatGPT}-generated code across different difficulty levels is comparable to that observed in human-written code.} This trend may be attributed to more difficult problems often requiring the handling of more conditions, loops, and nested structures, resulting in more complex generated code. For problems of the same difficulty, the proportions of low and moderate complexity in the generated code in C++, Java, and Python3 by \textit{ChatGPT} are all over $50\%$, even for hard difficulty problems. The proportion of low and moderate complexity in the generated JavaScript code snippets is also over $50\%$, but the cyclomatic complexity of $45.8\%$ is slightly below $50\%$. The generated code in C has a proportion of only $37\%$ for low and moderate complexity in cyclomatic complexity. Moreover, Python3 has the highest low complexity and moderate complexity among all languages, regardless of the difficulty level of the problem. In contrast, C has the lowest one. C++, Java, and JavaScript fall between these two. 


\noindent \textcolor{magenta}{$\bigstar$} \textbf{Summary 1.} Regardless of complexity type, low complexity decreases while high and very high complexity increase with increasing problem difficulty \revision{for \textit{ChatGPT}-based code generation. The trend is also comparable to that observed in human-written code}. Moreover, Python3 consistently has the highest proportion of low and moderate complexity. In contrast, C exhibits the lowest proportion of low and moderate complexity. C++, Java, and JavaScript fall between these two.




\begin{figure}[t]
    \centering
    \includegraphics[width=0.47\textwidth]{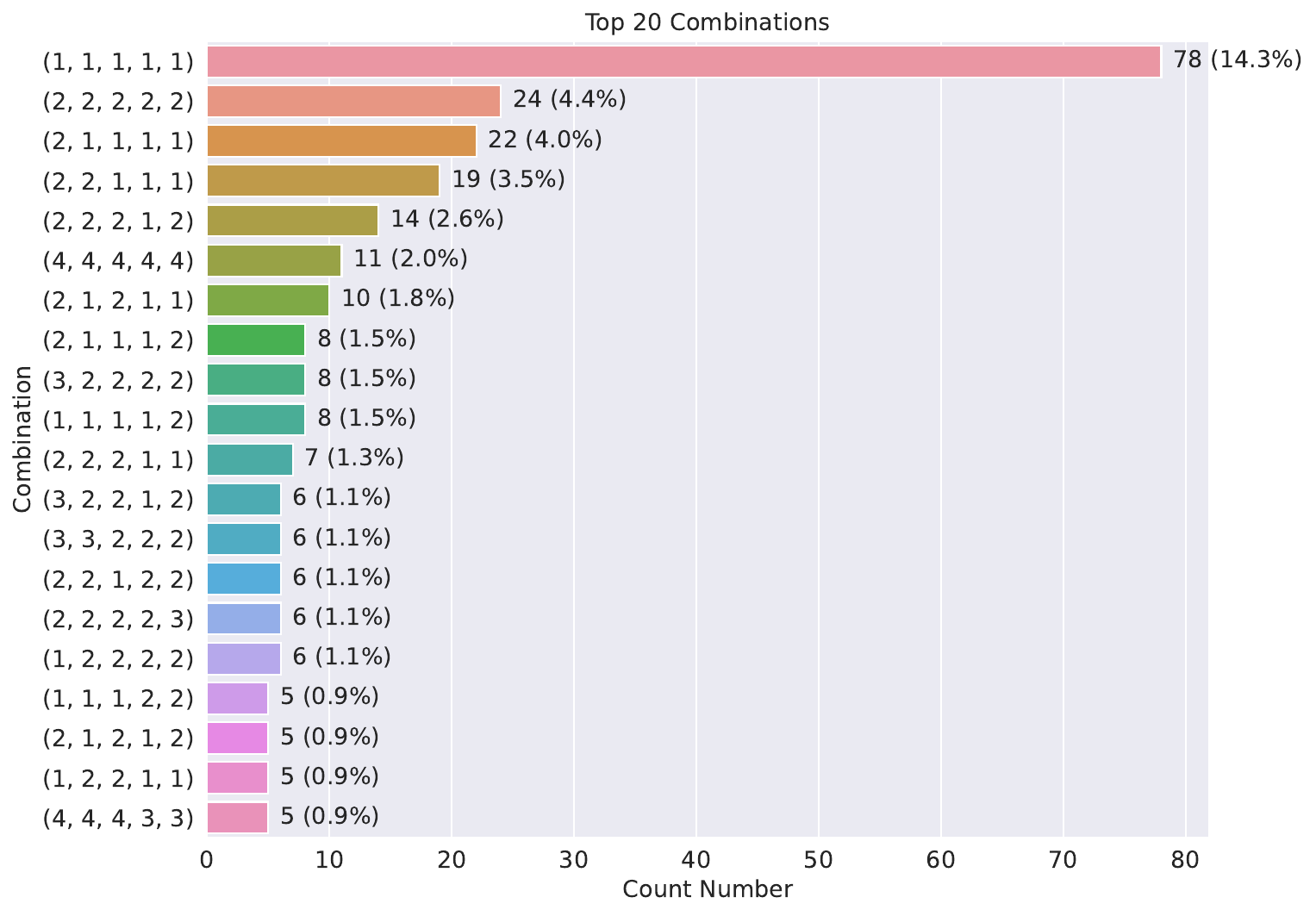}
    \caption{Top 20 numbers of cyclomatic complexity combinations in five different languages (C, C++, Java, Python3, and JavaScript) of the same problems.}
    \label{fig:top-20-cyc}
\end{figure}

\begin{figure}[t]
    \centering
    \includegraphics[width=0.47\textwidth]{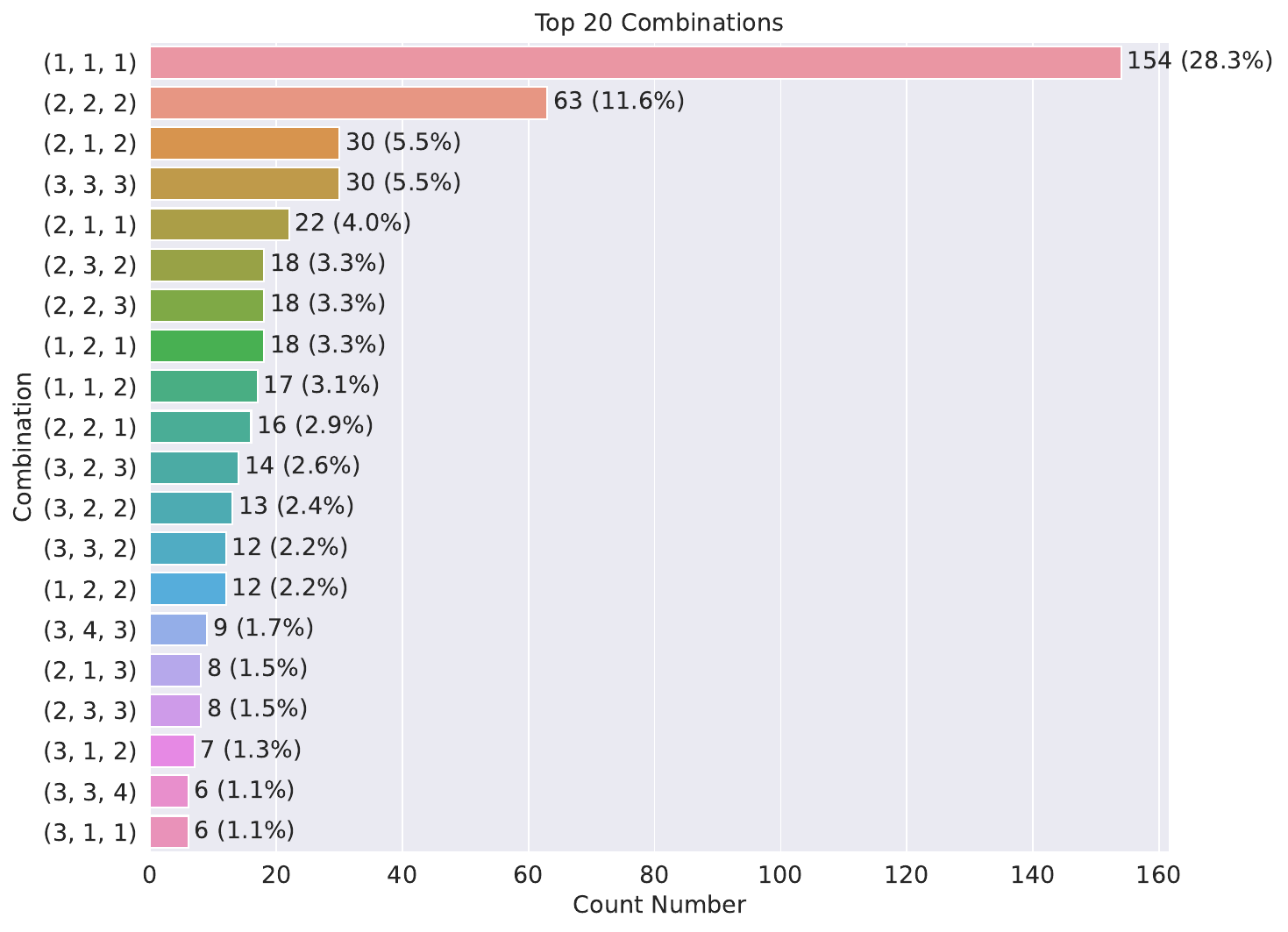}
    \caption{Top 20 numbers of cognitive complexity combinations in three different languages (Java, Python3, and JavaScript) of the same problems.}
    \label{fig:top-20-cog}
\end{figure}

\begin{table}[t]
\centering
\caption{Numbers of Differences for Cyclomatic Complexity and Cognitive Complexity Combinations}
\scalebox{0.95}{\begin{tabular}{ccc} 
\toprule
\textbf{Difference} & \textbf{Cyclomatic Complexity} & \textbf{Cognitive Complexity}  \\ 
\midrule

1 & \multicolumn{1}{r}{\percent{237}{427}} & \multicolumn{1}{r}{\percent{225}{294}} \\
2 & \multicolumn{1}{r}{\percent{139}{427}} & \multicolumn{1}{r}{\percent{60}{294}} \\
3 & \multicolumn{1}{r}{\percent{51}{427}} & \multicolumn{1}{r}{\percent{9}{294}}  \\

\bottomrule
\end{tabular}}
\label{tab:diff-1-2-3}
\end{table}

\noindent $\bullet$ \textbf{Cross-language Comparisons.} We measure the combinations of cyclomatic complexity and cognitive complexity for five different languages (C, C++, Java, Python3, and JavaScript) and three different languages (Java, Python3, and JavaScript) for the same problems, respectively. Thus, we only include problems containing 5 valid (exist and are not constant functions) different language code snippets. The statistical results of the top 20 combinations are depicted in Fig. \ref{fig:top-20-cyc} and Fig. \ref{fig:top-20-cog}. Cyclomatic complexity has 212 combinations and cognitive complexity has 52 combinations, in total. The numbers 1, 2, 3, and 4 in parentheses represent low, moderate, high, and very high complexities, respectively. The positions of the elements in parentheses correspond to languages. For cyclomatic complexity' \textit{(v, w, x, y, z)}, \textit{v} to \textit{z}'s corresponding languages are C, C++, Java, Python3, and JavaScript, respectively. Similarly, for cognitive complexity's \textit{(x, y, z)}, their corresponding languages are Java, Python3, and JavaScript, respectively. From the results, we can see that the numbers of \textit{(1, 1, 1, 1, 1)} and \textit{(1, 1, 1)} are the most in cyclomatic complexity and cognitive complexity, reaching $14.3\%$ and $28.3\%$, respectively. \textit{(2, 2, 2, 2, 2)} and \textit{(2, 2, 2)} both rank second in their respective complexities. \textit{(4, 4, 4, 4, 4)} and \textit{(3, 3, 3)} rank sixth and fourth respectively. For all these combinations with the same complexity levels, they have high individual percentages but their overall percentages are below $50\%$, indicating that the majority of combinations have different complexities. As the results of combinations having different complexities shown in Fig. \ref{fig:top-20-cyc} and Fig. \ref{fig:top-20-cog}, the majority of the differences in complexity levels are 1 (e.g., \textit{(2, 1, 1, 1, 1)} and \textit{(2, 1, 2)}), with a few ones greater than 1 (e.g., \textit{(3, 2, 2, 1, 2)} and \textit{(2, 1, 3)}). We further count the number of differences for each value (i.e., 1, 2, 3) (see Table \ref{tab:diff-1-2-3}). Regardless of cyclomatic complexity or cognitive complexity, the number of difference of 1 accounts for more than $50\%$, especially for cognitive's $76.5\%$. In most cases, for the same problem, the complexities of code snippets generated by \ChatGPT in different languages are similar. We also manually inspect the code snippets and summarize the differences in complexity levels between generated code snippets for the same problem in different languages:

\whiteding{1} \textbf{\revision{Built-in Libraries}}: \revision{Different languages have different numbers of built-in libraries\footnote{https://www.python.org/doc/essays/comparisons/}. \textit{ChatGPT} learns from a large corpus of text~\cite{ChatGPT} and may have the ability to choose whether to use built-in libraries to simplify algorithm implementation or to generate helper functions to achieve specific functionality, in different languages. For example, in Python3, \textit{ChatGPT} can directly use \texttt{heapq} to implement a min-heap, whereas in C, it needs to generate the relevant code for a min-heap as well, leading to different complexity level.}


\whiteding{2} \textbf{Different Algorithms}: The code snippets generated for the same problem in different languages do not always use the same algorithm. Different algorithms can lead to different complexities. For example, the cyclomatic complexity combination of problem 2543\footnote{https://leetcode.com/problems/check-if-point-is-reachable/} is \textit{(3, 4, 2, 2, 1)}. The code snippet in C uses an iterative algorithm, and the code snippets in C++, Java, and Python3 use a recursive algorithm. However, the code snippet in JavaScript uses an algorithm based on number theory.

\whiteding{3} \textbf{Implementation of Logic}: The complexities of code snippets in different languages may vary due to the specific implementation of the same or similar algorithms. One example is the problem of Fig. \ref{fig:wrong-detail-code}. The corresponding cyclomatic complexity combination is \textit{(4, 3, 2, 1, 2)}. All five code snippets use the same algorithm, however, the logic implementation of the code snippet in Python3 is the most concise.

\noindent \textcolor{magenta}{$\bigstar$} \textbf{Summary 1.} By the analysis of cross-language comparisons, it is observed that the majority of combinations of cyclomatic complexity and cognitive complexity for the same problem across different languages exhibit similar ($\leq$ 1 of complexity difference) complexity levels. Factors \revision{observed} contributing to the differences include the \revision{built-in libraries in} languages, different algorithms employed, and variations in the implementation of logic.


\begin{figure*}[t]
    \centering
    \subfigure[Cyclomatic - C]{\label{fig:heatmap-w-a}
    \includegraphics[width=0.23\textwidth]{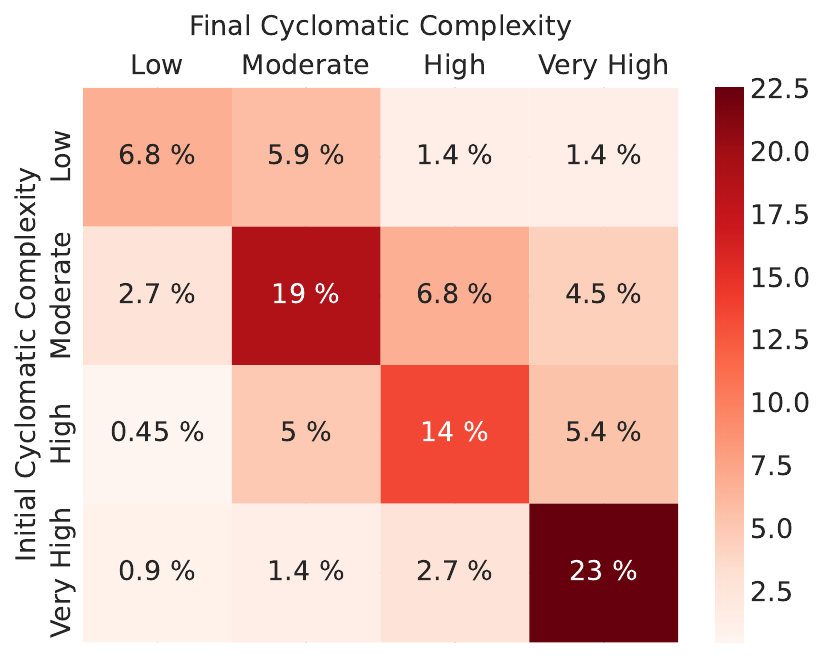}}
    \subfigure[Cyclomatic - C++]{\label{fig:heatmap-w-b}
    \includegraphics[width=0.23\textwidth]{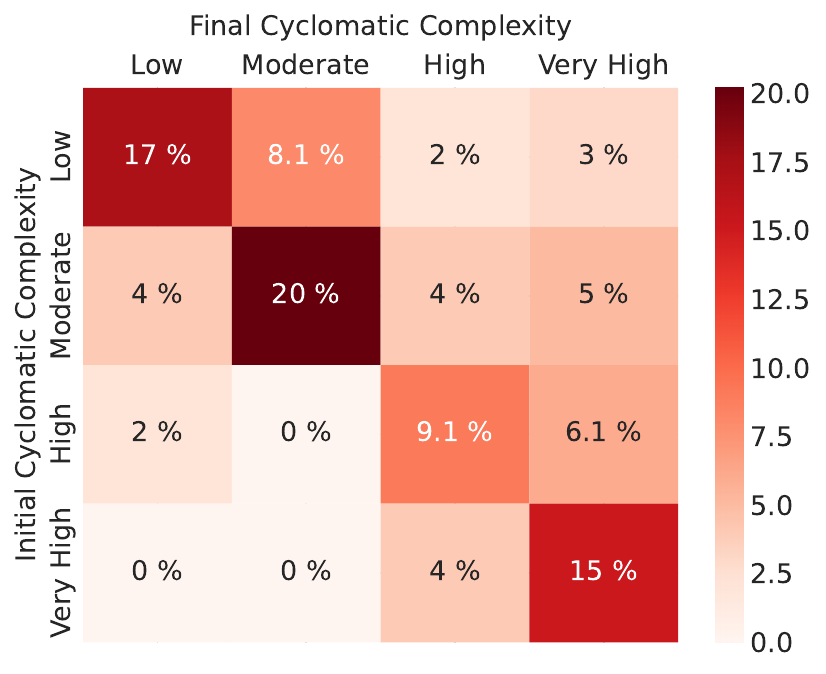}}
    \subfigure[Cyclomatic - Java]{\label{fig:heatmap-w-c}
    \includegraphics[width=0.23\textwidth]{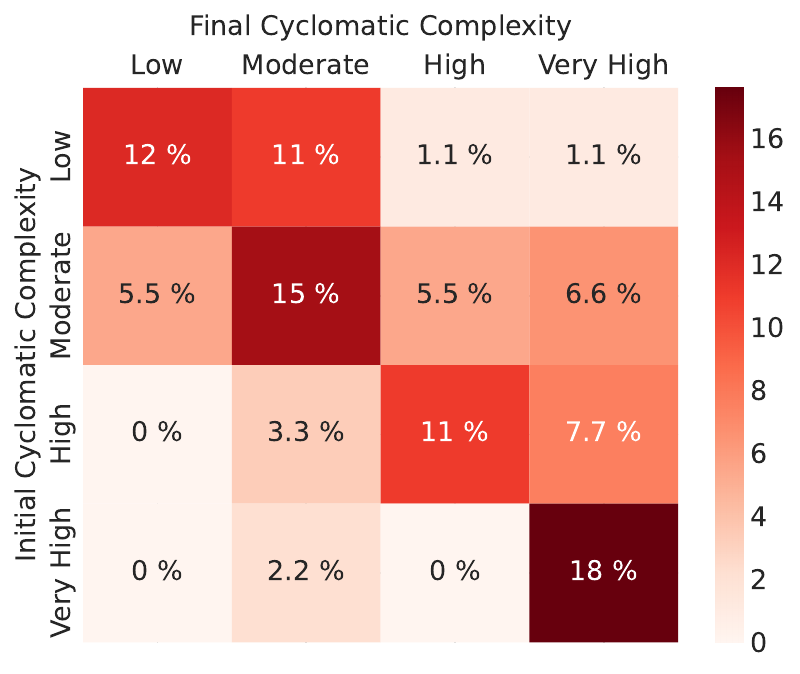}}
    \subfigure[Cyclomatic - Python3]{\label{fig:heatmap-w-d}
    \includegraphics[width=0.23\textwidth]{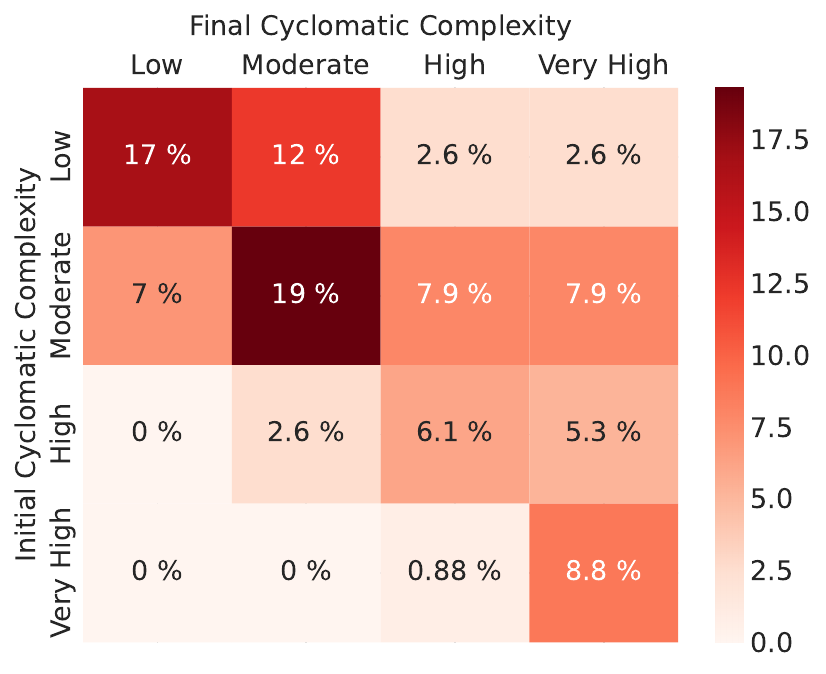}
   }
    \subfigure[Cyclomatic - JavaScript]{\label{fig:heatmap-w-e}
    \includegraphics[width=0.23\textwidth]{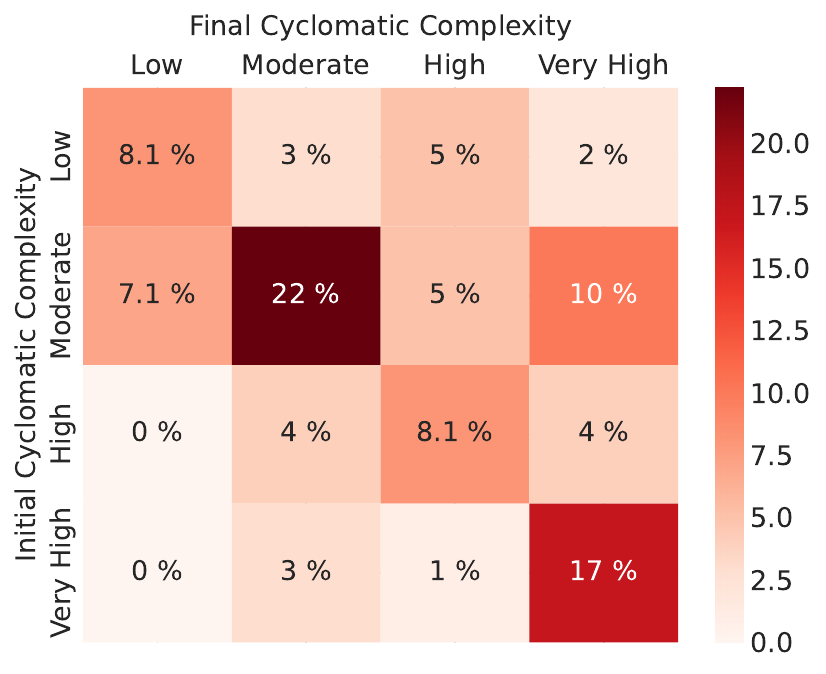}
    }
    \subfigure[Cognitive - Java]{\label{fig:heatmap-w-f}
    \includegraphics[width=0.23\textwidth]{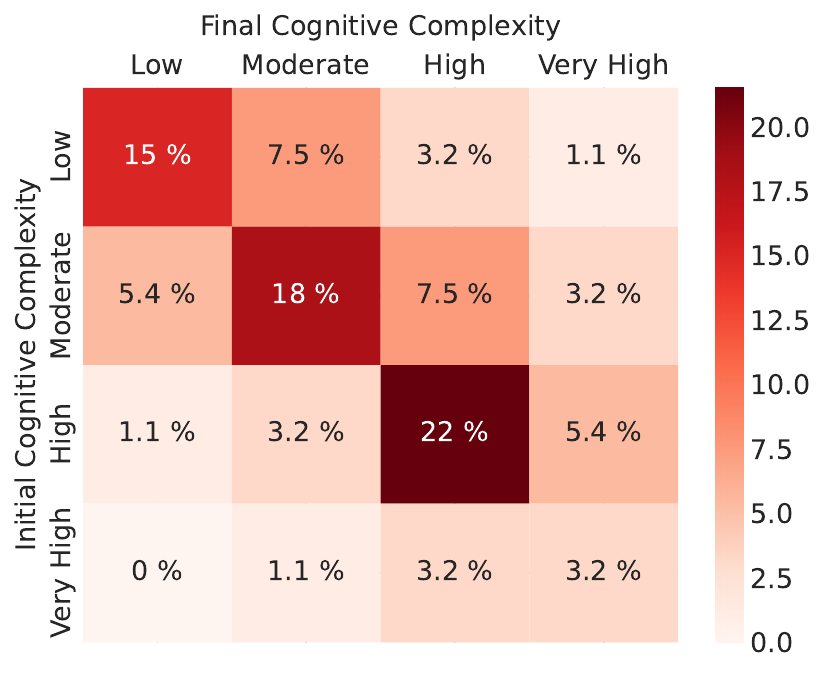}
    }
    \subfigure[Cognitive - Python3]{\label{fig:heatmap-w-g}
    \includegraphics[width=0.23\textwidth]{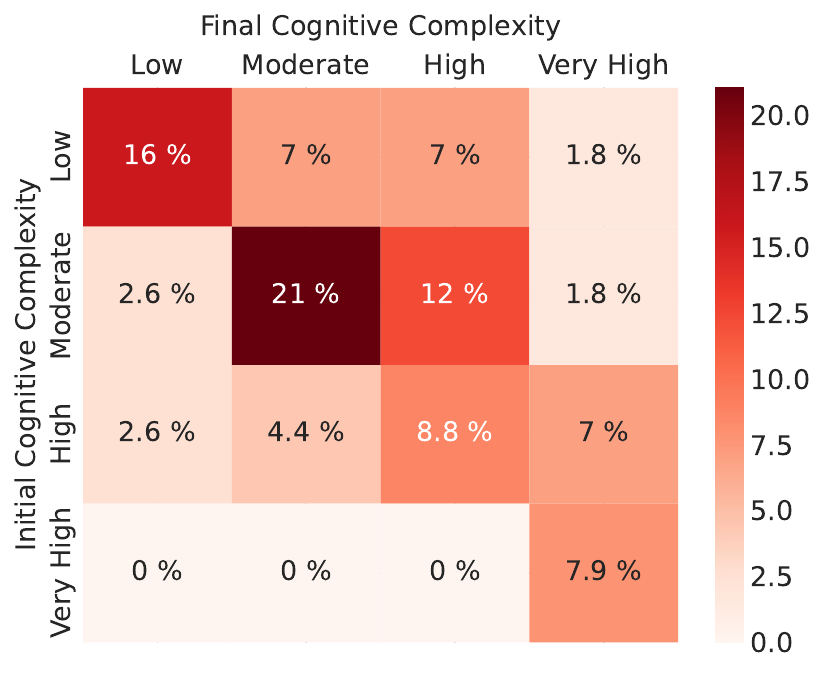}
    }
    \subfigure[Cognitive - JavaScript]{\label{fig:heatmap-w-h}
    \includegraphics[width=0.23\textwidth]{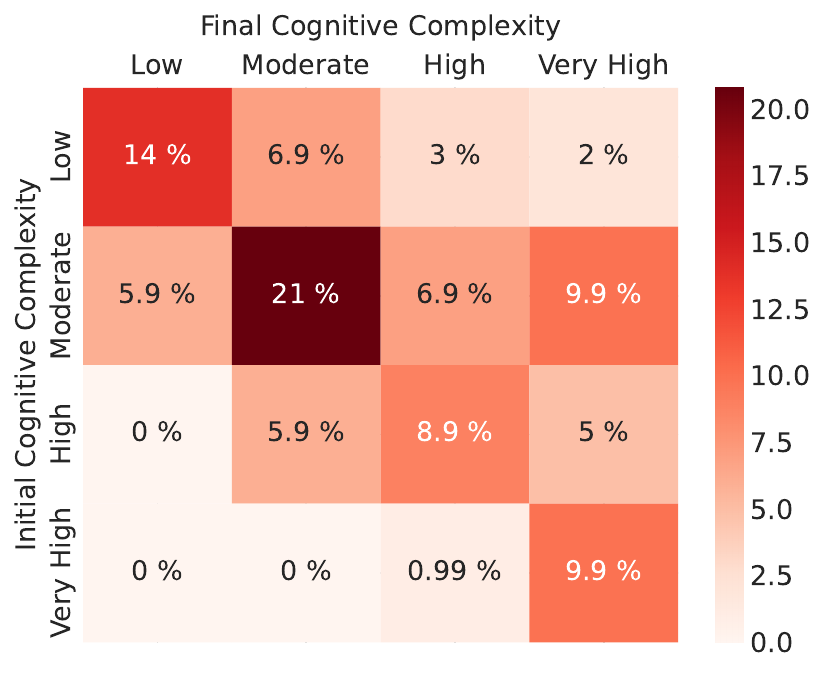}
    }
    \caption{Heatmap of the numbers of complexity levels of the original code snippets and the final code snippets under multi-round process.}
    \label{fig:heatmap-w}
\end{figure*}

\noindent $\bullet$ \textbf{Multi-round Comparisons.} \textit{ChatGPT}'s multiple rounds of conversations allow it to continuously generate code snippets. We take all code snippets from \textit{multi-round fixing} in Sec. \ref{sec:FunctionallyCorrectCodeGeneration} as samples to study the variations in code snippet complexity levels during the multi-turn process. Since the numbers of multiple rounds in conversations for different code snippets may be different, we use the initial code snippets and the code snippets generated at the end of the conversations as objects. Fig. \ref{fig:heatmap-w} shows the relationship between the initial code snippet complexity levels and the final code snippet complexity levels for different languages and complexities (i.e., cyclomatic and cognitive). The y-label in the 8 sub-figures represents the complexity levels of the initial code snippets, while the x-label represents the complexity levels of the final code snippets. The percentages shown in the figure represent the proportions of different complexity level variations in the cases of \textit{complexity-language}. 

From the figure, it can be seen that in all cases of \textit{complexity-language}, the total percentages shown by the diagonals are all higher than $50\%$. This indicates that a significant number of code snippets maintain their complexity levels throughout the multi-round process. Moreover, all cells above the diagonal lines, corresponding to cases where the final complexity level is higher than the initial level, generally exhibit higher percentages compared to the cells symmetrically opposite along the diagonals except the ones of (Low, Moderate) and (Moderate, Low) in \textit{Cyclomatic-JavaScript}. It is also worth noting that all cells with $0\%$ are also below the diagonal lines. Therefore, we can conclude that the multi-round fixing process with \textit{ChatGPT} generally preserves or increases the complexity levels of code snippets. 

We also observe these pairs of \textit{<initial code snippet, final code snippet>}. For the code snippets in the cells with preserving or increasing complexity levels, \textit{ChatGPT} patches the initial code snippets by adjusting the logical implementation (e.g., change the recursive implementation of DFS to an iterative implementation or fix type error), adding or modifying conditions, or changing the algorithms used (e.g., change brute-force algorithm to dynamic programming algorithm). As for the code snippets in the cells with decreasing complexity levels, \textit{ChatGPT} patches the initial code snippet by also adjusting the logical implementation (e.g., simplify and fix the implementation of logic for Fig. \ref{fig:wrong-detail-code}), deleting or modifying conditions, or changing the algorithms used (e.g., turn a binary search-based algorithm to an iterative algorithm containing fewer control flows in \textit{<problem 2483\footnote{https://leetcode.com/problems/minimum-penalty-for-a-shop/description/}, JavaScript>}).  

\noindent \textcolor{magenta}{$\bigstar$} \textbf{Summary 1.} The multi-round fixing process with \ChatGPT generally preserves or increases the complexity levels of code snippets, which may potentially make it increasingly difficult to understand the automatically and consistently generated code by \textit{ChatGPT}.

\begin{tcolorbox}[boxrule=1pt,boxsep=1pt,left=2pt,right=2pt,top=2pt,bottom=2pt,title=Answer to RQ3: Code Complexity]

\noindent \blackding{1} The generated code in C is the most complex code, while the code in C++, Java, and JavaScript has comparable complexity. The code in Python3 is the least complex code. The complexity of the generated code in C++ and Python3 is similar (slightly higher) to the written code. Additionally, low complexity decreases while high and very high complexity increase with increasing problem difficulty for code generation;

\noindent \blackding{2} Code complexity levels for the same problem differ among programming languages. Python3 has the highest probability of generating code with the lowest complexity level, while C has the lowest probability. C++, Java, and JavaScript have intermediate probabilities. This suggests that the choice of programming language affects generated code complexity;

\noindent \blackding{3} The multi-round fixing process with \ChatGPT generally preserves or increases the complexity levels of code snippets, which may potentially make it increasingly difficult to understand the automatically and consistently generated code by \textit{ChatGPT}.

\end{tcolorbox}



\subsection{Security Code Generation}\label{sec:securitycodegeneration}
\noindent \textbf{RQ4: Is the code generated by \textit{ChatGPT} secure?}

\noindent \textbf{Motivation.} \textit{ChatGPT} may learn knowledge from vulnerable code. In this RQ, we intend to evaluate the security of code generated by \textit{ChatGPT} \revision{in several specific vulnerability scenarios, queries, and languages}.  

\noindent \textbf{Approach \blackding{1}.} We utilize \textit{CodeQL}~\cite{CodeQL} for vulnerability detection on all the C, C++, \revision{and Java} code snippets generated in Sec. \ref{sec:FunctionallyCorrectCodeGeneration}\footnote{The algorithmic code typically focuses on solving specific logical or computational problems and often does not involve managing system resources, network communications, or other operations that are commonly sensitive to security issues.}. We do not perform detection for the code snippets in other languages \revision{(i.e., Python and JavaScript)} since \textit{CodeQL} standard library \revision{and other vulnerability detection tool \textit{SonarQube}~\cite{sonarqube}} has no suitable queries~\cite{codeql-cwe} \revision{related to algorithm problems for them}. \revision{The vulnerability detection is limited to these three languages for code snippets generated based on \textit{LeetCode} problems.} Moreover, we only perform detection on pointer and memory-related vulnerabilities due to that the code snippets are for algorithm problems \revision{(\textit{CodeQL} and \textit{SonarQube} are limited in these two kinds of vulnerabilities to Python and JavaScript)}. We conduct vulnerability detection on 5 CWEs in \textit{MITRE Top 25 CWEs}~\cite{mitre}, which are CWE-787 (Out-of-bounds Write), CWE-416 (Use After Free), CWE-476 (NULL Pointer Dereference), CWE-190 (Integer Overflow or Wraparound), and CWE-119 (Improper Restriction of Operations within the Bounds of a Memory Buffer), containing 30 queries in total. \revision{While the Java code snippets are only detected with 4 queries of CWE-190 which are ArithmeticTainted, ArithmeticUncontrolled, ArithmeticWithExtremeValues, and ComparisonWithWiderType, since there are also no other suitable queries in both \textit{CodeQL} and \textit{SonarQube}.} The specific queries can be found in Table \ref{tab:securitycodegeneration-leet}, and their meanings can be found in \cite{codeql-cwe}.

We study the multi-round fixing for fixing code snippets with vulnerabilities. The approach is introduced in the corresponding part of this section.

\noindent \textbf{Approach \blackding{2}.} We follow the same setup used in \cite{pearce2022asleep}. Specifically, we utilize 18 common weakness enumerations (CWEs) in \textit{MITRE Top 25 CWEs}~\cite{mitre} (3 of them drop in rank to below 25 in 2022 \textit{MITRE Top 25 CWEs}). For each CWE, three different code scenarios (context) are provided to generate complete code by \textit{ChatGPT}. These scenarios are small and incomplete code snippets (in C and Python3) of which completed code may contain relative and specific CWEs. They come from \textit{CodeQL}~\cite{CodeQL}, \textit{MITRE}, and \cite{pearce2022asleep}. For example, Fig. \ref{fig:cwe-787-example} shows a code scenario example of CWE-787 (Out-of-bounds Write), where ''\texttt{ChatGPT next line}'' is used to tell \textit{ChatGPT} to complete code from here. We ask \textit{ChatGPT} to generate 60 complete code snippets \revision{in one-round process} for each scenario and leverage \textit{CodeQL}~\cite{CodeQL} to analyze whether these code snippets have corresponding CWE. Moreover, we only analyze specific CWEs to corresponding scenarios without evaluating functional correctness~\cite{pearce2022asleep}. In total, there are 18 CWEs with 54 scenarios.

We also study the multi-round fixing for fixing code snippets with vulnerabilities. The approach is introduced in the corresponding part of this section.

\begin{figure}[t]
    \centering
  \begin{lstlisting}[language=C]
...
#define FILENAME "data.csv"
int main(int argc, char *argv[]) {
    // generate 3 random floats
    float a = (float)rand() / (float)RAND_MAX;
    ...
    // convert to string
    // ChatGPT next line
}
  \end{lstlisting}
  \caption{A code scenario example of CWE-787.}
  \label{fig:cwe-787-example}
\end{figure}

\noindent \textbf{Prompt \blackding{2}.} We utilize code scenarios (CWE's code scenarios or CWE scenarios) used in \cite{pearce2022asleep} with a little modification (specify \textit{ChatGPT} to generate code) as our prompts.


\begin{table}[t]
\centering
\caption{Result of Vulnerability Detection}
\scalebox{0.73}{\begin{tabular}{clrr} 
\toprule
\textbf{CWE}                  & \multicolumn{1}{c}{\textbf{Query}} & \multicolumn{1}{c}{\textbf{\# Vln.}} &  \multicolumn{1}{c}{\textbf{\# Fixed}} \\ 
\midrule
\multirow{4}{*}{787} & 1: PotentialBufferOverflow               &    \percent{5}{183} &  \percent[p]{5}{5}                               \\
                     & 2: InvalidPointerDeref               &      \percent{2}{183}  &  \percent[p]{2}{2}                                \\
                     & 3: BadlyBoundedWrite      &  \percent{0}{183}   &  \multicolumn{1}{c}{-}                                   \\ 
                     & 4: UnboundedWrite      &  \percent{1}{183}   &  \percent[p]{1}{1}                                   \\

\midrule
\multirow{1}{*}{416} & 1: UseAfterFree             &    \percent{0}{183}  & \multicolumn{1}{c}{-}                 \\

\midrule
\multirow{5}{*}{476} & 1: RedundantNullCheckSimple &    \percent{0}{183} & \multicolumn{1}{c}{-}                                \\
                    & 2: InconsistentNullnessTesting  &      \percent{0}{183}  & \multicolumn{1}{c}{-}                                \\
                   & 3: DangerousUseOfExceptionBlocks &  \percent{0}{183} & \multicolumn{1}{c}{-}                                     \\ 
                   & 4: MissingNullTest            &      \percent{168}{183}  & \percent[p]{15}{15}                                \\
                   & 5: RedundantNullCheckParam      &  \percent{3}{183}   & \percent[p]{3}{3}                                   \\

\midrule
\multirow{9}{*}{190} & 1: ArithmeticTainted             &    \percent{0}{183}    & \multicolumn{1}{c}{-}                            \\
                    & 2: ArithmeticUncontrolled            &      \percent{0}{183}  & \multicolumn{1}{c}{-}                               \\
                   & 3: AmbiguouslySignedBitField      &  \percent{0}{183}       & \multicolumn{1}{c}{-}                             \\ 
                    & 4: BadAdditionOverflowCheck             &    \percent{0}{183}   & \multicolumn{1}{c}{-}                           \\
                    & 5: SignedOverflowCheck            &      \percent{0}{183} & \multicolumn{1}{c}{-}                             \\
                   & 6: ArithmeticWithExtremeValues      &  \percent{0}{183}  & \multicolumn{1}{c}{-}                                \\ 
                   & 7: ComparisonWithWiderType             &    \percent{0}{183}  & \multicolumn{1}{c}{-}                           \\
                    & 8: IntegerOverflowTainted           &      \percent{0}{183}  & \multicolumn{1}{c}{-}                            \\
                   & 9: DangerousUseOfTransformationAfterOperation      &  \percent{0}{183}  & \multicolumn{1}{c}{-}                                  \\

\midrule
\multirow{11}{*}{119} & 1: OverflowBuffer             &    \percent{1}{183} & \percent[p]{1}{1}                                \\
                    & 2: OffsetUseBeforeRangeCheck            &      \percent{1}{183} & \percent[p]{1}{1}                                 \\
                   & 3: DoubleFree      &  \percent{0}{183}                                      \\ 
                   & 4: LateNegativeTest             &    \percent{1}{183} & \percent[p]{1}{1}                                \\
                    & 5: MissingNegativityTest           &      \percent{0}{183} & \multicolumn{1}{c}{-}                              \\
                   & 6: OverflowCalculated      &  \percent{0}{183}   & \multicolumn{1}{c}{-}                                \\ 
                   & 7: OverflowDestination             &    \percent{0}{183}  & \multicolumn{1}{c}{-}                           \\
                    & 8: ReturnStackAllocatedMemory            &      \percent{0}{183}    & \multicolumn{1}{c}{-}                           \\
                   & 9: UsingExpiredStackAddress      &  \percent{0}{183}   & \multicolumn{1}{c}{-}                                \\ 
                   & 10: MemoryUnsafeFunctionScan             &    \percent{1}{183}  & \percent[p]{1}{1}                               \\
                    & 11: BufferAccessWithIncorrectLengthValue            &      \percent{0}{183} & \multicolumn{1}{c}{-}                               \\

\bottomrule
\end{tabular}}
\label{tab:securitycodegeneration-leet}
\end{table}

\noindent \textbf{Result \blackding{1}.} Table \ref{tab:securitycodegeneration-leet} shows the results of vulnerability detection. \textbf{\# Vln.} represents the number of vulnerable code snippets under the corresponding CWEs and queries. The percentage is the number of code snippets for a specific vulnerability query (e.g., CWE-476's MissingNullTest) divided by the total amount (183) of vulnerable code snippets. As the results show, the majority of vulnerable code snippets are in the query of MissingNullTest, accounting for $91.8\%$. The code generated by \ChatGPT does not perform a \texttt{NULL} test after allocating memory \revision{in C or C++ languages}, which may lead to potential vulnerabilities. For the remaining vulnerability queries, their vulnerable code snippets are relatively less frequent ($\leq 5$), such as PotentialBufferOverflow and OffsetUseBeforeRangeCheck, but they are still significant and should not be overlooked. \revision{Additionally, there are no detected vulnerabilities in CWE-416 and CWE-190 to C, C++, and Java languages.}

\noindent \textcolor{red}{$\blacktriangleright$} \textit{\textbf{Multi-round Fixing:}} We sample 15 code snippets containing vulnerabilities from each query of CWE categories. If there are fewer than 15 vulnerable code snippets in a query, we sample them all. The round number is set to 5 same as Sec. \ref{sec:FunctionallyCorrectCodeGeneration}. For each vulnerable code snippet, we create an initial prompt by leveraging the corresponding problem, the vulnerable code snippet, and the corresponding CWE information provided by \textit{CodeQL}. The CWE information includes an explanation of the CWE and the
vulnerable locations in the code. One example is shown below:

\begin{lstlisting}[
    basicstyle=\ttfamily\footnotesize,
    xleftmargin=0.5ex,
    backgroundcolor=\color{bg},
    breaklines=true,
    numbers=none,
    escapeinside=||
]
|\underline{\textbf{Prompt}}:|
|\textbf{The \textit{k-beauty} of an integer \code{num} is defined as the number...}|

The code in |\textbf{C}| below has |\textbf{CWE-787 (PotentialBufferOverflow)}| vulnerability:
```
|\textbf{int isDivisor(char *str, int len, int num) \{...\}}|
```

Vulnerability Message:
|\textbf{"Potential buffer overflow", ...}|
|\textbf{"This conversation may yield a string of length 12, ...", ...}|

Fix the vulnerable code and generate the fixed code.
\end{lstlisting}

\noindent If the newly generated code snippet is still vulnerable to the same vulnerability, the corresponding CWE information (i.e., vulnerability message) returned by \textit{CodeQL} is taken directly as a new prompt provided to \ChatGPT to fix and generate a new code snippet, in the same conversation. The whole process continues for a maximum of five rounds if the generated code is never fixed. \revision{Furthermore, the strategy of mitigating token limitation follows the setting in W.A. multi-round fixing (Sec. \ref{sec:code-with-wrong-answer}).}

The results of multi-round fixing are shown in Table \ref{tab:securitycodegeneration-leet}'s \textbf{\# Fixed} column. All vulnerable code snippets are fixed. The fix for these code snippets is straightforward since it only requires some additional statements for checking corresponding vulnerabilities. For instance, by including statements to test for \texttt{NULL} or perform boundary checks. Thus, in general, \textit{ChatGPT} performs well in this multi-round fixing process for the code snippets in Sec. \ref{sec:FunctionallyCorrectCodeGeneration}.

\noindent \textcolor{magenta}{$\bigstar$} \textbf{Summary 1.} The majority of vulnerable code snippets generated by \ChatGPT are related to the MissingNullTest query, accounting for $91.8\%$ of the total. These code snippets fail to perform \texttt{NULL} tests after memory allocation, potentially leading to security vulnerabilities. Although the remaining vulnerability queries, such as PotentialBufferOverflow and OffsetUseBeforeRangeCheck, are less frequent, they are still significant and should not be disregarded. By applying multi-round fixing, all sampled vulnerable code snippets are fixed. \ChatGPT performs well for the vulnerable code snippets in the scenario of algorithm problems.

\begin{table}[t]
\centering
\caption{Result of Security Code Generation}
\scalebox{0.72}{\begin{tabular}{cclccrr} 
\toprule
\textbf{R.}               & \textbf{CWE}                  & \multicolumn{1}{c}{\textbf{Code Scenario}} & \textbf{Lg.} & \textbf{Ori.} & \multicolumn{1}{c}{\textbf{\# Vdn.}} & \multicolumn{1}{c}{\textbf{\# Vln.}}  \\ 
\midrule
\multirow{3}{*}{1} & \multirow{3}{*}{787} & 1: \textit{MT.}-2             & C        & \textit{MT.}  & \percent[p]{53}{54}                           &    \cellcolor{yellow!50}{\percent[p]{14}{53}}                                 \\
                   &                      & 2: \textit{MT.}-5             & C        & \textit{MT.}  & \percent[p]{53}{53}                            &      \cellcolor{yellow!50}{\percent[p]{16}{53}}                                  \\
                   &                      & 3: PotentialBufferOverflow      & C        & \textit{CQ.} & \percent[p]{48}{48}                            &  \cellcolor{red!50}{\percent[p]{40}{48}}                                      \\ 
\midrule
\multirow{3}{*}{2} & \multirow{3}{*}{79}  & 1: ExternalAPISinkExample       & C        & \textit{CQ.} & \percent[p]{53}{57}                            &      \cellcolor{yellow!50}{\percent[p]{8}{53}}                                  \\
                   &                      & 2: ReflectedXss                 & Py.  & \textit{CQ.} & \percent[p]{52}{52}                            &      \cellcolor{yellow!50}{\percent[p]{4}{52}}                                  \\
                   &                      & 3: Jinja2WithoutEscaping        & Py.  & \textit{CQ.} & \percent[p]{56}{56}                            &          \cellcolor{green!30}{\percent[p]{0}{56}}                              \\

                   \midrule
\multirow{3}{*}{3} & \multirow{3}{*}{89}  & 1: SqlInjection-a       & Py.        & \textit{CQ.} & \percent[p]{47}{47}                            &   \cellcolor{yellow!50}{\percent[p]{9}{47}}                                     \\
                   &                      & 2: \cite{pearce2022asleep}-1                 & Py.  & \cite{pearce2022asleep} & \percent[p]{54}{54}                            &         \cellcolor{yellow!50}{\percent[p]{11}{54}}                               \\
                   &                      & 3: \cite{pearce2022asleep}-2        & Py.  & \cite{pearce2022asleep} & \percent[p]{55}{55}                            &   \cellcolor{yellow!50}{\percent[p]{6}{55}}                                     \\


                   \midrule
\multirow{3}{*}{4} & \multirow{3}{*}{20}  & 1: IncompleteUrlSubstringSanitization       & Py.        & \textit{CQ.} & \percent[p]{55}{56}                            &   \cellcolor{orange!50}{\percent[p]{1}{55}}                                     \\
                   &                      & 2: IncompleteHostnameRegExp                 & Py.  & \textit{CQ.} & \percent[p]{58}{58}                            &    \cellcolor{green!30}{\percent[p]{0}{58}}                                    \\
                   &                      & \cellcolor{blue!30}{3: \cite{pearce2022asleep}-1}                 & C  & \cite{pearce2022asleep} & \percent[p]{58}{58}                            &    \cellcolor{red!50}{\percent[p]{39}{58}}                                    \\

\midrule
\multirow{3}{*}{5} & \multirow{3}{*}{125}  & 1: \textit{MT.}-1       & C        & \textit{MT.} & \percent[p]{56}{56}                            &    \cellcolor{green!30}{\percent[p]{0}{56}}                                    \\
                   &                      & 2: \cite{pearce2022asleep}-1                 & C  & \cite{pearce2022asleep} & \percent[p]{58}{58}                            & \cellcolor{green!30}{\percent[p]{0}{58}}                                       \\
                   &                      & 3: \cite{pearce2022asleep}-2        & C  & \cite{pearce2022asleep} & \percent[p]{56}{56}                            &        \cellcolor{green!30}{\percent[p]{0}{56}}                                \\

\midrule
\multirow{3}{*}{6} & \multirow{3}{*}{78}  & 1: ExecTainted       & C        & \textit{CQ.} & \percent[p]{59}{59}                            &  \cellcolor{red!50}{\percent[p]{59}{59}}                                      \\
                   &                      & 2: CommandInjection                 & Py.  & \textit{CQ.} & \percent[p]{56}{57}                            &   \cellcolor{yellow!50}{\percent[p]{19}{56}}                                     \\
                   &                      & 3: \cite{pearce2022asleep}-1        & C  & \cite{pearce2022asleep} & \percent[p]{56}{56}                            &  \cellcolor{red!50}{\percent[p]{56}{56}}                                      \\

\midrule
\multirow{3}{*}{7} & \multirow{3}{*}{416}  & \cellcolor{blue!30}{1: \textit{MT.}-2}       & C        & \textit{MT.} & \percent[p]{59}{60}                            &       \cellcolor{red!50}{\percent[p]{55}{59}}                                 \\
& & 2: UseAfterFree       & C        & \textit{CQ.} & \percent[p]{57}{57}                            &       \cellcolor{green!30}{\percent[p]{0}{57}}                                 \\
                   &                      & 3: \cite{pearce2022asleep}-1                 & C  & \cite{pearce2022asleep} & \percent[p]{60}{60}                            &   \cellcolor{green!30}{\percent[p]{0}{60}}                                     \\

\midrule
\multirow{3}{*}{8} & \multirow{3}{*}{22}  & 1: TaintedPath       & C        & \textit{CQ.} & \percent[p]{59}{59}                            &      \cellcolor{red!50}{\percent[p]{59}{59}}                            \\
                   &                      & 2: TaintedPath                 & Py.  & \textit{CQ.} & \percent[p]{58}{59}                            &     \cellcolor{yellow!50}{\percent[p]{25}{58}}                                   \\
                   &                      & 3: TarSlip        & Py.  & \textit{CQ.} & \percent[p]{59}{59}                            &     \cellcolor{red!50}{\percent[p]{47}{59}}                                    \\

\midrule
\multirow{3}{*}{10} & \multirow{3}{*}{434}  & \cellcolor{blue!30}{1: \cite{pearce2022asleep}-1}       & Py.        & \cite{pearce2022asleep} & \percent[p]{52}{52}                            &      \cellcolor{red!50}{\percent[p]{31}{52}}                            \\
                   &                      & \cellcolor{blue!30}{2: \cite{pearce2022asleep}-2}                 & Py.  & \cite{pearce2022asleep} & \percent[p]{57}{57}                            &     \cellcolor{yellow!50}{\percent[p]{5}{57}}                                   \\
                   &                      & \cellcolor{blue!30}{3: \cite{pearce2022asleep}-3}        & Py.  & \cite{pearce2022asleep} & \percent[p]{49}{49}                            &     \cellcolor{green!30}{\percent[p]{0}{49}}                                    \\

\midrule
\multirow{3}{*}{11} & \multirow{3}{*}{476}  & 1: MissingNullTest-a       & C        & \textit{CQ.} & \percent[p]{57}{57}                            & \cellcolor{red!50}{\percent[p]{57}{57}}                                       \\
                   &                      & 2: MissingNullTest-b                 & C  & \textit{CQ.} & \percent[p]{57}{57}                            &   \cellcolor{red!50}{\percent[p]{57}{57}}                                     \\
                   &                      & 3: MissingNullTest-c        & C & \textit{CQ.} & \percent[p]{52}{52}                            &       \cellcolor{red!50}{\percent[p]{52}{52}}                                 \\

\midrule
\multirow{3}{*}{12} & \multirow{3}{*}{502}  & 1: UnsafeDeserialization-a       & Py.        & \textit{CQ.} & \percent[p]{57}{59}                            &    \cellcolor{yellow!50}{\percent[p]{4}{57}}                                  \\
                   &                      & 2: UnsafeDeserialization-b                 & Py.  & \textit{CQ.} & \percent[p]{56}{56}                            &    \cellcolor{yellow!50}{\percent[p]{25}{56}}                                    \\
                   &                      & 3: UnsafeDeserialization-c       & Py. & \textit{CQ.} & \percent[p]{59}{59}                            &      \cellcolor{yellow!50}{\percent[p]{5}{59}}                                  \\

\midrule
\multirow{3}{*}{13} & \multirow{3}{*}{190}  & 1: \textit{MT.}-4       & C        & \textit{MT.} & \percent[p]{54}{54}                            &   \cellcolor{red!50}{\percent[p]{54}{54}}                                     \\
                   &                      & 2: ArithmeticTainted                 & C  & \textit{CQ.} & \percent[p]{55}{55}                            &      \cellcolor{red!50}{\percent[p]{51}{55}}                               \\
                   &                      & 3: ArithmeticUncontrolled       & C & \textit{CQ.} & \percent[p]{50}{51}                            &       \cellcolor{green!30}{\percent[p]{0}{50}}                                 \\

\midrule
\multirow{3}{*}{15} & \multirow{3}{*}{798}  & 1: HardcodedCredentials-a       & Py.        & \textit{CQ.} & \percent[p]{48}{50}                            &     \cellcolor{orange!50}{\percent[p]{1}{48}}                                     \\
                   &                      & 2: HardcodedCredentials-b                 & Py.  & \textit{CQ.} & \percent[p]{51}{51}                            &   \cellcolor{green!30}{\percent[p]{0}{51}}                                  \\
                   &                      & 3: HardcodedCredentials-c       & Py. & \textit{CQ.} & \percent[p]{54}{55}                            &     \cellcolor{yellow!50}{\percent[p]{7}{54}}                                   \\

\midrule
\multirow{3}{*}{18} & \multirow{3}{*}{306}  & \cellcolor{blue!30}{1: \cite{pearce2022asleep}-1}       & Py.        & \cite{pearce2022asleep} & \percent[p]{54}{54}                            &     \cellcolor{orange!50}{\percent[p]{1}{54}}                                     \\
                   &                      & \cellcolor{blue!30}{2: \cite{pearce2022asleep}-2}                 & Py.  & \cite{pearce2022asleep} & \percent[p]{59}{59}                            &   \cellcolor{green!30}{\percent[p]{0}{59}}                                  \\
                   &                      & \cellcolor{blue!30}{3: \cite{pearce2022asleep}-3}       & Py. & \cite{pearce2022asleep} & \percent[p]{51}{59}                            &     \cellcolor{green!30}{\percent[p]{0}{51}}  \\

\midrule
\multirow{3}{*}{19} & \multirow{3}{*}{119}  & 1: \textit{MT.}-3       & C        & \textit{MT.} & \percent[p]{57}{58}                            &     \cellcolor{red!50}{\percent[p]{48}{57}}                                   \\
                   &                      & 2: OverflowBuffer                 & C  & \textit{CQ.} & \percent[p]{59}{59}                            &   \cellcolor{green!30}{\percent[p]{0}{59}}                                  \\
                   &                      & 3: \cite{pearce2022asleep}-1       & C & \cite{pearce2022asleep} & \percent[p]{59}{59}                            &  \cellcolor{red!50}{\percent[p]{59}{59}}                                      \\

\midrule
\multirow{3}{*}{30} & \multirow{3}{*}{732}  & 1: DoNotCreateWorldWriteable-a       & C        & \textit{CQ.} & \percent[p]{59}{59}                            &        \cellcolor{green!30}{\percent[p]{0}{59}}                                \\
                   &                      & 2: DoNotCreateWorldWriteable-b                 & C  & \textit{CQ.} & \percent[p]{58}{58}                            &      \cellcolor{green!30}{\percent[p]{0}{58}}                               \\
                   &                      & 3: WeakFilePermissions       & Py. & \textit{CQ.} & \percent[p]{58}{58}                            &    \cellcolor{green!30}{\percent[p]{0}{58}}                                    \\

\midrule
\multirow{3}{*}{33} & \multirow{3}{*}{200}  & \cellcolor{blue!30}{1: \textit{MT.}-1}       & Py.        & \textit{MT.} & \percent[p]{54}{58}                            &        \cellcolor{green!30}{\percent[p]{0}{54}}                                \\
                   &                      & \cellcolor{blue!30}{2: \textit{MT.}-2}                 & Py.  & \textit{MT.} & \percent[p]{55}{58}                            &      \cellcolor{yellow!50}{\percent[p]{13}{55}}                               \\
                   &                      & \cellcolor{blue!30}{3: \textit{MT.}-6}       & Py. & \textit{MT.} & \percent[p]{59}{59}                            &    \cellcolor{orange!50}{\percent[p]{1}{59}}                                    \\

\midrule
\multirow{3}{*}{38} & \multirow{3}{*}{522}  & \cellcolor{blue!30}{1: \cite{pearce2022asleep}-1-a}       & Py.        & \cite{pearce2022asleep} & \percent[p]{53}{53}                            &        \cellcolor{red!50}{\percent[p]{52}{53}}                                \\
                   &                      & \cellcolor{blue!30}{2: \cite{pearce2022asleep}-1-b}                 & Py.  & \cite{pearce2022asleep} & \percent[p]{51}{51}                            &      \cellcolor{yellow!50}{\percent[p]{3}{51}}                               \\
                   &                      & \cellcolor{blue!30}{3: \cite{pearce2022asleep}-1-c}       & Py. & \cite{pearce2022asleep} & \percent[p]{54}{54}                            &    \cellcolor{green!30}{\percent[p]{0}{54}}                                    \\

\bottomrule
\end{tabular}}
\label{tab:securitycodegeneration}
\end{table}

\noindent \textbf{Result \blackding{2}.} Table \ref{tab:securitycodegeneration} shows the results of security code generation. \textbf{R.} represents the ranking of CWEs in 2022 \textit{MITRE}. \textbf{Lg.} is the language used where \textbf{Py.} represents Python3 for short. \textbf{Ori.} is the source of code scenarios from \textit{MITRE} (\textit{MT.}), \textit{CodeQL} (\textit{CQ.}), and \cite{pearce2022asleep}. \textbf{\# Vdn.} specifies the valid (compilable and syntactically compliant) number and percentage of generated code by \textit{ChatGPT}. \textbf{\# Vln.} represents the number and percentage (\textbf{\# Vln.} number $/$ \textbf{\# Vdn.} number) of vulnerable code snippets in the corresponding code scenarios and CWEs. Note that though CWE-732, CWE-200, and CWE-522 are down to rank 30, 33, and 38 in 2022 \textit{MITRE}, we still include them to be consistent with \cite{pearce2022asleep}. They are also highlighted by 2022 MITRE~\cite{mitre}. We also mark $0\%$, $(0\%, 5\%]$, $(5\%, 50\%]$, and $(50\%, 100\%]$ as green, orange, yellow, and red in \textbf{\# Vln.} cells. The code scenarios marked as blue are checked by the authors manually.

As shown in the table, \textit{ChatGPT} generates 2,983 valid code snippets achieving a $99.07\%$ valid rate on average, where 994 ($33.32\%$) of them are vulnerable. Broken down into languages, there are 1,402 ($47\%$) valid code snippets in C containing 724 ($51.64\%$) vulnerable ones, and 1,581 ($53\%$) valid code snippets in Python3 containing 270 ($17.08\%$) vulnerable ones. Moreover, there are 18 ($33\%$), 4 ($7\%$), 16 ($30\%$), and 16 ($30\%$) code scenarios marked as green, orange, yellow, and red, respectively. The maximum and minimum \textbf{\# Vln}. percentages are $100\%$ and $0\%$, respectively. 

We divide the 18 CWEs into 6 groups according to their relationships and descriptions~\cite{mitre}, which are shown in Table \ref{tab:cwe-category}.

\begin{table}[t]
\centering
\caption{Group Categories and Descriptions for the 18 CWEs}
\scalebox{0.75}{\begin{tabular}{m{2.8cm}<{\centering}|l|l} 
\toprule
\multicolumn{1}{c|}{\textbf{Group Category}}     & \multicolumn{1}{c|}{\textbf{CWE Group}} & \multicolumn{1}{c}{\textbf{Group Description}}  \\ 
\midrule
\textbf{Overflow}                          & \multicolumn{1}{m{3.9cm}|}{CWE-787, CWE-125, CWE-190, CWE-119}      &     \multicolumn{1}{m{4cm}}{Relate to buffer overflow and integer overflow}                                      \\ 
\midrule
\textbf{Data Neutralization Issues}        & \multicolumn{1}{m{3.9cm}|}{CWE-79, CWE-89, CWE-78}                  &      \multicolumn{1}{m{4cm}}{Create or neutralize data using an incorrect format}                                     \\ 
\midrule
\textbf{Input Validation}                    & \multicolumn{1}{m{3.9cm}|}{CWE-20, CWE-22, CWE-434, CWE-502}        &      \multicolumn{1}{m{4cm}}{Vulnerabilities caused by improper input without appropriate validation}                                     \\ 
\midrule
\textbf{Unsafe Memory Operation}          & \multicolumn{1}{m{3.9cm}|}{CWE-416, CWE-476}                        &        \multicolumn{1}{m{4cm}}{Relate to the use of pointers}                                   \\ 
\midrule
\textbf{Unauthenticated Access}       & \multicolumn{1}{m{3.9cm}|}{CWE-798, CWE-306, CWE-522}               &    \multicolumn{1}{m{4cm}}{Improper use, authentication, and protection of credentials}                                       \\ 
\midrule
\textbf{Information Leakage} & \multicolumn{1}{m{3.9cm}|}{CWE-732, CWE-200}                        &    \multicolumn{1}{m{4cm}}{Incorrect permission assignments and unauthorized actions to critical resources}                                       \\
\bottomrule
\end{tabular}}
\label{tab:cwe-category}
\end{table}

\begin{figure}[t]
    \centering
  \begin{lstlisting}[language=C]
// generate 3 random floats
...
// convert to string
char str_a[20], str_b[20], str_c[20];
sprintf(str_a, "%f", a);
sprintf(str_b, "%f", b);
sprintf(str_c, "%f", c);
  \end{lstlisting}
  \vspace{-1em}
  \caption{Example of generated code in CWE-787 code scenario 3.}
  \label{fig:cwe-787-example-code-generated}
\end{figure}

$\vartriangleright$ \textbf{Overflow:} This group is related to buffer overflow and integer overflow. Out of the 12 code scenarios, 7 of them are marked as red (5) and yellow (2), and the remaining are marked as green. Where CWE-787, CWE-125, and CWE-119 are related to buffer overflow, and CWE-190 is data overflow. For each CWE in buffer overflow, it has at least one red mark. Fig. \ref{fig:cwe-787-example-code-generated} shows an example of vulnerable code generated in CWE-787 code scenario 3. The scenario asks \textit{ChatGPT} to convert float variables to string formats, but the generated code sets the length of an array in \texttt{char} less than 48 (the maximum number of characters output is $47 + 1$ for the \texttt{'\textbackslash 0'} by \texttt{sprintf}). However, CWE-190 has three green marks. \textit{ChatGPT} performs poorly on CWE-787, CWE-125 and CWE-119 but well on CWE-190 code scenarios. 

\begin{figure}[t]
    \centering
  \begin{lstlisting}[language=C]
// get the username from the command line
char *username = argv[1];
// get the userinfo from the system
char command[100] = {0};
sprintf(command, "getent passwd %s", username);
system(command);
  \end{lstlisting}
  \vspace{-1em}
  \caption{Example of generated code in CWE-78 code scenario 1.}
  \label{fig:cwe-78-example-code-generated}
\end{figure}

$\vartriangleright$ \textbf{Data Neutralization Issues:} This group is related to creating or neutralizing data using an incorrect format (e.g., SQL Injection). Out of the 9 scenarios, 2, 6, and 1 of them are marked as red, yellow, and green. Both 2 red marks are in CWE-78 achieving $100\%$ \textbf{\# Vln.} percentages. CWE-78 is related to OS command injection. An example of generated code in scenario 1 is shown in Fig. \ref{fig:cwe-78-example-code-generated}. The scenario asks \textit{ChatGPT} to retrieve the user's information using given \texttt{username}, but it does not check whether \texttt{username} is valid, which lets attackers be able to inject arbitrary commands. For CWE-79 and CWE-89, their yellow marks are all lower than $21\%$, indicating that most of the generated code in CWE-79 and CWE-89 code scenarios is non-vulnerable.

\begin{figure}[t]
    \centering
  \begin{lstlisting}[language=C]
char *reqFileName = argv[1];
int reqFileNameLen = strlen(reqFileName);
char *safeDirectory = "/safe/";
// read the requested file from the safe directory
char fileNameBuffer[256];
if (reqFileNameLen + strlen(safeDirectory) > 256) {
    ...
}
// concatenate safeDirectory with reqFileName
FILE *fp = fopen(fileNameBuffer, "r");
  \end{lstlisting}
  \vspace{-1em}
  \caption{Example of generated code in CWE-22 code scenario 1.}
  \label{fig:cwe-22-example-code-generated}
\end{figure}

$\vartriangleright$ \textbf{Input Validation:} This group is related to Vulnerabilities caused by improper input without appropriate validation. An improper input may lead to data leakage or execution of arbitrary commands. Out of the 12 code scenarios, 4 of them are marked as red belonging to CWE-20, CWE-22, and CWE-434. The remaining's \textbf{\# Vln.} percentages are all lower than $10\%$ (except both CWE-22 and CWE-502 code scenario 2), indicating that the majority of generated code in various scenarios is non-vulnerable.  Fig. \ref{fig:cwe-22-example-code-generated} shows an example of generated vulnerable code in CWE-22 scenario 1. It asks \textit{ChatGPT} to read files by using given file names, but \textit{ChatGPT} does not check whether file names contain any special elements such as '\texttt{..}' and '\texttt{/}'. Interestingly, the same code scenario 2 of CWE-22 but in a Python3 web app has a \textbf{\# Vln.} percentage of $43.1\%$ much lower than CWE-22 scenario 1's $100\%$. The reason may be the given context (e.g., \texttt{Flask} module) or the training code seen forcing \textit{ChatGPT} to generate secure code as much as possible. For CWE-502 code scenario 2 achieving a yellow mark close to a \textbf{\# Vln.} percentage of $45\%$, it asks \textit{ChatGPT} to deserialize \texttt{YAML} type file in Python3, however, 26 code snippets use \texttt{yaml.load} (the method can deserialize Python objects) for deserialization instead of \texttt{yaml.safe\_load}, which may allow attackers to execute arbitrary code\footnote{The risks of insecure YAML deserialization in Python: https://excellalabs.com/the-risks-of-insecure-yaml-deserialization-in-python/.}.


\begin{figure}[t]
    \centering
  \begin{lstlisting}[language=C, 
  %linebackgroundcolor={\ifnum\value{lstnumber}=5\color{red!35}\fi}
  ]
char array[ARRAY_SIZE];
char abort_flag = 0;
// allocate a temporary buffer
char *tmp = (char *)malloc(sizeof(char) * ARRAY_SIZE);
...
// store characters from stdin in tmp
...
// no character received
if (i == 0) {
    free(tmp);
    abort_flag = 1;
}
// characters received, store them in the array
else {...}
// tidy up
free(tmp);
return 0;
  \end{lstlisting}
  \vspace{-1em}
  \caption{Example of generated code in CWE-416 code scenario 1.}
  \label{fig:cwe-416-example-code-generated}
\end{figure}

$\vartriangleright$ \textbf{Unsafe Memory Operation:} This group is related to pointer-related problems including referencing memory after free (CWE-416) and dereferencing \texttt{NULL} pointer (CWE-476). Out of the 6 code scenarios, 4 and 2 of them are marked as red and green, respectively. CWE-416 code scenario 1 asks \textit{ChatGPT} to free temporary memory according to the context. Fig. \ref{fig:cwe-416-example-code-generated} shows an example of this scenario. The generated code frees \texttt{tmp} without checking \texttt{abort\_flag} which may lead to being freed twice. For CWE-476, all its scenarios are marked as red. One example is already shown in Fig. \ref{fig:code-heap-buffer-overflow} at lines 7, 9, and 15. The code generated by \textit{ChatGPT} does not check whether a pointer is \texttt{NULL}. Overall, \textit{ChatGPT} performs poorly in this group.

\begin{figure}[t]
    \centering
  \begin{lstlisting}[language=Python]
password = request.form['password']
hashed_password = hashlib.sha256(password.encode()).hexdigest()
  \end{lstlisting}
  \vspace{-1em}
  \caption{Example of generated code in CWE-522 code scenario 1.}
  \label{fig:cwe-522-example-code-generated}
\end{figure}

$\vartriangleright$ \textbf{Unauthenticated Access:} This group is related to improper use (CWE-798), authentication (CWE-306), and protection (CWE-522) of credentials. Out of the 9 code scenarios, only 1 is marked as red in CWE-522. The remaining's \textbf{\# Vln.} percentages are all lower than $6\%$ (except CWE-798 code scenario 3 with $13\%$). \textit{ChatGPT} performs well in this group. For CWE-522 code scenario 1, it asks \textit{ChatGPT} to hash the password given before storing it. One example is shown in Fig. \ref{fig:cwe-522-example-code-generated} using one round SHA-256, a secure hash function, to directly hash the password which is still susceptible to brute-force attacks, and vulnerable to collision attacks~\cite{stallings2012computer}. Interestingly, the other code scenarios in CWE-522 are the same as scenario 1 with additional either \texttt{salt=os.urandom(16)} statement or \texttt{bcrypt} module, avoiding generated code with the vulnerability with a high probability.

$\vartriangleright$ \textbf{Information Leakage:} This group is related to incorrect permission assignments (CWE-732) and unauthorized actions (CWE-200) to critical resources. Out of the 6 code scenarios, none of them are marked as red, and only CWE-200 code scenario 2 is marked as yellow with a \textbf{\# Vln.} percentage of $23.6\%$. The remaining are marked as orange (1) and green (4). The yellow mark scenario asks \textit{ChatGPT} to connect to a database in a \texttt{try/catch} structure in a Python3 web app, expecting the generated code not to return database error messages. The majority of generated code avoids the vulnerability. Thus, we can conclude that \textit{ChatGPT} performs well in these code scenarios.

\noindent \textcolor{red}{$\blacktriangleright$} \textit{\textbf{Multi-round Fixing:}} We sample 5 code snippets containing vulnerabilities from each code scenario of CWE categories and input these code snippets into \textit{ChatGPT} for fixing. If there are fewer than 5 vulnerable code snippets in a code scenario, we sample them all. The round number is set to 5 same as Sec. \ref{sec:FunctionallyCorrectCodeGeneration}. For each vulnerable code snippet, we create an initial prompt by leveraging the vulnerable code snippet and the corresponding CWE information provided by \textit{CodeQL}, or the corresponding CWE descriptions from~\cite{mitre} for code snippets checked by authors. One example is shown below:

\begin{lstlisting}[
    basicstyle=\ttfamily\footnotesize,
    xleftmargin=0.5ex,
    backgroundcolor=\color{bg},
    breaklines=true,
    numbers=none,
    escapeinside=||
]
|\underline{\textbf{Prompt}}:|
The code below has |\textbf{CWE-79}| vulnerability:
```
|\textbf{Vunerable code snippet}|
```

Vulnerability Message:
|\textbf{"Untrusted data passed to external API", ...}|
|\textbf{"Call to free with untrusted data from...", ...}|

Fix the vulnerable code and generate the fixed code.
\end{lstlisting}

\noindent If the newly generated code snippet is still vulnerable to the same vulnerability (e.g., CWE-79's ExternalAPISink), the corresponding CWE information (i.e., vulnerability message) returned by \textit{CodeQL} is taken directly as a new prompt provided to \ChatGPT to fix and generate a new code snippet, in the same conversation. For the code snippets checked by authors, we tell \textit{ChatGPT} that the newly generated code snippets still have the same CWE vulnerabilities (e.g., "the newly generated code snippet still contains the CWE-787 vulnerability."). The whole process continues for a maximum of five rounds if the generated code is never fixed. \revision{Furthermore, the strategy of mitigating token limitation follows the setting in W.A. multi-round fixing (Sec. \ref{sec:code-with-wrong-answer}).}

\begin{table}[t]
\centering
\caption{Result of Multi-round Vulnerable Code Fixing}
\scalebox{0.73}{\begin{tabular}{cclcr} 
\toprule
\textbf{R.}               & \textbf{CWE}                  & \multicolumn{1}{c}{\textbf{Code Scenario}} & \textbf{Lg.} & \multicolumn{1}{c}{\textbf{\# Fixed}}  \\ 
\midrule
\multirow{3}{*}{1} & \multirow{3}{*}{787} & 1: \textit{MT.}-2 & C & \percent[r]{5}{5} \\
                   &                      & 2: \textit{MT.}-5 & C  & \percent[r]{5}{5} \\
                   &                      & 3: PotentialBufferOverflow  & C & \percent[r]{5}{5} \\ 
\midrule
\multirow{2}{*}{2} & \multirow{2}{*}{79}  & 1: ExternalAPISinkExample       & C        & \percent[r]{5}{5}      \\
     &    & 2: ReflectedXss    & Py.  & \percent[r]{4}{4}     \\
\midrule
\multirow{3}{*}{3} & \multirow{3}{*}{89}  & 1: SqlInjection-a       & Py.        & \percent[r]{5}{5}   \\
                   &   & 2: \cite{pearce2022asleep}-1   & Py.  & \percent[r]{5}{5}  \\
                   &    & 3: \cite{pearce2022asleep}-2  & Py.  & \percent[r]{5}{5} \\
\midrule
\multirow{2}{*}{4} & \multirow{2}{*}{20}  & 1: IncompleteUrlSubstringSanitization       & Py.        & \percent[r]{1}{1}  \\

                   &                      & 3: \cite{pearce2022asleep}-1                 & C  & \percent[r]{0}{5} \\
\midrule
\multirow{3}{*}{6} & \multirow{3}{*}{78}  & 1: ExecTainted       & C        & \percent[r]{4}{5}  \\
                   &    & 2: CommandInjection                 & Py.  & \percent[r]{5}{5}  \\
                   &    & 3: \cite{pearce2022asleep}-1        & C  & \percent[r]{5}{5} \\
\midrule
\multirow{1}{*}{7} & \multirow{1}{*}{416}  & 1: \textit{MT.}-2       & C        & \percent[r]{4}{5} \\
                
\midrule
\multirow{3}{*}{8} & \multirow{3}{*}{22}  & 1: TaintedPath       & C        & \percent[r]{5}{5} \\
                   &                      & 2: TaintedPath                 & Py.  & \percent[r]{5}{5} \\
                   &                      & 3: TarSlip        & Py.  & \percent[r]{5}{5}  \\

\midrule
\multirow{2}{*}{10} & \multirow{2}{*}{434}  & 1: \cite{pearce2022asleep}-1       & Py.        & \percent[r]{5}{5}    \\
                   &                      & 2: \cite{pearce2022asleep}-2                 & Py.  & \percent[r]{3}{5} \\

\midrule
\multirow{3}{*}{11} & \multirow{3}{*}{476}  & 1: MissingNullTest-a       & C        & \percent[r]{5}{5} \\
                   &        & 2: MissingNullTest-b                 & C  & \percent[r]{5}{5}  \\
                   &          & 3: MissingNullTest-c        & C & \percent[r]{5}{5}  \\

\midrule
\multirow{3}{*}{12} & \multirow{3}{*}{502}  & 1: UnsafeDeserialization-a       & Py.        & \percent[r]{4}{4} \\
                   &        & 2: UnsafeDeserialization-b                 & Py.  & \percent[r]{5}{5} \\
                   &          & 3: UnsafeDeserialization-c       & Py. & \percent[r]{5}{5}  \\

\midrule
\multirow{2}{*}{13} & \multirow{2}{*}{190}  & 1: \textit{MT.}-4     & C    & \percent[r]{5}{5} \\
                   &    & 2: ArithmeticTainted    & C  & \percent[r]{5}{5} \\

\midrule
\multirow{2}{*}{15} & \multirow{2}{*}{798}  & 1: HardcodedCredentials-a       & Py.        & \percent[r]{1}{1} \\
                   &                      & 3: HardcodedCredentials-c       & Py. & \percent[r]{5}{5} \\

\midrule
\multirow{1}{*}{18} & \multirow{1}{*}{306}  & 1: \cite{pearce2022asleep}-1       & Py.        & \percent[r]{1}{1}  \\

\midrule
\multirow{2}{*}{19} & \multirow{2}{*}{119}  & 1: \textit{MT.}-3       & C        & \percent[r]{5}{5}  \\
    &    & 3: \cite{pearce2022asleep}-1       & C & \percent[r]{5}{5} \\

\midrule
\multirow{2}{*}{33} & \multirow{2}{*}{200}  & 2: \textit{MT.}-2  & Py.  & \percent[r]{0}{5}  \\
                   &                      & 3: \textit{MT.}-6       & Py. & \percent[r]{1}{1}  \\

\midrule
\multirow{2}{*}{38} & \multirow{2}{*}{522}  & 1: \cite{pearce2022asleep}-1-a  & Py.  & \percent[r]{2}{5}  \\
                   &    & 2: \cite{pearce2022asleep}-1-b  & Py.  & \percent[r]{3}{3} \\

\midrule

Total & -  & -  & -  & \percent[r]{143}{160}  \\

\bottomrule
\end{tabular}}
\label{tab:securitycodegeneration-fix}
\end{table}

The results of multi-round fixing are shown in Table \ref{tab:securitycodegeneration-fix}, where \textbf{\# Fixed} represents the fixed numbers for corresponding code scenarios. Out of 160 vulnerable code snippets, 143 of them can be fixed. Moreover, there are 30 code scenarios where all code snippets are fixed, 4 code scenarios where code snippets are partially fixable, and 2 code scenarios where all code snippets are not fixable. 

$\vartriangleright$ \textbf{Overflow:} All code snippets in CWE-787, CWE-190, and CWE-119 can be fixed. Although the first generated code snippets by \textit{ChatGPT} have overflow vulnerability, providing these code snippets with corresponding CWE information helps \textit{ChatGPT} to successfully fix simple overflow problems. For instance, for Fig. \ref{fig:cwe-787-example-code-generated}, \textit{ChatGPT} turns \texttt{sprintf} to \texttt{snprintf}\footnote{https://learn.microsoft.com/en-us/cpp/c-runtime-library/reference/snprintf-snprintf-snprintf-l-snwprintf-snwprintf-l?view=msvc-170}, preventing buffer overflow.

$\vartriangleright$ \textbf{Data Neutralization Issues:} All code snippets in CWE-79, CWE-89, and CWE-78'S code scenarios 2 and 3 can be fixed. One code snippet in CWE-78's code scenario 1 is still vulnerable. Most vulnerable code snippets can be fixed by providing corresponding CWE information. For the vulnerable one, \textit{ChatGPT} fails to fix the code for checking the external input of \texttt{username}, which may lead to OS command injection.

\begin{figure}[t]
    \centering
  \begin{lstlisting}[language=C, 
  %linebackgroundcolor={\ifnum\value{lstnumber}=5\color{red!35}\fi}
  ]
// Returns the market price of a stock
// Returns -1 if an error occurred
extern float get_market_price(const char *symbol);
// ...
// Returns:  (float) - Value of the sale (always positive or 0)
//           0 if an error occurred
float calculate_sale_value(const char *symbol, float quantity) {
    float sale_value = 0.0f;
    ...
    // Copy the symbol to the new memory location
    strncpy(symbol_copy, symbol, symbol_len);
    ...
    // Get the market price of the stock
    float market_price = get_market_price(symbol_copy);
    if (market_price != -1) {
        sale_value = quantity * market_price;
    }
    ...
    return sale_value;
}
  \end{lstlisting}
  \vspace{-1em}
  \caption{Example of the final generated code in CWE-20 code scenario 3.}
  \label{fig:cwe-20-example-code-generated-w}
\end{figure}

$\vartriangleright$ \textbf{Input Validation:} All code snippets in CWE-20's code scenario 1, CWE-22, CWE-434's code scenario 1, and CWE-502 can be fixed. Partial code snippets in CWE-434's code scenario 2 can be fixed. None of the code snippets in CWE-20's code scenario 3 can be fixed. The fixing performance of \textit{ChatGPT} in this group category is poor. For the 2 unfixed vulnerable code snippets in CWE-434's code scenario 2 (the requested images should be saved in the database as base64 encoded, and their types must be JPG and sizes should be less than 1,024KB), both of them satisfy the first requirement but do not meet the second requirement at all, missing alignment. As for the 5 unfixed vulnerable code snippets in CWE-20's code scenario 3 (generate the values of a share sale where the price comes from an external function. The values should $\geq 0$), all of them conduct a lot of necessary checks but they overlook checking the input values of the function as well as the values of output (i.e., the values of a share sale). For instance (see Fig. \ref{fig:cwe-20-example-code-generated-w}), the final generated code snippet does not check the input parameter \texttt{quantity}, which may result in the function \texttt{calculate\_sale\_value}'s return value being less than 0 when \texttt{quantity} is less than 0.

$\vartriangleright$ \textbf{Unsafe Memory Operation:} All code snippets in CWE-476 can be fixed and only one code snippet in CWE-416 is still vulnerable. In general, \textit{ChatGPT} performs well in this group category. For the vulnerable code snippet in CWE-416 code scenario 1, it still contains the problem of being freed twice (i.e., Fig. \ref{fig:cwe-416-example-code-generated}).

$\vartriangleright$ \textbf{Unauthenticated Access:} All code snippets in CWE-798, CWE-306, and CWE-522's code scenario 1 can be fixed. 3 of 5 code snippets in CWE-522's code scenario 1 are still vulnerable. \textit{ChatGPT} performs well in this group category. For the 3 vulnerable code snippets, they still use \texttt{hashlib.sha256} method one time rather than a more secure way (e.g., use slow hash method \texttt{bcrypt.hashpw}).

$\vartriangleright$ \textbf{Information Leakage:} The only one code snippet in CWE-200's code scenario 3 is fixed, but the other 5 code snippets in CWE-200's code scenario 2 are all still vulnerable, though the code scenario 2 is marked as yellow in Table \ref{tab:securitycodegeneration} ($23.6\%$ vulnerability rate). The final code snippets generated by \ChatGPT still return database error messages by exception handler. In general, \textit{ChatGPT} performs poorly in this group category.

\noindent \textcolor{magenta}{$\bigstar$} \textbf{Summary 1.} \textit{ChatGPT} generates 2,983 ($99.07\%$) valid code snippets successfully where 994 (33.32\%) are vulnerable. Moreover, the vulnerable code snippet percentage in C ($51.64\%$) is much higher than the one in Python3 ($17.08\%$), indicating that developers should be more aware of the security of code generated by \textit{ChatGPT} in C than in Python3. The reason for the result can be the context of the provided code scenarios and the quality of the code in C and Python seen in the training set.

\noindent \textcolor{magenta}{$\bigstar$} \textbf{Summary 2.} \textit{ChatGPT} has different performances under different groups, CWEs, and code scenarios in security code generation. Overall, no code scenarios are marked as red for the group of \textbf{Information Leakage}, while the remaining 5 groups have at least one code scenario marked as red. Where the group of  \textbf{Unsafe Memory Operation} has 4/6 (the highest percentage) code scenarios marked as red. Among all CWEs, 10 of them have at least one code scenario marked as red, but only 3 CWEs have scenarios only marked as green or orange. Among all code scenarios, there are 18 ($33\%$), 4 ($7\%$), 16 ($30\%$), and 16 ($30\%$) code scenarios marked as green, orange, yellow, and red, respectively.

\noindent \textcolor{magenta}{$\bigstar$} \textbf{Summary 3.} The multi-round fixing process for vulnerable code snippets shows promising results, with a high percentage ($89.4\%$) of vulnerabilities successfully addressed. Most vulnerabilities related to \textbf{Overflow}, \textbf{Data Neutralization Issues}, \textbf{Unsafe Memory Operations}, and \textbf{Unauthenticated Access} can be fixed through multi-round fixing, demonstrating the ability of \textit{ChatGPT} to generate fixed code by incorporating prompts based on corresponding CWE information. However, the performance in fixing vulnerabilities of \textbf{Input Validation} and \textbf{Information Leakage} is relatively weak, indicating room for improvement.

\vspace{-0.5em}
\begin{tcolorbox}[boxrule=1pt,boxsep=1pt,left=2pt,right=2pt,top=2pt,bottom=2pt,title=Answer to RQ4: Security Code Generation]
\noindent \blackding{1} In most scenarios including the scenario of algorithm problems and CWE scenarios, the code generated by \textit{ChatGPT} has relevant vulnerabilities such as \textbf{Overflow}, \textbf{Unsafe Memory Operation} (e.g., MissingNullTest) and so on;

\noindent \blackding{2} The multi-round fixing process for vulnerable code snippets demonstrates promising results, with a high percentage ($100\%$ and $89.4\%$) of vulnerabilities successfully addressed. The experiment result indicates that combining \ChatGPT with vulnerability detection tools can mitigate the presence of vulnerabilities in the code generated by \textit{ChatGPT}.
\end{tcolorbox}

\vspace{-1em}
\subsection{\revision{Non-determinism of \textit{ChatGPT}}}\label{sec:nondeterminism}
\noindent \textbf{\revision{RQ5: How does the non-deterministic output of \textit{ChatGPT} affect code generation?}}

\noindent \textbf{\revision{Motivation.}} \revision{LLMs like \textit{ChatGPT} have a non-deterministic nature~\cite{ChatGPT} typically due to the sampling methods such as top-k sampling~\cite{krishna2022rankgen}, which means they can produce various responses to the same input~\minorrevision{\cite{pearce2022asleep, pearce2022examining}}. In this RQ, we intend to investigate the non-deterministic output of \textit{ChatGPT}.}  

\noindent \textbf{\revision{Approach.}} \revision{We randomly and respectively select 9 problems from Aft. problems and Bef. problems, and leverage \textit{ChatGPT} to generate code for each of these 18 problems with five languages 10 times \minorrevision{across 2 temperatures of 0.7 (the default value used in the paper) and 0 (for stabilizing output~\cite{ChatGPT})}. The generated code snippets from these repeated trials are compared across functional correctness, complexity, and security. \minorrevision{Additionally, we sample 20 CWE code scenarios and use \ChatGPT to generate code snippets 10 times at temperature 0. These code snippets are compared in security. Moreover, the multi-round fixing process is included at temperature settings of 0.7 and 0 where each sampled error code or vulnerable code is fixed 5 times. The maximum round number is set to 5.}}

\begin{table*}
\centering
\caption{Functional Correctness, Complexity, and Security for Aft. Problems in 10 Trials at Temperature 0.7}
\scalebox{0.77}{\begin{tabular}{cccccccccccccccc}
\toprule
\multirow{2}{*}{\textbf{Problem Id}}        & \multirow{2}{*}{\textbf{Language}}   & \multirow{2}{*}{\textbf{A.}} & \multirow{2}{*}{\textbf{W.A.}} & \multirow{2}{*}{\textbf{C.E.}} & \multirow{2}{*}{\textbf{T.L.E.}} & \multirow{2}{*}{\textbf{R.E.}} & \multicolumn{4}{c}{\textbf{Cyclomatic}} & \multicolumn{4}{c}{\textbf{Cognitive}} & \multirow{2}{*}{\textbf{CWE}}  \\

\cline{8-15}

& & & & & & &\textbf{L}&\textbf{M}&\textbf{H}&\textbf{V}& \textbf{L}&\textbf{M} &\textbf{H} & \textbf{V} &  \\

\midrule
\multirow{5}{*}{2124} & C      &  $30.0\%$  &  $70.0\%$ &   $0.0\%$&   $0.0\%$ &   $0.0\%$&  $0.0\%$&$100.0\%$&$0.0\%$&$0.0\%$  &  \multicolumn{4}{c}{-}  &  $0.0\%$  \\
                  & C++        &  $30.0\%$  &  $70.0\%$ & $0.0\%$  & $0.0\%$   &  $0.0\%$ & $0.0\%$&$90.0\%$&$10.0\%$&$0.0\%$   &   \multicolumn{4}{c}{-}    &  $0.0\%$ \\
                  & Java       & $0.0\%$ &  $100.0\%$ &  $0.0\%$ &  $0.0\%$  &  $0.0\%$ &  $10.0\%$&$90.0\%$&$0.0\%$&$0.0\%$  & $0.0\%$&$90.0\%$&$10.0\%$&$0.0\%$   & $0.0\%$ \\
                  & Python3    &  $20.0\%$  & $80.0\%$  &   -   &  $0.0\%$  & $0.0\%$  &  $40.0\%$&$60.0\%$&$0.0\%$&$0.0\%$  & $20.0\%$&$60.0\%$&$20.0\%$&$0.0\%$&  -   \\
                  & JavaScript & $0.0\%$ &  $100.0\%$ &  - &  $0.0\%$  &  $0.0\%$ &  $0.0\%$&$100.0\%$&$0.0\%$&$0.0\%$ &$40.0\%$&$50.0\%$&$10.0\%$&$0.0\%$&  -  \\
\midrule

\multirow{5}{*}{2224} & C      &$30.0\%$ &  $30.0\%$  & $40.0\%$  & $0.0\%$   &   $0.0\%$   & $0.0\%$&$70.0\%$&$10.0\%$&$20.0\%$ &  \multicolumn{4}{c}{-}    & $0.0\%$ \\
                  & C++        & $60.0\%$   & $0.0\%$  & $40.0\%$ & $0.0\%$  & $0.0\%$ & $0.0\%$&$100.0\%$&$0.0\%$&$0.0\%$ &   \multicolumn{4}{c}{-}  &  $0.0\%$ \\
                  & Java       & $80.0\%$ & $10.0\%$  &  $0.0\%$    &  $10.0\%$  & $0.0\%$ &$20.0\%$&$70.0\%$&$10.0\%$&$0.0\%$&$20.0\%$&$70.0\%$&$10.0\%$&$0.0\%$&  $0.0\%$ \\
                  & Python3    & $70.0\%$ & $30.0\%$  &  -   &  $0.0\%$  & $0.0\%$ & $90.0\%$&$10.0\%$&$0.0\%$&$0.0\%$ &$20.0\%$&$80.0\%$&$0.0\%$&$0.0\%$&   -   \\
                  & JavaScript & $50.0\%$ & $10.0\%$  &  -   &  $10.0\%$  & $30.0\%$& $30.0\%$&$50.0\%$&$10.0\%$&$10.0\%$ &$10.0\%$&$70.0\%$&$20.0\%$&$0.0\%$&  -    \\

\midrule

\multirow{5}{*}{2227} & C      & $0.0\%$ & $50.0\%$  &$30.0\%$ &  $0.0\%$  &  $20.0\%$ & $0.0\%$&$0.0\%$&$0.0\%$&$100.0\%$ & \multicolumn{4}{c}{-}   & $100\%$     \\
                  & C++        &  $0.0\%$ & $70.0\%$ & $20.0\%$ & $0.0\%$   & $10.0\%$  & $75.0\%$&$0.0\%$&$0.0\%$&$25.0\%$ &   \multicolumn{4}{c}{-}  & $0.0\%$  \\
                  & Java       &$0.0\%$ & $80.0\%$ & $20.0\%$ &  $0.0\%$   & $0.0\%$  & $0.0\%$&$0.0\%$&$0.0\%$&$100.0\%$ &$0.0\%$&$0.0\%$&$70.0\%$&$30.0\%$&  $0.0\%$ \\
                  & Python3    &$0.0\%$ &  $70.0\%$ &  -  & $0.0\%$  & $30.0\%$  &  $0.0\%$&$0.0\%$&$30.0\%$&$70.0\%$&$0.0\%$&$10.0\%$&$70.0\%$&$20.0\%$&  -    \\
                  & JavaScript & $0.0\%$ & $100.0\%$ &  -  & $0.0\%$  &  $0.0\%$ & $0.0\%$&$0.0\%$&$80.0\%$&$20.0\%$ &$0.0\%$&$30.0\%$&$70.0\%$&$0.0\%$&  -    \\

\midrule

\multirow{5}{*}{2264} & C      & $30.0\%$ & $30.0\%$& $30.0\%$&  $0.0\%$  & $10.0\%$ & $25.0\%$&$75.0\%$&$0.0\%$&$0.0\%$ &  \multicolumn{4}{c}{-}  &   $42.85\%$   \\
                  & C++        &  $0.0\%$  &  $60.0\%$& $40.0\%$ & $0.0\%$  & $0.0\%$ & $16.7\%$&$33.3\%$&$16.7\%$&$33.3\%$ &  \multicolumn{4}{c}{-}  &  $0.0\%$ \\
                  & Java       &$30.0\%$ & $40.0\%$ & $30.0\%$ &  $0.0\%$ & $0.0\%$ & $60.0\%$&$40.0\%$&$0.0\%$&$0.0\%$&$60.0\%$&$40.0\%$&$0.0\%$&$0.0\%$&  $0.0\%$ \\
                  & Python3    & $40.0\%$ & $60.0\%$ &  -  & $0.0\%$ &  $0.0\%$ & $40.0\%$&$60.0\%$&$0.0\%$&$0.0\%$&$20.0\%$&$80.0\%$&$0.0\%$&$0.0\%$&    -  \\
                  & JavaScript & $30.0\%$ & $70.0\%$ &  -  & $0.0\%$  &  $0.0\%$ & $40.0\%$&$60.0\%$&$0.0\%$&$0.0\%$ &$40.0\%$&$60.0\%$&$0.0\%$&$0.0\%$&   -   \\

\midrule

\multirow{5}{*}{2304} & C      & $10.0\%$ & $50.0\%$ & $40.0\%$ &  $0.0\%$  &  $0.0\%$& $0.0\%$&$87.5\%$&$12.5\%$&$0.0\%$ &  \multicolumn{4}{c}{-} &  $0.0\%$  \\
                  & C++        & $50.0\%$& $10.0\%$ &  $20.0\%$ &  $0.0\%$  & $20.0\%$ & $0.0\%$&$100.0\%$&$0.0\%$&$0.0\%$ &   \multicolumn{4}{c}{-}  & $0.0\%$  \\
                  & Java       & $30.0\%$& $10.0\%$ & $60.0\%$ & $0.0\%$ & $0.0\%$& $60.0\%$&$40.0\%$&$0.0\%$&$0.0\%$& $60.0\%$&$40.0\%$&$0.0\%$&$0.0\%$&   $0.0\%$ \\
                  & Python3    & $50.0\%$ &$50.0\%$ &  -  &  $0.0\%$&  $0.0\%$ &$70.0\%$&$30.0\%$&$0.0\%$&$0.0\%$ & $10.0\%$&$90.0\%$&$0.0\%$&$0.0\%$&   -   \\
                  & JavaScript & $90.0\%$& $10.0\%$&  -   &  $0.0\%$  &$0.0\%$ & $0.0\%$&$100.0\%$&$0.0\%$&$0.0\%$ &$0.0\%$&$100.0\%$&$0.0\%$&$0.0\%$&   -   \\

\midrule

\multirow{5}{*}{2400} & C      & $0.0\%$& $70.0\%$ &  $0.0\%$ &  $0.0\%$ & $30.0\%$ & $70.0\%$&$30.0\%$&$0.0\%$&$0.0\%$ & \multicolumn{4}{c}{-}  &  $0.0\%$ \\
                  & C++        & $0.0\%$& $10.0\%$ &  $60.0\%$ &  $0.0\%$  & $30.0\%$ & $80.0\%$&$20.0\%$&$0.0\%$&$0.0\%$& \multicolumn{4}{c}{-}& $0.0\%$  \\
                  & Java       & $0.0\%$& $70.0\%$  & $10.0\%$ &  $0.0\%$  & $20.0\%$ & $50.0\%$&$50.0\%$&$0.0\%$&$0.0\%$ &$40.0\%$&$50.0\%$&$10.0\%$&$0.0\%$ &  $0.0\%$ \\
                  & Python3    & $0.0\%$& $40.0\%$  &  -  &  $0.0\%$  & $60.0\%$ & $70.0\%$&$30.0\%$&$0.0\%$&$0.0\%$ &$60.0\%$&$40.0\%$&$0.0\%$&$0.0\%$&   -   \\
                  & JavaScript & $0.0\%$&  $100.0\%$  &  -  &  $0.0\%$  &  $0.0\%$ & $60.0\%$&$40.0\%$&$0.0\%$&$0.0\%$ & $60.0\%$&$30.0\%$&$10.0\%$&$0.0\%$ &   -   \\

\midrule

\multirow{5}{*}{2435} & C      & $50.0\%$& $40.0\%$  &  $0.0\%$ &  $0.0\%$ & $10.0\%$ & $10.0\%$&$80.0\%$&$10.0\%$&$0.0\%$ & \multicolumn{4}{c}{-}  &  $0.0\%$  \\
                  & C++        & $0.0\%$ & $40.0\%$ & $60.0\%$ &$0.0\%$   &  $0.0\%$& $0.0\%$&$0.0\%$&$75.0\%$&$25.0\%$&   \multicolumn{4}{c}{-}&  $0.0\%$ \\
                  & Java       & $0.0\%$& $90.0\%$ &   $10.0\%$   & $0.0\%$ & $0.0\%$     & $10.0\%$&$0.0\%$&$80.0\%$&$10.0\%$ & $10.0\%$&$0.0\%$&$90.0\%$&$0.0\%$ &  $0.0\%$\\
                  & Python3    & $0.0\%$ & $90.0\%$ &   -  &  $0.0\%$   &  $10.0\%$ & $10.0\%$&$10.0\%$&$70.0\%$&$10.0\%$ & $10.0\%$&$70.0\%$&$20.0\%$&$0.0\%$ &  -    \\
                  & JavaScript & $0.0\%$ & $100.0\%$ &   -  &  $0.0\%$  & $0.0\%$ &  $10.0\%$&$10.0\%$&$60.0\%$&$20.0\%$ & $10.0\%$&$10.0\%$&$70.0\%$&$10.0\%$ &   -   \\

\midrule

\multirow{5}{*}{2523} & C      & $10.0\%$ & $40.0\%$ &  $10.0\%$  &  $40.0\%$  & $0.0\%$  & $0.0\%$&$0.0\%$&$0.0\%$&$100.0\%$& \multicolumn{4}{c}{-}  & $100\%$  \\
                  & C++        & $20.0\%$& $20.0\%$ &  $50.0\%$ & $0.0\%$   &  $10.0\%$  &$0.0\%$&$20.0\%$&$40.0\%$&$40.0\%$&  \multicolumn{4}{c}{-} & $0.0\%$  \\
                  & Java       & $10.0\%$& $10.0\%$ & $60.0\%$ &  $20.0\%$  &  $0.0\%$& $60.0\%$&$0.0\%$&$30.0\%$&$10.0\%$ &  $60.0\%$&$0.0\%$&$30.0\%$&$10.0\%$  & $0.0\%$\\
                  & Python3    & $0.0\%$ & $0.0\%$ &  -  &  $100.0\%$ &  $0.0\%$  & $0.0\%$&$0.0\%$&$100.0\%$&$0.0\%$ & $0.0\%$&$10.0\%$&$90.0\%$&$0.0\%$ &  -    \\
                  & JavaScript & $20.0\%$ & $40.0\%$  &  -  &  $40.0\%$  & $0.0\%$  & $20.0\%$&$0.0\%$&$40.0\%$&$40.0\%$& $20.0\%$&$0.0\%$&$60.0\%$&$20.0\%$&  -    \\

\midrule

\multirow{5}{*}{2532} & C      & $0.0\%$ & $10.0\%$  & $80.0\%$ &  $10.0\%$  & $0.0\%$  & $14.3\%$&$0.0\%$&$0.0\%$&$85.7\%$ & \multicolumn{4}{c}{-}  & $0.0\%$ \\
                  & C++        &$0.0\%$ & $20.0\%$ & $70.0\%$ &  $0.0\%$ & $10.0\%$ & $0.0\%$&$33.3\%$&$66.7\%$&$0.0\%$ & \multicolumn{4}{c}{-}  &  $0.0\%$ \\
                  & Java       & $0.0\%$ & $0.0\%$ &$90.0\%$ & $10.0\%$  &$0.0\%$& $90.0\%$&$0.0\%$&$0.0\%$&$10.0\%$ &$90.0\%$&$0.0\%$&$10.0\%$&$0.0\%$ & $0.0\%$ \\
                  & Python3    & $0.0\%$& $30.0\%$ &  -  &$10.0\%$  & $60.0\%$ &$50.0\%$&$20.0\%$&$10.0\%$&$20.0\%$ & $30.0\%$&$30.0\%$&$30.0\%$&$10.0\%$ &   -   \\
                  & JavaScript & $0.0\%$ & $70.0\%$&  -  &  $20.0\%$  & $10.0\%$& $30.0\%$&$0.0\%$&$10.0\%$&$60.0\%$ & $30.0\%$&$0.0\%$&$20.0\%$&$50.0\%$ &  -    \\

\bottomrule
\end{tabular}}
\label{tab:af-non-determinism}
\end{table*}

\begin{table*}
\centering
\caption{Functional Correctness, Complexity, and Security for Bef. Problems in 10 Trials at Temperature 0.7}
\scalebox{0.77}{\begin{tabular}{cccccccccccccccc}
\toprule
\multirow{2}{*}{\textbf{Problem Id}}        & \multirow{2}{*}{\textbf{Language}}   & \multirow{2}{*}{\textbf{A.}} & \multirow{2}{*}{\textbf{W.A.}} & \multirow{2}{*}{\textbf{C.E.}} & \multirow{2}{*}{\textbf{T.L.E.}} & \multirow{2}{*}{\textbf{R.E.}} & \multicolumn{4}{c}{\textbf{Cyclomatic}} & \multicolumn{4}{c}{\textbf{Cognitive}} & \multirow{2}{*}{\textbf{CWE}}  \\

\cline{8-15}

& & & & & & &\textbf{L}&\textbf{M}&\textbf{H}&\textbf{V}& \textbf{L}&\textbf{M} &\textbf{H} & \textbf{V} &  \\

\midrule
\multirow{5}{*}{9} & C          & $30.0\%$ & $0.0\%$  &  $0.0\%$& $0.0\%$  & $70.0\%$ &  $70.0\%$&$30.0\%$&$0.0\%$&$0.0\%$ &     \multicolumn{4}{c}{-}     & $0.0\%$  \\
                  & C++        & $40.0\%$&  $0.0\%$ & $0.0\%$     & $0.0\%$  &$60.0\%$  & $60.0\%$&$40.0\%$&$0.0\%$&$0.0\%$ &  \multicolumn{4}{c}{-}       & $0.0\%$  \\
                  & Java       & $100.0\%$ & $0.0\%$  &  $0.0\%$ &   $0.0\%$  & $0.0\%$ &  $90.0\%$&$10.0\%$&$0.0\%$&$0.0\%$& $100.0\%$&$0.0\%$&$0.0\%$&$0.0\%$ &  $0.0\%$ \\
                  & Python3    & $100.0\%$ & $0.0\%$ &   -   &  $0.0\%$ & $0.0\%$  &  $100.0\%$&$0.0\%$&$0.0\%$&$0.0\%$  &$100.0\%$&$0.0\%$&$0.0\%$&$0.0\%$ &  -    \\
                  & JavaScript & $100.0\%$&  $0.0\%$ &   -   &  $0.0\%$ & $0.0\%$ & $90.0\%$&$10.0\%$&$0.0\%$&$0.0\%$&$100.0\%$&$0.0\%$&$0.0\%$&$0.0\%$ &   -   \\
\midrule

\multirow{5}{*}{70} & C        & $100.0\%$ & $0.0\%$& $0.0\%$  &  $0.0\%$  &   $0.0\%$& $60.0\%$&$40.0\%$&$0.0\%$&$0.0\%$  &  \multicolumn{4}{c}{-}      & $0.0\%$ \\
                  & C++        & $100.0\%$ & $0.0\%$& $0.0\%$ & $0.0\%$  & $0.0\%$ & $90.0\%$&$10.0\%$&$0.0\%$&$0.0\%$ &    \multicolumn{4}{c}{-}    &  $0.0\%$ \\
                  & Java       & $100.0\%$ &  $0.0\%$& $0.0\%$ & $0.0\%$   &$0.0\%$  &  $100.0\%$&$0.0\%$&$0.0\%$&$0.0\%$ & $100.0\%$&$0.0\%$&$0.0\%$&$0.0\%$ &  $0.0\%$ \\
                  & Python3    & $100.0\%$&  $0.0\%$&   -   &   $0.0\%$ & $0.0\%$  & $100.0\%$&$0.0\%$&$0.0\%$&$0.0\%$ &$100.0\%$&$0.0\%$&$0.0\%$&$0.0\%$ &   -   \\
                  & JavaScript & $100.0\%$&  $0.0\%$ &   -   &  $0.0\%$  & $0.0\%$  &$100.0\%$&$0.0\%$&$0.0\%$&$0.0\%$ & $100.0\%$&$0.0\%$&$0.0\%$&$0.0\%$&   -   \\

\midrule

\multirow{5}{*}{304} & C       & $80.0\%$& $0.0\%$  &  $20.0\%$  & $0.0\%$  &  $0.0\%$& $0.0\%$&$90.0\%$&$10.0\%$&$0.0\%$ &   \multicolumn{4}{c}{-}     &  $100\%$    \\
                  & C++        & $100.0\%$ & $0.0\%$  & $0.0\%$ &  $0.0\%$  & $0.0\%$  &  $60.0\%$&$40.0\%$&$0.0\%$&$0.0\%$ &  \multicolumn{4}{c}{-}     & $0.0\%$ \\
                  & Java       & $100.0\%$& $0.0\%$ &  $0.0\%$ &  $0.0\%$ & $0.0\%$ & $50.0\%$&$40.0\%$&$10.0\%$&$0.0\%$ &$80.0\%$&$20.0\%$&$0.0\%$&$0.0\%$ &  $0.0\%$ \\
                  & Python3    & $100.0\%$& $0.0\%$ &  -  &  $0.0\%$  & $0.0\%$ &  $70.0\%$&$30.0\%$&$0.0\%$&$0.0\%$ & $100.0\%$&$0.0\%$&$0.0\%$&$0.0\%$ &   -   \\
                  & JavaScript & $100.0\%$& $0.0\%$ &  -   &  $0.0\%$  & $0.0\%$ & $0.0\%$&$100.0\%$&$0.0\%$&$0.0\%$ &$100.0\%$&$0.0\%$&$0.0\%$&$0.0\%$ &  -    \\

\midrule

\multirow{5}{*}{363} & C       & $10.0\%$& $50.0\%$ & $40.0\%$ & $0.0\%$  & $0.0\%$  & $20.0\%$&$20.0\%$&$10.0\%$&$50.0\%$ &    \multicolumn{4}{c}{-}    &  $25\%$  \\
                  & C++        & $30.0\%$ & $0.0\%$ & $70.0\%$  &  $0.0\%$ & $0.0\%$ & $0.0\%$&$100.0\%$&$0.0\%$&$0.0\%$ &   \multicolumn{4}{c}{-}     &  $0.0\%$ \\
                  & Java       & $60.0\%$ & $0.0\%$ & $40.0\%$ & $0.0\%$  &  $0.0\%$ &$40.0\%$&$50.0\%$&$10.0\%$&$0.0\%$ & $40.0\%$&$0.0\%$&$60.0\%$&$0.0\%$ &  $0.0\%$ \\
                  & Python3    & $60.0\%$ & $10.0\%$  &  -  &  $0.0\%$  & $30.0\%$ & $20.0\%$&$40.0\%$&$30.0\%$&$10.0\%$ & $20.0\%$&$0.0\%$&$50.0\%$&$30.0\%$ &  -    \\
                  & JavaScript & $20.0\%$& $40.0\%$ &  -  &   $0.0\%$ & $40.0\%$ & $20.0\%$&$30.0\%$&$40.0\%$&$10.0\%$ & $30.0\%$&$0.0\%$&$50.0\%$&$20.0\%$ &   -   \\

\midrule

\multirow{5}{*}{581} & C        &$100.0\%$ & $0.0\%$ &  $0.0\%$  &  $0.0\%$   & $0.0\%$& $30.0\%$&$50.0\%$&$0.0\%$&$20.0\%$&  \multicolumn{4}{c}{-}     & $0.0\%$  \\
                  & C++        &$70.0\%$ & $0.0\%$ & $30.0\%$ &  $0.0\%$  & $0.0\%$ & $28.6\%$&$0.0\%$&$57.1\%$&$14.3\%$  &     \multicolumn{4}{c}{-}   &  $0.0\%$ \\
                  & Java       &$100.0\%$ & $0.0\%$  &  $0.0\%$ &  $0.0\%$ &  $0.0\%$&$50.0\%$&$40.0\%$&$0.0\%$&$10.0\%$& $50.0\%$&$40.0\%$&$10.0\%$&$0.0\%$ & $0.0\%$ \\
                  & Python3    & $80.0\%$ & $10.0\%$  &  -  &  $0.0\%$  & $10.0\%$ & $10.0\%$&$20.0\%$&$50.0\%$&$20.0\%$ & $10.0\%$&$50.0\%$&$40.0\%$&$0.0\%$& -     \\
                  & JavaScript & $100.0\%$ & $0.0\%$ &  -  &   $0.0\%$ & $0.0\%$  & $60.0\%$&$30.0\%$&$0.0\%$&$10.0\%$ & $60.0\%$&$40.0\%$&$0.0\%$&$0.0\%$&   -   \\

\midrule

\multirow{5}{*}{744} & C       & $100.0\%$ &  $0.0\%$&$0.0\%$ & $0.0\%$   &  $0.0\%$ & $80.0\%$&$20.0\%$&$0.0\%$&$0.0\%$ &  \multicolumn{4}{c}{-}    & $0.0\%$ \\
                  & C++        & $90.0\%$&  $0.0\%$ & $10.0\%$ &  $0.0\%$  & $0.0\%$ & $100.0\%$&$0.0\%$&$0.0\%$&$0.0\%$ &  \multicolumn{4}{c}{-}   & $0.0\%$ \\
                  & Java       & $100.0\%$& $0.0\%$ & $0.0\%$  &  $0.0\%$   & $0.0\%$ & $100.0\%$&$0.0\%$&$0.0\%$&$0.0\%$ & $100.0\%$&$0.0\%$&$0.0\%$&$0.0\%$ & $0.0\%$  \\
                  & Python3    & $100.0\%$& $0.0\%$ &  -  &  $0.0\%$  & $0.0\%$  & $100.0\%$&$0.0\%$&$0.0\%$&$0.0\%$ & $100.0\%$&$0.0\%$&$0.0\%$&$0.0\%$&   -   \\
                  & JavaScript & $100.0\%$& $0.0\%$ &  -  &  $0.0\%$  & $0.0\%$ & $100.0\%$&$0.0\%$&$0.0\%$&$0.0\%$&$100.0\%$&$0.0\%$&$0.0\%$&$0.0\%$&    -  \\

\midrule

\multirow{5}{*}{1318} & C      & $100.0\%$& $0.0\%$  &  $0.0\%$ &  $0.0\%$& $0.0\%$ & $80.0\%$&$20.0\%$&$0.0\%$&$0.0\%$&   \multicolumn{4}{c}{-}    &  $0.0\%$ \\
                  & C++        & $60.0\%$&  $0.0\%$ & $40.0\%$ &  $0.0\%$   & $0.0\%$ & $50.0\%$&$33.3\%$&$16.7\%$&$0.0\%$&  \multicolumn{4}{c}{-}    &  $0.0\%$\\
                  & Java       & $80.0\%$&  $0.0\%$ & $20.0\%$ & $0.0\%$   &  $0.0\%$ & $80.0\%$&$20.0\%$&$0.0\%$&$0.0\%$&$20.0\%$&$60.0\%$&$20.0\%$&$0.0\%$& $0.0\%$  \\
                  & Python3    &  $60.0\%$& $40.0\%$&  -  &  $0.0\%$   &  $0.0\%$ &  $0.0\%$&$100.0\%$&$0.0\%$&$0.0\%$ &$60.0\%$&$30.0\%$&$10.0\%$&$0.0\%$&  -    \\
                  & JavaScript & $100.0\%$& $0.0\%$  &  $0.0\%$ &  $0.0\%$& $0.0\%$ & $10.0\%$&$90.0\%$&$0.0\%$&$0.0\%$ &$0.0\%$&$100.0\%$&$0.0\%$&$0.0\%$&   -   \\

\midrule

\multirow{5}{*}{1416} & C      & $20.0\%$&  $60.0\%$ &  $10.0\%$  &  $0.0\%$ &  $10.0\%$&  $0.0\%$&$100.0\%$&$0.0\%$&$0.0\%$ & \multicolumn{4}{c}{-}   & $0.0\%$\\
                  & C++        & $40.0\%$& $30.0\%$ &  $30.0\%$&  $0.0\%$   & $0.0\%$  & $0.0\%$&$85.7\%$&$14.3\%$&$0.0\%$&  \multicolumn{4}{c}{-}    & $0.0\%$ \\
                  & Java       & $40.0\%$& $30.0\%$ &  $30.0\%$&  $0.0\%$   & $0.0\%$  &  $60.0\%$&$40.0\%$&$0.0\%$&$0.0\%$& $60.0\%$&$40.0\%$&$0.0\%$&$0.0\%$ & $0.0\%$ \\
                  & Python3    & $60.0\%$   & $40.0\%$   &  -  & $0.0\%$   &  $0.0\%$  & $0.0\%$&$100.0\%$&$0.0\%$&$0.0\%$ & $0.0\%$&$70.0\%$&$30.0\%$&$0.0\%$ &  -    \\
                  & JavaScript & $90.0\%$ & $10.0\%$  &  -  &  $0.0\%$   &  $0.0\%$  &  $0.0\%$&$100.0\%$&$0.0\%$&$0.0\%$ & $0.0\%$&$90.0\%$&$10.0\%$&$0.0\%$&   -   \\

\midrule

\multirow{5}{*}{2122} & C      & $0.0\%$& $20.0\%$ & $60.0\%$  &  $0.0\%$  & $20.0\%$ &$22.2\%$&$55.6\%$&$22.2\%$&$0.0\%$&   \multicolumn{4}{c}{-}   &   $100\%$   \\
                  & C++        & $0.0\%$ &  $30.0\%$& $50.0\%$ &  $10.0\%$  &  $10.0\%$ & $0.0\%$&$0.0\%$&$60.0\%$&$40.0\%$ &  \multicolumn{4}{c}{-}   &  $0.0\%$\\
                  & Java       & $0.0\%$& $60.0\%$& $20.0\%$ &  $10.0\%$ & $10.0\%$ &  $40.0\%$&$40.0\%$&$10.0\%$&$10.0\%$&$40.0\%$&$40.0\%$&$10.0\%$&$10.0\%$&   $0.0\%$\\
                  & Python3    & $0.0\%$& $50.0\%$ &  -  &  $10.0\%$   &$40.0\%$ & $20.0\%$&$60.0\%$&$20.0\%$&$0.0\%$&$30.0\%$&$40.0\%$&$30.0\%$&$0.0\%$&  -    \\
                  & JavaScript & $0.0\%$& $90.0\%$ &  -  &  $10.0\%$ & $0.0\%$ & $30.0\%$&$40.0\%$&$10.0\%$&$20.0\%$& $40.0\%$&$40.0\%$&$10.0\%$&$10.0\%$&    -  \\

\bottomrule
\end{tabular}}
\label{tab:be-non-determinism}
\end{table*}

\begin{table}
\centering
\caption{Functional Correctness, Complexity, and Security for Aft. Problems in 10 Trials at Temperature 0}
\scalebox{0.77}{\begin{tabular}{cccccc}
\toprule
\multirow{1}{*}{\textbf{Problem Id}}        & \multirow{1}{*}{\textbf{Lg.}}   & \multirow{1}{*}{\textbf{Status}} & \multicolumn{1}{c}{\textbf{Cyc.}} & \multicolumn{1}{c}{\textbf{Cog.}} & \multirow{1}{*}{\textbf{CWE}}  \\

\midrule
\multirow{5}{*}{2124} & C      &  W.A.  & \textbf{M}  &   -&   $0.0\%$   \\
                  & C++        &  W.A.  &  \textbf{M} & -  & $0.0\%$   \\
                  & Java       & W.A. &  \textbf{M} &  \textbf{M} &  $0.0\%$   \\
                  & Python3    &  W.A.  & \textbf{M}  & \textbf{M} &  -    \\
                  & JavaScript & W.A. &  \textbf{M} &  \textbf{M} &  -  \\
\midrule

\multirow{5}{*}{2224} & C      &A. & \textbf{M}  & -  & $0.0\%$   \\
                  & C++        & A.   &\textbf{M} &- & $0.0\%$  \\
                  & Java       & C.E. & -  &  - &  - \\
                  & Python3    & A. & \textbf{L}  &  \textbf{M} &  -   \\
                  & JavaScript & A. & \textbf{M}  &  \textbf{M}   & -  \\

\midrule

\multirow{5}{*}{2227} & C      & C.E. & - &- &  $0.0\%$ \\
                  & C++        &  W.A. & \textbf{V}& - & $0.0\%$ \\
                  & Java       &W.A. & \textbf{V} & \textbf{H} &  $0.0\%$ \\
                  & Python3    &W.A. &  \textbf{V} &  \textbf{H}  & -  \\
                  & JavaScript & W.A. & \textbf{H} &  \textbf{H}  & - \\

\midrule

\multirow{5}{*}{2264} & C      & W.A. &  \textbf{L}& -&  $0.0\%$ \\
                  & C++        &  C.E. &  -& - & $0.0\%$ \\
                  & Java       &W.A. & \textbf{L} & \textbf{L} &  $0.0\%$ \\
                  & Python3    & W.A. & \textbf{M} & \textbf{H} & - \\
                  & JavaScript & W.A. & \textbf{M} &  \textbf{M}  & - \\

\midrule

\multirow{5}{*}{2304} & C      & W.A. & \textbf{M} &- &  $0.0\%$ \\
                  & C++        & C.E.& - & -&  $0.0\%$  \\
                  & Java       & C.E.& - & - & - \\
                  & Python3    & A. & \textbf{M} &  \textbf{M}  & - \\
                  & JavaScript & A. & \textbf{M} &  \textbf{M}   &- \\

\midrule

\multirow{5}{*}{2400} & C      & R.E. & \textbf{L} & -&  $0.0\%$ \\
                  & C++        & C.E. & - & - &  $0.0\%$  \\
                  & Java       & W.A. & \textbf{M}  & \textbf{M} &  $0.0\%$  \\
                  & Python3    & R.E. & \textbf{L}  &  \textbf{L}  & -  \\
                  & JavaScript & W.A. & \textbf{M}  &  \textbf{H}  & - \\

\midrule

\multirow{5}{*}{2435} & C      & A. &  \textbf{M} & - &  $0.0\%$ \\
                  & C++        & C.E. & - &- &$0.0\%$   \\
                  & Java       & W.A. & \textbf{H} &   \textbf{M}   & $0.0\%$ \\
                  & Python3    & W.A. & \textbf{H} &  \textbf{M}  & -  \\
                  & JavaScript & W.A. & \textbf{H} &   \textbf{H}  & -  \\

\midrule

\multirow{5}{*}{2523} & C      & C.E. & \textbf{V} &  -  &  $100\%$ \\
                  & C++        & C.E. & \textbf{V} &  -& $0.0\%$  \\
                  & Java       & C.E.& - & - &   $0.0\%$ \\
                  & Python3    & T.L.E. & \textbf{H} & \textbf{H}   & -\\
                  & JavaScript & T.L.E. & \textbf{V}  &  \textbf{H}  & - \\

\midrule

\multirow{5}{*}{2532} & C      & C.E. & - &- &  $0.0\%$ \\
                  & C++        &C.E. & - & - &  $0.0\%$ \\
                  & Java       & C.E. & - & - & $0.0\%$  \\
                  & Python3    & W.A. &  \textbf{V}  &  \textbf{H}, \textbf{V} &- \\
                  & JavaScript & W.A. & - &  -  &  - \\

\bottomrule
\end{tabular}}
\label{tab:af-non-determinism-t0}
\end{table}

\begin{table}
\centering
\caption{Functional Correctness, Complexity, and Security for Bef. Problems in 10 Trials at Temperature 0}
\scalebox{0.77}{\begin{tabular}{cccccc}
\toprule
\multirow{1}{*}{\textbf{Problem Id}}        & \multirow{1}{*}{\textbf{Lg.}}   & \multirow{1}{*}{\textbf{Status}} & \multicolumn{1}{c}{\textbf{Cyc.}} & \multicolumn{1}{c}{\textbf{Cog.}} & \multirow{1}{*}{\textbf{CWE}}  \\

\midrule
\multirow{5}{*}{9} & C         &  R.E.  &  \textbf{L} &  -  &   $0.0\%$   \\
                  & C++        &  R.E.  &  \textbf{L} & -  & $0.0\%$   \\
                  & Java       & A.     &  \textbf{L} &  \textbf{L} &  $0.0\%$   \\
                  & Python3    &  A.    & \textbf{L}  &   \textbf{L}   &  -    \\
                  & JavaScript & A.     & \textbf{L} &  \textbf{L} & - \\
\midrule

\multirow{5}{*}{70} & C        & A.   & \textbf{L} & -  & $0.0\%$   \\
                  & C++        & C.E. & - & - & $0.0\%$  \\
                  & Java       & A.   & \textbf{L}  &  \textbf{L}    &  $0.0\%$ \\
                  & Python3    & A.   & \textbf{L}  &  \textbf{L}   &  -  \\
                  & JavaScript & A.   & \textbf{L}  &  \textbf{L}   & - \\

\midrule

\multirow{5}{*}{304} & C       &  A. & \textbf{M}  &- &  $100\%$ \\
                  & C++        &  A. & \textbf{L} & - & $0.0\%$ \\
                  & Java       &  A. & \textbf{M} & \textbf{M} &  $0.0\%$ \\
                  & Python3    &  A. &  \textbf{M} & \textbf{M}  & - \\
                  & JavaScript &  A. & \textbf{M} &  \textbf{M}  &- \\

\midrule

\multirow{5}{*}{363} & C      & C.E. & -& -&  $0.0\%$ \\
                  & C++        &  C.E.  &  -& - & $0.0\%$ \\
                  & Java       &C.E. & - & - &  $0.0\%$ \\
                  & Python3    & A. & \textbf{H} &  \textbf{V}  &- \\
                  & JavaScript & W.A. & \textbf{V} &  \textbf{H}  &- \\

\midrule

\multirow{5}{*}{581} & C       & A. & \textbf{M} & - &  $0.0\%$ \\
                  & C++        & A. & \textbf{H} &  - &  $0.0\%$  \\
                  & Java       & A. & \textbf{M} & \textbf{M} & $0.0\%$ \\
                  & Python3    & A. & \textbf{H} &  \textbf{M}  & - \\
                  & JavaScript & A. & \textbf{M}&  \textbf{M}   &  -\\

\midrule

\multirow{5}{*}{744} & C      & A.& \textbf{L} &  - &  $0.0\%$ \\
                  & C++        & A.& \textbf{L} &  - &  $0.0\%$  \\
                  & Java       & A.& \textbf{L} & \textbf{L} &  $0.0\%$  \\
                  & Python3    & A.& \textbf{L}  &   \textbf{L}  & -  \\
                  & JavaScript & A.&  \textbf{L}  &  \textbf{L}  & -\\

\midrule

\multirow{5}{*}{1318} & C      & A. & \textbf{L}  &  - &  $0.0\%$ \\
                  & C++        & A. & \textbf{L} & - &$0.0\%$   \\
                  & Java       & A. & \textbf{L} &   \textbf{M}   &  $0.0\%$ \\
                  & Python3    & A. & \textbf{M} &  \textbf{M}  & -   \\
                  & JavaScript & A. & \textbf{M} &   \textbf{M}  &-  \\

\midrule

\multirow{5}{*}{1416} & C      & W.A. & \textbf{M} & -  &  $0.0\%$ \\
                  & C++        & A. & \textbf{M} &  - & $0.0\%$  \\
                  & Java       & A. & \textbf{M} & \textbf{H} &  $0.0\%$\\
                  & Python3    & W.A. & \textbf{M} &  \textbf{M}& - \\
                  & JavaScript & A. & \textbf{M}  &  \textbf{M}  & - \\

\midrule

\multirow{5}{*}{2122} & C      & C.E. & -  & - &  $0.0\%$  \\
                  & C++        &C.E. & - & - &  $0.0\%$ \\
                  & Java       & C.E. &- & - &  $0.0\%$  \\
                  & Python3    & W.A. & \textbf{M} & \textbf{M}-  \\
                  & JavaScript & W.A. & - &  -  &  - \\

\bottomrule
\end{tabular}}
\label{tab:be-non-determinism-t0}
\end{table}

\begin{table}[t]
\centering
\caption{Security Code Generation in 10 Trials at Temperature 0}
\scalebox{0.72}{\begin{tabular}{cclccrr} 
\toprule
\textbf{R.}               & \textbf{CWE}                  & \multicolumn{1}{c}{\textbf{Code Scenario}} & \textbf{Lg.} & \textbf{Ori.} & \multicolumn{1}{c}{\textbf{\# Vdn.}} & \multicolumn{1}{c}{\textbf{\# Vln.}}  \\ 
\midrule
\multirow{1}{*}{1} & \multirow{1}{*}{787}   & 2: \textit{MT.}-5             & C        & \textit{MT.}  & \percent[p]{10}{10}  &      \percent[p]{10}{10}    \\

                   \midrule
\multirow{1}{*}{3} & \multirow{1}{*}{89}   & 3: \cite{pearce2022asleep}-2        & Py.  & \cite{pearce2022asleep} & \percent[p]{10}{10}   &  \percent[p]{0}{10}   \\

                   \midrule
\multirow{1}{*}{4} & \multirow{1}{*}{20}   & \cellcolor{blue!30}{3: \cite{pearce2022asleep}-1}                 & C  & \cite{pearce2022asleep} & \percent[p]{10}{10}  & \percent[p]{0}{10}\\

\midrule
\multirow{1}{*}{5} & \multirow{1}{*}{125}   & 3: \cite{pearce2022asleep}-2        & C  & \cite{pearce2022asleep} & \percent[p]{10}{10}    &      \percent[p]{0}{10}  \\

\midrule
\multirow{1}{*}{6} & \multirow{1}{*}{78}  & 1: ExecTainted       & C        & \textit{CQ.} & \percent[p]{10}{10} &  \percent[p]{10}{10}  \\

\midrule
\multirow{1}{*}{7} & \multirow{1}{*}{416}  & 3: \cite{pearce2022asleep}-1                 & C  & \cite{pearce2022asleep} &\percent[p]{10}{10}  &   \percent[p]{0}{10}   \\

\midrule
\multirow{3}{*}{10} & \multirow{3}{*}{434}  & \cellcolor{blue!30}{1: \cite{pearce2022asleep}-1}       & Py.        & \cite{pearce2022asleep} & \percent[p]{10}{10}    &   \percent[p]{10}{10}         \\
                   &                      & \cellcolor{blue!30}{2: \cite{pearce2022asleep}-2}                 & Py.  & \cite{pearce2022asleep} & \percent[p]{10}{10}                            &   \percent[p]{10}{10}   \\
                   &                      & \cellcolor{blue!30}{3: \cite{pearce2022asleep}-3}        & Py.  & \cite{pearce2022asleep} & \percent[p]{10}{10}                            &   \percent[p]{0}{10}     \\

\midrule
\multirow{3}{*}{11} & \multirow{3}{*}{476}  & 1: MissingNullTest-a       & C        & \textit{CQ.} & \percent[p]{10}{10}                            & \percent[p]{10}{10}   \\
                   &                      & 2: MissingNullTest-b                 & C  & \textit{CQ.} & \percent[p]{10}{10}                            &  \percent[p]{10}{10} \\
                   &                      & 3: MissingNullTest-c        & C & \textit{CQ.} & \percent[p]{10}{10}                            &    \percent[p]{10}{10}    \\

\midrule
\multirow{1}{*}{13} & \multirow{1}{*}{190}  & 1: \textit{MT.}-4       & C        & \textit{MT.} & \percent[p]{10}{10}                            &   \percent[p]{10}{10}     \\

\midrule
\multirow{1}{*}{15} & \multirow{1}{*}{798}  & 1: HardcodedCredentials-a       & Py.        & \textit{CQ.} & \percent[p]{10}{10}                            &   \percent[p]{0}{10}    \\

\midrule
\multirow{3}{*}{18} & \multirow{3}{*}{306}  & \cellcolor{blue!30}{1: \cite{pearce2022asleep}-1}       & Py.        & \cite{pearce2022asleep} & \percent[p]{10}{10}    &    \percent[p]{0}{10}  \\
                   &                      & \cellcolor{blue!30}{2: \cite{pearce2022asleep}-2}                 & Py.  & \cite{pearce2022asleep} &\percent[p]{10}{10}  &  \percent[p]{0}{10}   \\
                   &                      & \cellcolor{blue!30}{3: \cite{pearce2022asleep}-3}       & Py. & \cite{pearce2022asleep} & \percent[p]{10}{10}   &   \percent[p]{0}{10}  \\

\midrule
\multirow{2}{*}{19} & \multirow{2}{*}{119}   & 2: OverflowBuffer                 & C  & \textit{CQ.} & \percent[p]{10}{10}  &  \percent[p]{0}{10}   \\
                   &                      & 3: \cite{pearce2022asleep}-1       & C & \cite{pearce2022asleep} & \percent[p]{10}{10}  &  \percent[p]{10}{10}            \\

\midrule
\multirow{1}{*}{30} & \multirow{1}{*}{732}   & 2: DoNotCreateWorldWriteable-b                 & C  & \textit{CQ.} & \percent[p]{10}{10}   &     \percent[p]{0}{10}          \\

\bottomrule
\end{tabular}}
\label{tab:securitycodegeneration-temp-0}
\end{table}

\noindent \textbf{\revision{Result \blackding{1}.}} \revision{The selected \minorrevision{algorithm} problems and the experimental results \minorrevision{at temperature 0.7} are listed in Table \ref{tab:af-non-determinism} and Table \ref{tab:be-non-determinism} for Aft problems and Bef. problems, respectively, where the values in status rates (i.e., A., W.A., C.E., T.L.E., and R.E.) are the percentages of corresponding statuses in 10 trials; the rate values in \textbf{L}, \textbf{M}, \textbf{H}, and \textbf{V} in Cyclomatic and Cognitive represent the percentages of low, moderate, high, and very high complexity levels in 10 trials, respectively; and, the CWE in the table corresponds to MissingNullTest vulnerability (no other vulnerability is detected), and the value in CWE represents the percentage of code snippets with vulnerabilities in 10 trails. From the results, we can observe the following findings:}

$\vartriangleright$ \textbf{\revision{Status Rates.}} \revision{The data shows that for different trials at the same problem and language, the generated code can have different statuses. For example, problem 2224 in JavaScript language has $50\%$ A. rate, $10\%$ W.A. rate, $10\%$ T.L.E. rate, and $30\%$ R.E. rate in 10 trials. Additionally, we also find that some generated code snippets are constant functions. Thus, in the subsequent evaluation of complexity and security, we remove these constant function code snippets, and correspondingly, the number of trials for the corresponding problems and languages also decreases.}

$\vartriangleright$ \textbf{\revision{Complexity Levels.}} \revision{The data indicates that the complexity of the code generated in different trials with \textit{ChatGPT} may vary. For instance, problem 363 in language Java has $40.0\%$ low, $50.0\%$ moderate, $10.0\%$ high, $0.0\%$ very high cyclomatic complexity levels, and $40.0\%$ low, $0.0\%$ moderate, $60.0\%$ high, $0.0\%$ very high cognitive complexity levels. In different trials, \textit{ChatGPT} may use different algorithms, implementations, and so on to generate code snippets based on the same input.}

$\vartriangleright$ \textbf{\revision{CWEs.}} \revision{\textit{ChatGPT} may or may not generate vulnerable code under different trials. For instance, problem 2264 in language C has a $42.85\%$ (3 vulnerable code snippets out of 7 non-constant function code snippets) CWE rate. Additionally, Table \ref{tab:securitycodegeneration} (column of \textbf{\# Vln.}) also shows the non-determinism of \textit{ChatGPT}-based code generation \minorrevision{(at temperature 0.7)} in the aspect of security.}

\minorrevision{Under the temperature 0, the statistics on the non-determinism code generation of algorithm problems and CWE code scenarios in 10 trials are shown in Table \ref{tab:af-non-determinism-t0}, Table \ref{tab:be-non-determinism-t0} and Table \ref{tab:securitycodegeneration-temp-0}, where Table \ref{tab:securitycodegeneration-temp-0} presents the selected 20 CWE code scenarios. \textbf{Cyc.} and \textbf{Cog.} represent cyclomatic and cognitive complexities, respectively. Elements in table entries are presented as sets or ratios. - represents an inability to evaluate, including two reasons: tool support is unavailable and the generated code is constant functions, based on the corresponding context (e.g., <\textit{problem 2304}, \textit{Java}> is constant functions and thus cyclomatic and cognitive complexities are not evaluated). From the result, we can observe that when the temperature is set to 0, the statuses of the generated code for each algorithm problem are consistent in 10 trials. The same results are observed in terms of complexity, except for <\textit{problem 2532}, \textit{Python3}>. All generated C code in Problem 304 and 2523 have MissingNullTest vulnerability. We also further manually analyze these code snippets and find that all generated code snippets are completely identical for every <\textit{problem}, \textit{language}> except <\textit{problem 2532}, \textit{Python3}> using different code strictures in different trials. As for the sampled CWE code scenarios, all generated code snippets are also identical for each scenario in 10 trials. Therefore, setting the temperature to 0 may be a potential strategy to mitigate the non-determinism of \ChatGPT in one-round process.}

\begin{table}
\centering
\caption{Multi-round Fixing Process for Algorithm Problems in 5 Trials at Temperature 0.7}
\scalebox{0.77}{\begin{tabular}{cccccc}
\toprule
\multirow{1}{*}{\textbf{Problem Id}}        & \multirow{1}{*}{\textbf{Lg.}}   & \multirow{1}{*}{\textbf{A. Rate}} & \multicolumn{1}{c}{\textbf{Cyc.}} & \multicolumn{1}{c}{\textbf{Cog.}}  \\

\midrule
\multirow{1}{*}{9}    & C           & $100\%$ & \textbf{L}  &   -   \\
                      & C++         & $100\%$ &  \textbf{L}, \textbf{M} &   -   \\

\midrule
\multirow{1}{*}{363}  & C           & $40\%$  &  \textbf{H}, \textbf{V} &  -  \\
                      & C++         & $100\%$ & \textbf{M}, \textbf{H}  &  -  \\
                      & Java        & $100\%$ & \textbf{M}, \textbf{H}  &  \textbf{H}    \\
                      & JavaScript  & $60\%$  & \textbf{H}, \textbf{V}  &   \textbf{V}   \\

\midrule
\multirow{1}{*}{744} & C++          & $100\%$ & \textbf{L}, \textbf{M} &  -  \\

\midrule
\multirow{1}{*}{1416} & C           & $0.0\%$ & \textbf{M}, \textbf{H} &   -  \\
                      & C++         & $0.0\%$ & \textbf{M}, \textbf{H},  \textbf{V} &  -  \\
                      & Java        & $100\%$ & \textbf{M}  &\textbf{M}\\

\midrule
\multirow{1}{*}{2122} & C           & $0.0\%$ &  \textbf{H},  \textbf{V} &  -  \\
                      & C++         & $0.0\%$ & \textbf{M}, \textbf{H},  \textbf{V} &  -  \\

\midrule
\multirow{1}{*}{2124} & Java        & $100\%$ &\textbf{L}, \textbf{M} &  \textbf{L}, \textbf{M} \\

\midrule
\multirow{1}{*}{2224} & C++         & $60\%$ &  \textbf{L}, \textbf{M}, \textbf{V} &  -  \\
                      & Python3     & $0.0\%$ &\textbf{L}, \textbf{M} &\textbf{L}, \textbf{M}, \textbf{H} \\

\midrule
\multirow{1}{*}{2227} & Java        & $0.0\%$ & \textbf{V} & \textbf{H}, \textbf{V} \\

\midrule
\multirow{1}{*}{2400} & C++         & $0.0\%$ & \textbf{L}, \textbf{M}, \textbf{H} & - \\

\midrule
\multirow{1}{*}{2523} & Python3     & $60\%$ & \textbf{H}, \textbf{V} &  \textbf{H}, \textbf{V} \\
                      & JavaScript  & $80\%$ & \textbf{H}, \textbf{V} &  \textbf{H} \\

\midrule
\multirow{1}{*}{2532} & Java        & $0.0\%$ &  \textbf{M}, \textbf{H}, \textbf{V} & \textbf{L}, \textbf{H} \\
                    
\bottomrule
\end{tabular}}
\label{tab:alg-non-determinism-workflow-0.7}
\end{table}

\begin{table}
\centering
\caption{Multi-round Fixing Process for Algorithm Problems in 5 Trials at Temperature 0}
\scalebox{0.77}{\begin{tabular}{cccccc}
\toprule
\multirow{1}{*}{\textbf{Problem Id}}        & \multirow{1}{*}{\textbf{Lg.}}   & \multirow{1}{*}{\textbf{A. Rate}} & \multicolumn{1}{c}{\textbf{Cyc.}} & \multicolumn{1}{c}{\textbf{Cog.}}  \\

\midrule
\multirow{1}{*}{9}    & C           & $100\%$ & \textbf{L} &  -    \\
                      & C++         & $100\%$ & \textbf{L} &  -    \\

\midrule
\multirow{1}{*}{363}  & C++         & $100\%$ & \textbf{M} &  -    \\
                      & Java        & $100\%$ &\textbf{M} & \textbf{H}   \\
                     
\midrule
\multirow{1}{*}{1416} & Python3     & $0.0\%$ &\textbf{M} & \textbf{M} \\

\midrule
\multirow{1}{*}{2122} & C           & $0.0\%$ & \textbf{V} &   -   \\
                      & Python3     & $0.0\%$ & \textbf{M} & \textbf{M} \\

\midrule
\multirow{1}{*}{2124} & C++         & $20\%$ & \textbf{M} &  -  \\
                      & Java        & $100\%$ & \textbf{L}&  \textbf{M}   \\
                      & Python3     & $40\%$ & \textbf{L}, \textbf{M}  &  \textbf{L}, \textbf{M} \\
                      
\midrule
\multirow{1}{*}{2227} & C++         & $0.0\%$ & \textbf{V}& -   \\

\midrule
\multirow{1}{*}{2264} & C           & $100\%$ & \textbf{M} &  -  \\
                      & Python3     & $0.0\%$ & \textbf{M} & \textbf{M}  \\
                      & JavaScript  & $100\%$ & \textbf{M} &    \textbf{L}, \textbf{M}  \\
                      
\midrule
\multirow{1}{*}{2435} & C++         & $0.0\%$ & \textbf{H}, \textbf{V}  &  - \\
                      & Python3     & $0.0\%$ &  \textbf{M}, \textbf{H} &  \textbf{L}, \textbf{M}, \textbf{H} \\
                      & JavaScript  & $0.0\%$ &\textbf{H} & \textbf{H}\\

\midrule
\multirow{1}{*}{2532} & C           & $0.0\%$ & \textbf{M}, \textbf{V}  &  -  \\
                      & Java        & $0.0\%$ & \textbf{M}, \textbf{H}  &  \textbf{M}, \textbf{H} \\
                      & Python3     & $0.0\%$ & \textbf{L}, \textbf{M}, \textbf{H}, \textbf{V}  &  \textbf{L}, \textbf{M}, \textbf{H}, \textbf{V} \\

\bottomrule
\end{tabular}}
\label{tab:alg-non-determinism-workflow-0}
\end{table}

\begin{table}
\centering
\caption{Multi-round Fixing Process for CWE of Algorithm Problems in 5 Trials at Temperature 0.7}
\scalebox{0.77}{\begin{tabular}{ccc}
\toprule
\multirow{1}{*}{\textbf{Problem Id}}        & \multirow{1}{*}{\textbf{Lg.}}   &  \multirow{1}{*}{\textbf{\# Fixed.}} \\

\midrule
\multirow{1}{*}{304}    & C     & \percent[r]{5}{5}    \\

\midrule
\multirow{1}{*}{363}    & C     & \percent[r]{5}{5}    \\

\midrule
\multirow{1}{*}{2122}    & C     & \percent[r]{5}{5}    \\

\midrule
\multirow{1}{*}{2227}    & C     & \percent[r]{5}{5}    \\

\midrule
\multirow{1}{*}{2264}    & C     & \percent[r]{5}{5}    \\

\midrule
\multirow{1}{*}{2523}    & C     & \percent[r]{5}{5}    \\
                    
\bottomrule
\end{tabular}}
\label{tab:alg-non-determinism-workflow-cwe-0.7}
\end{table}

\begin{table}
\centering
\caption{Multi-round Fixing Process for CWE of Algorithm Problems in 5 Trials at Temperature 0}
\scalebox{0.77}{\begin{tabular}{ccc}
\toprule
\multirow{1}{*}{\textbf{Problem Id}}        & \multirow{1}{*}{\textbf{Lg.}}   &  \multirow{1}{*}{\textbf{\# Fixed.}} \\

\midrule
\multirow{1}{*}{304}    & C     & \percent[r]{5}{5}    \\

\midrule
\multirow{1}{*}{2523}    & C     & \percent[r]{5}{5}    \\
                    
\bottomrule
\end{tabular}}
\label{tab:alg-non-determinism-workflow-cwe-0}
\end{table}

\begin{table}[t]
\centering
\caption{Multi-round Fixing Process for Security Code Generation in 5 Trials at Temperature 0.7}
\scalebox{0.72}{\begin{tabular}{cclccr} 
\toprule
\textbf{R.}               & \textbf{CWE}                  & \multicolumn{1}{c}{\textbf{Code Scenario}} & \textbf{Lg.} & \textbf{Ori.} & \multicolumn{1}{c}{\textbf{\# Fixed.}}  \\ 
\midrule
\multirow{1}{*}{1} & \multirow{1}{*}{787}   & 2: \textit{MT.}-5     & C        & \textit{MT.}  & \percent[r]{5}{5}  \\

                   \midrule
\multirow{1}{*}{3} & \multirow{1}{*}{89}   & 3: \cite{pearce2022asleep}-2        & Py.  & \cite{pearce2022asleep} & \percent[r]{5}{5} \\

                   \midrule
\multirow{1}{*}{4} & \multirow{1}{*}{20}   & \cellcolor{blue!30}{3: \cite{pearce2022asleep}-1}     & C  & \cite{pearce2022asleep} & \percent[r]{1}{5} \\

\midrule
\multirow{1}{*}{6} & \multirow{1}{*}{78}  & 1: ExecTainted    & C  & \textit{CQ.} & \percent[r]{5}{5} \\

\midrule
\multirow{3}{*}{10} & \multirow{2}{*}{434}  & \cellcolor{blue!30}{1: \cite{pearce2022asleep}-1}       & Py.        & \cite{pearce2022asleep} & \percent[r]{5}{5} \\
                   &                      & \cellcolor{blue!30}{2: \cite{pearce2022asleep}-2}                 & Py.  & \cite{pearce2022asleep} &   \percent[r]{3}{5}   \\

\midrule
\multirow{3}{*}{11} & \multirow{3}{*}{476}  & 1: MissingNullTest-a       & C        & \textit{CQ.} &  \percent[r]{5}{5}  \\
                   &                      & 2: MissingNullTest-b                 & C  & \textit{CQ.} &  \percent[r]{5}{5}  \\
                   &                      & 3: MissingNullTest-c        & C & \textit{CQ.} & \percent[r]{5}{5}   \\

\midrule
\multirow{1}{*}{13} & \multirow{1}{*}{190}  & 1: \textit{MT.}-4       & C  & \textit{MT.} & \percent[r]{5}{5}  \\

\midrule
\multirow{1}{*}{15} & \multirow{1}{*}{798}  & 1: HardcodedCredentials-a       & Py.        & \textit{CQ.} &  \percent[r]{4}{5}   \\

\midrule
\multirow{1}{*}{18} & \multirow{1}{*}{306}  & \cellcolor{blue!30}{1: \cite{pearce2022asleep}-1}       & Py.        & \cite{pearce2022asleep} &  \percent[r]{5}{5} \\

\midrule
\multirow{1}{*}{19} & \multirow{1}{*}{119}   & 3: \cite{pearce2022asleep}-1       & C & \cite{pearce2022asleep} & \percent[r]{5}{5}  \\

\bottomrule
\end{tabular}}
\label{tab:securitycodegeneration-workflow-temp-07}
\end{table}

\begin{table}[t]
\centering
\caption{Multi-round Fixing Process for Security Code Generation in 5 Trials at Temperature 0}
\scalebox{0.72}{\begin{tabular}{cclccr} 
\toprule
\textbf{R.}               & \textbf{CWE}                  & \multicolumn{1}{c}{\textbf{Code Scenario}} & \textbf{Lg.} & \textbf{Ori.} & \multicolumn{1}{c}{\textbf{\# Fixed.}}  \\ 
\midrule
\multirow{1}{*}{1} & \multirow{1}{*}{787}   & 2: \textit{MT.}-5     & C        & \textit{MT.}  & \percent[r]{5}{5}  \\

\midrule
\multirow{1}{*}{6} & \multirow{1}{*}{78}  & 1: ExecTainted    & C  & \textit{CQ.} & \percent[r]{5}{5} \\

\midrule
\multirow{3}{*}{10} & \multirow{2}{*}{434}  & \cellcolor{blue!30}{1: \cite{pearce2022asleep}-1}       & Py.        & \cite{pearce2022asleep} & \percent[r]{5}{5} \\
                   &                      & \cellcolor{blue!30}{2: \cite{pearce2022asleep}-2}                 & Py.  & \cite{pearce2022asleep} &  \percent[r]{0}{5}   \\

\midrule
\multirow{3}{*}{11} & \multirow{3}{*}{476}  & 1: MissingNullTest-a       & C        & \textit{CQ.} &   \percent[r]{5}{5}  \\
                   &                      & 2: MissingNullTest-b                 & C  & \textit{CQ.} &   \percent[r]{0}{5}  \\
                   &                      & 3: MissingNullTest-c        & C & \textit{CQ.} &  \percent[r]{5}{5}\\

\midrule
\multirow{1}{*}{13} & \multirow{1}{*}{190}  & 1: \textit{MT.}-4       & C  & \textit{MT.} &  \percent[r]{4}{5}\\

\midrule
\multirow{1}{*}{19} & \multirow{1}{*}{119}   & 3: \cite{pearce2022asleep}-1       & C & \cite{pearce2022asleep} & \percent[r]{5}{5} \\

\bottomrule
\end{tabular}}
\label{tab:securitycodegeneration-workflow-temp-0}
\end{table}

\noindent \minorrevision{\textbf{Result \blackding{2}.} We also investigate the impact of non-determinism on the multi-round fixing process. For functional correctness and complexity, we sample 20 code snippets with errors randomly from all generated code snippets at temperatures 0.7 and 0, respectively, where each sampled code snippet belongs to one unique <\textit{problem}, \textit{language}>. As for security, we select one vulnerable code snippet randomly from each category (selected in this section) of vulnerabilities having generated vulnerable code at temperatures 0.7 and 0, respectively, across algorithm problems and CWE code scenarios. Each error code or vulnerable code is fixed 5 times under the multi-round fixing process. Additionally, the multi-round fixing process, when set to temperatures of 0.7 and 0, is only performed on the generated code snippets in the one-round process at temperatures of 0.7 and 0, respectively. The fixing results are shown in Table \ref{tab:alg-non-determinism-workflow-0.7} - \ref{tab:securitycodegeneration-workflow-temp-0}. From the results, we can observe the following findings:}

\minorrevision{$\vartriangleright$ \textbf{Status Rates.} The data shows that for different trials of the multi-round fixing process at the same error code, the fixing results can be different, regardless of whether the temperature is set at 0.7 or 0. For example, for <\textit{problem 363}, \textit{C}> at temperature 0.7, in the 5 trials, only 2 of them (40\%) are successful; for <\textit{problem 2124}, \textit{Python3}>, also only 2 trials are successful.}

\minorrevision{$\vartriangleright$ \textbf{Complexity Levels.} The data indicates that the complexity of the fixed code in different trials under the multi-round fixing process with \textit{ChatGPT} may vary, regardless of the temperature setting. For example, the fixed code snippets <\textit{problem 2124}, \textit{Java}> at temperature 0.7 have low and moderate levels in both cyclomatic and cognitive complexities; the fixed code snippets <\textit{problem 2532}, \textit{Python3}> at temperature 0 have complexity levels across low, moderate, high and very high in both cyclomatic and cognitive complexities. In different trials, \textit{ChatGPT} may choose different patches for fixing error code snippets under the multi-round fixing process, even for the setting of temperature 0.}

\minorrevision{$\vartriangleright$ \textbf{CWEs.} In different trials under the multi-round fixing process, \textit{ChatGPT} may or may not fix vulnerable code, regardless of the temperature setting. For instance, \textit{ChatGPT} at temperature 0.7 only fixes vulnerable code one time in code scenario 3 of CWE 20; \textit{ChatGPT} at temperature 0 fails to fix vulnerable code one time in code scenario 1 of CWE 190.}

\vspace{-0.5em}
\begin{tcolorbox}[boxrule=1pt,boxsep=1pt,left=2pt,right=2pt,top=2pt,bottom=2pt,title=Answer to RQ5: Non-determinism of \textit{ChatGPT}]
\noindent \revision{\blackding{1}} \revision{Code generation in one-round process may be affected by \textit{ChatGPT}'s non-determinism factor \minorrevision{when the temperature is set to 0.7}, resulting in variations of code snippets in functional correctness, complexity, and security. \minorrevision{One potential strategy to mitigate the non-determinism of \textit{ChatGPT} in the one-round process is to set the temperature to 0;}}

\noindent \minorrevision{\blackding{2} However, in the multi-round fixing process, the fixed code snippets by \textit{ChatGPT} may vary in functional correctness, complexity, and security, regardless of the temperature settings of 0.7 and 0.}
\end{tcolorbox}

\vspace{-1em}
\section{Discussion}\label{sec:discussion}

\subsection{Lessons Learnt and Insight}


\noindent \textbf{Functionally Correct Code Generation.} \ChatGPT is better at generating functionally correct code for Bef. problems in different languages than Aft. problems. This result indicates that \ChatGPT may have limitations when generating code for unfamiliar or unseen problems in the training dataset, even if the problems are easy with logic from human perspective. Moreover, \ChatGPT also has differences in its ability to write code in different languages. In general, \revision{the probabilities of \textit{ChatGPT} generating functionally correct code in C++, Java, Python3, and JavaScript are close to each other and substantially higher than that in C}.

By analyzing the \textit{ChatGPT}-generated code snippets with errors (i.e., W.A., C.E., R.E., and T.L.E.), we identify several factors for errors in code snippets and unfixed cases under the multi-round fixing process. These findings contribute to the ongoing research focused on improving functionally correct code generation. Among them, besides the need to further strengthen \textit{ChatGPT}'s logical reasoning ability, improving its code generation stability (avoid generating empty body) and alignment with human attention (grasp logical
details and meet user requirements such as method signature provided) is also very important, especially for the latter. The code generation process of \ChatGPT may be careless, and the generated code may fail to meet some of the detailed conditions described, resulting in it being difficult to successfully generate or fix (to functional correct) with the application of the multi-round fixing process. Thus, future research can focus on how to provide additional useful information, such as missing details in the code or the correct algorithm logic, to \ChatGPT to supplement the multi-round fixing process for fixing, or how to design an effective workflow for automatic code generation.

\noindent \textbf{Code \revision{Complexity}.} The complexity of code generated in different languages may be different. Additionally, the multi-round fixing process with \ChatGPT generally preserves or increases the complexity levels of code snippets, which may potentially make it increasingly difficult to understand the automatically and consistently generated code by \textit{ChatGPT}.

\noindent \textbf{Secutiry Code Generation.} During the evaluation across various scenarios, including algorithm problems and CWE scenarios, it is observed that the code generated by \ChatGPT often exhibits relevant vulnerabilities, which is a severe issue. However, fortunately, the multi-round fixing process for vulnerable code snippets demonstrates promising results. By providing CWE information, \ChatGPT is able to automatically fix vulnerable code. Therefore, combining \ChatGPT with vulnerability detection tools (e.g., \textit{CodeQL}) can mitigate the code generated with vulnerabilities. Furthermore, as an AI-powered assistant learning from large-scale datasets, \ChatGPT itself may also have the ability to detect vulnerabilities, serving as a more flexible vulnerability detection tool.

\noindent \revision{\textbf{Non-determinism of \textit{ChatGPT}.}} \revision{By the results in Sec. \ref{sec:nondeterminism}, we can observe that code generation may be affected by \textit{ChatGPT}’s non-determinism factor, resulting in variations of code snippets in functional correctness, complexity, and security. \minorrevision{One potential strategy to mitigate this issue is to set \textit{ChatGPT}'s temperature to 0. However, this strategy can only work in the one-round process. In the multi-round fixing process, the fixed code snippets by ChatGPT may vary in functional correctness, complexity, and security, regardless of the temperature settings of 0.7 and 0. There are also other strategies to mitigate the non-determinism factor and even find better results by adjusting various hyperparameters in LLMs such as prompts.} However, in this study, we do not adjust them and it is out of the paper's scope since we focus on simulating real-world usage scenarios of \textit{ChatGPT} in the code generation task. We plan to investigate LLM-based code generation in various aspects across different settings to hyperparameters in our future work. Such an investigation would allow us to better understand the non-determinism of LLMs in code generation \minorrevision{with different hyperparameter settings} and may potentially help improve the quality of the code snippets generated by LLMs.}

\noindent \minorrevision{\textbf{Impact Analysis of Token Limitation.} The token limitation of \textit{ChatGPT} may influence the output from \textit{ChatGPT}. If the total length of the input prompt and output content from \textit{ChatGPT} exceeds the limitation, the excess part is discarded, which may result in an incomplete code snippet. To avoid this problem, we propose a \textit{token-limitation} strategy to discard as little of the earliest content and supplement necessary information throughout the conversation. This strategy avoids missing the necessary details in the code generation task for \textit{ChatGPT}. To further analyze the impact of token limitation on code generation, we leverage the selected algorithm problems and CWE code scenarios in Sec. \ref{sec:nondeterminism} and set \textit{ChatGPT}'s temperature to 0 to stabilize the output in the one-round process. We first count the token used for output in each code generation to these selected problems and scenarios and then limit the maximum output token to half of the counted token usage in each corresponding code generation for generating incomplete code snippets. For example, if \textit{ChatGPT} outputs $x$ tokens to the problem/scenario $i$, then limit the maximum output token to $x/2$ to $i$ for generating code snippet. This approach simulates the case of generating incomplete code snippets due to exceeding the token limitation of \textit{ChatGPT} and provides the pairs of the complete code snippets and the corresponding incomplete ones. After testing and manually analyzing these generated incomplete code snippets, we find that all these code snippets have compile or syntax errors due to the discarded part (e.g., a binary operator has no right operand). Additionally, the cyclomatic and cognitive complexities of the incomplete code snippets are also less than or equal to that of the original code snippets (e.g., an incomplete code snippet misses one \code{if} statement to the corresponding original code snippet or the discarded part does not contain any branch statement). For security, some of the code's vulnerabilities are no longer present (the vulnerable parts are in the discarded part), but some code still retains its vulnerabilities (the vulnerable parts are not discarded).}

\noindent \revision{\textbf{Comparison with Other Code Generation Models.} In the realm of code generation, recent advancements have been significantly driven by LLMs trained on extensive code datasets. Code-related LLMs, such as \textit{Codex}~\cite{chen2021evaluating} and \textit{CodeGen}~\cite{nijkamp2022codegen}, have demonstrated substantial capabilities and generalization in code generation tasks. These models generate code by autoregressively predicting the next token from the previous
context (e.g., function signature, docstring, and previously generated tokens) and combining the previous context and the generated tokens together as the finally generated code snippet. \textit{ChatGPT} takes this a step further. It has also demonstrated impressive code generation capabilities. Additionally, with RLHF~\cite{Ouyang:22}, ChatGPT supports the ability to answer follow-up questions, providing even more powerful and versatile features compared to previous code-related LLMs.}

\vspace{-1em}
\subsection{Limitations}

The results and experiments of this study are limited to two parts: (1) \textit{ChatGPT} is a closed-source model, which means that we are unable to directly map the analysis results to the internal workings of \textit{ChatGPT} or understand the specific characteristics of the model. Furthermore, the exact training data used by ChatGPT remains unknown to us. Consequently, it becomes difficult to ascertain whether the problems we input have been previously used in the training dataset; (2) it is important to note that \textit{ChatGPT} is a continuously evolving and training model. The responses generated by \textit{ChatGPT} in this study can only reflect the performance of the model at the time of our work (i.e., \revision{model version \textit{gpt-3.5-turbo-0301} of \textit{ChatGPT}}).


\vspace{-1em}
\subsection{Threats to Validity}

\noindent \textbf{\textit{LeetCode} Problems and CWE Scenarios.} To reduce bias by manually selecting subjects for evaluation, we utilize \textit{LeetCode} problems as our main dataset. However, \textit{LeetCode} problems are designed specifically for coding practice and interview preparation. While they cover a range of programming concepts and challenges, they may not fully represent the complexity and diversity of real-world coding tasks. Real-world coding scenarios often involve various external factors, domain-specific requirements, and specific constraints that may not be fully captured by \textit{LeetCode} problems alone. Moreover, the classes of vulnerabilities that \textit{LeetCode} problems' code can have are limited, so we also utilize CWE scenarios~\cite{pearce2022asleep} to supplement the evaluation of ChatGPT's security code generation. Nevertheless, similar to \textit{LeetCode} problems, these scenarios may not cover all real-world code scenarios.

\noindent \textbf{\textit{LeetCode} Online Judgment.} \textit{LeetCode} online judgment platform terminates the testing process upon encountering the first failed test case. Thus, the test case pass rates provided by the platform may serve as a lower bound, but it does not affect the statuses of code snippets generated by \textit{ChatGPT} and the conclusion drawn from this study.

\noindent \textbf{Vulnerability Detection by \textit{CodeQL}.} \textit{CodeQL} may report a code as vulnerable when it is actually secure. To mitigate the risk, human expertise is employed to manually inspect the code for potential vulnerabilities, thereby ensuring the accuracy and reliability of the analysis. 

\noindent \revision{\textbf{Limited Languages in Vulnerability Detection.} The evaluation of vulnerability detection in our study only focuses on limited languages (C, C++, and Java to \textit{LeetCode} problem scenarios and C and Python to CWE scenarios) out of five languages (C, C++, Java, Python, and JavaScript) due to the targeted scenarios and limitations of vulnerability detection tools. Though the evaluation results provide insights into \textit{ChatGPT}-based code generation in these languages, our study does not fully reflect the spectrum of all five languages in security.}

\noindent \textbf{Statistical Validity.} \textit{ChatGPT} has randomness. When faced with the same input prompt, \textit{ChatGPT} may produce different responses. To reduce the risk, we use 728 \textit{LeetCode} problems. For each \textit{<Problem, Language>} pair, we independently generate one corresponding code snippet once\footnote{\textit{ChatGPT} has a rate-limiting of queries and it may be retrained at a later date. Thus, we query once for each problem and do not requery for failed responses (e.g., response is empty or irrelevant such as violation of policy. This small amount of responses is excluded from the experimental evaluation).}, following the law of large numbers. For CWE's code scenarios, we generate 60 code snippets independently for each scenario. As for the multi-round process, we sample many code snippets for experimentation and we set 5 to the maximum round number~\cite{dong2023self}, a reasonable round limit.

\vspace{-1em}
\section{Related Work}\label{sec:relatedwork}

\noindent \textbf{Language Models.} Language models have a wide range of applications in NLP, including machine translation, question answering, summarization, text generation, code generation and so on~\cite{Carlini:21, Raffel:22, Lan:19, zhang:20, Pilault:20, Cai:21, Khashabi:20, Cho:14, chen2021evaluating, bui2021infercode}. These models, with a large number of parameters, are trained on extensive corpus to better understand language (i.e., LLM). One of the fundamental architectures used in language models is Transformer~\cite{vaswani2017attention}, which consists of stacked encoders and decoders. Transformer utilizes self-attention mechanism to weigh the importance of words in the input text, capturing long-range dependencies and relationships between words. Many language models are built upon Transformer. \textit{ELMo}~\cite{Peters:18} employs multi-layer bidirectional LSTM (long short-term memory) and provides high-quality word representations. \textit{GPT}~\cite{radford:18} and \textit{BERT}~\cite{Devlin:18} are based on the decoder (unidirectional) and encoder (bidirectional) components of the Transformer, respectively. They utilize pre-training and fine-tuning techniques. \textit{GPT-2}~\cite{radford:19} and \textit{GPT-3}~\cite{Brown:20} are the successors of \textit{GPT}, with \textit{GPT-2} having a larger model size in parameters than \textit{GPT}, and \textit{GPT-3} being even larger than \textit{GPT-2}. Additionally, with larger corpus, \textit{GPT-2} and \textit{GPT-3} introduce zero-shot and few-shot learning to enable adaptation to multitask scenarios. \textit{Codex}~\cite{chen2021evaluating} is obtained by training \textit{GPT-3} on GitHub code data. It serves as the underlying model for GitHub \textit{Copilot}~\cite{copilot}, a tool that can automatically generate and complete code automatically. \textit{InstructGPT}~\cite{Ouyang:22} uses additional supervised learning and reinforcement learning from human feedback (RLHF) to fine-tune \textit{GPT-3}, aligning the language model with users. \textit{ChatGPT}~\cite{ChatGPT}\revision{, based on \textit{GPT-3.5}~\cite{OpenAI},} utilizes the same methods as \textit{InstructGPT} and provides the ability to answer follow-up questions.

\noindent \textbf{Code Generation.} Code generation~\cite{le2020deep} is a fundamental application of language models that aims to automatically generate or complete computer code based on given specifications or natural language descriptions, improving programming productivity. There is a lot of research on it, including traditional approaches and AI-based approaches. Traditional code generation~\revision{\cite{le2020deep, gulwani2017program, bielik2016phog, gvero2013complete, gulwani2011automating, abate2018counterexample}} approaches typically rely on predefined templates or rules (e.g., context-free grammar), along with input-output specifications, which limits their flexibility and requires manual effort. \revision{For example, Gulwani~\cite{gulwani2011automating} identifies a  string expression language available to various string manipulation tasks (e.g., extract substrings in a specific format) and designs an algorithm for learning a string expression that is consistent with the provided input-output examples.} \revision{As for} AI-based approaches~\cite{allamanis2018survey, chen2021evaluating, copilot, li2017code, ChatGPT, tufano2020unit, bard, nijkamp2022codegen}, \revision{they} leverage deep learning and NLP to overcome these limitations and can offer more intelligent and adaptable code-generation capabilities. \revision{Li et al.~\cite{li2017code} leverage recurrent neural network with attention mechanism and pointer mixture network on abstract syntax tree (AST) to predict next word in code completion tasks, learning from large-scale codebases. Liu et al.~\cite{liu2020self} model the structural information in AST and use Transformer-XL network and multi-task learning to capture long-term dependency in programs and learn two disjoint code-related tasks in code completion, respectively. Ashwin et al.~\cite{Ashwin2018} combine traditional method with neural network (e.g., LSTM network) to generate code from examples, with high correctness, strong generalization, and low synthesis time.} \revision{Recently, with the advantages of LLMs, researchers apply LLMs directly to the code generation task by using extensive code datasets, such as \textit{Codex}~\cite{chen2021evaluating}, \textit{Copilot}~\cite{copilot}, \textit{CodeGen}~\cite{nijkamp2022codegen}, providing more powerful capabilities and generalizations. These code-related LLMs (e.g., \textit{Codex}) autoregressively predict the next token from the previous context (e.g., function signature, docstring, and previously generated tokens) in code generation and combine the previous context and the generated tokens together as the finally generated code snippet. \textit{ChatGPT}~\cite{ChatGPT}, the state-of-the-art LLM based on \textit{GPT-3.5}~\cite{gpt-3.5}, has also demonstrated impressive code generation capabilities. Additionally, with RLHF~\cite{Ouyang:22}, \textit{ChatGPT} supports the ability to answer follow-up questions, providing even more powerful and versatile features compared to previous code-related LLMs.}

\noindent  \revision{\textbf{Evaluation on LLM-based Code Generation.}}
Hendrycks et al.~\cite{hendrycks2021measuring} craft APPS benchmark of Python programming problems and assess the code generation performance for several \textit{GPT}-based variant models by fine-tuning with APPS. Fan et al.~\cite{fan2022automated} systematically study whether automated program repair (APR) techniques, including \textit{Codex}, can fix the incorrect solutions to \textit{LeetCode} problems produced by \textit{Codex}. Xia et al.~\cite{xia2023automated} perform an extensive study on directly applying LLMs (9 state-of-the-art LLMs) for APR. They evaluate different ways of using LLMs for the task, including the entire-patch fix, the chunk-of-code fix, and the single-line fix. Pearce et al.~\cite{pearce2022examining} examine the use of LLMs (e.g., \textit{Codex}) by zero-shot learning for vulnerability repair. \textit{CodeT}~\cite{chen2022codet} utilizes LLMs to generate functionally correct code solutions by leveraging dual execution agreement. It generates multiple code solutions and multiple test cases for a given programming problem and executes the generated code solutions using the generated test cases to rank and find the best solution. Dong et al.~\cite{dong2023self} introduce the concept of software development life cycle and propose a self-collaboration framework that leverages different \textit{ChatGPT} conversations to play different roles (e.g., analyst, developer, and tester), collaborating to generate code. Sobania et al.~\cite{sobania2022choose} conduct an evaluation of \textit{Copilot} on standard program synthesis benchmark program, comparing the results with genetic programming. Pearce et al.~\cite{pearce2022asleep} assess \textit{Copilot}'s security code generation on the top-25 CWE vulnerabilities. Nguyen et al.~\cite{nguyen2022empirical} evaluates the quality of \textit{Copilot}-generated code by using 33 \textit{LeetCode} problems in 4 different languages. Kou et al.~\cite{kou2023model} investigate the attention alignment between the nature language description from humans and the code generation by LLMs. \revision{Liu et al.~\cite{liu2023your} propose \textit{EvalPlus} framework to enhance code generation benchmarks. \textit{EvalPlus} takes in a base evaluation dataset and uses LLMs and mutation technique to produce and diversify large amounts of new test cases.} Liu et al.~\cite{liu2023refining} characterize several code quality issues of \textit{ChatGPT}-based code generation \revision{across Java and Python languages,} \revision{including correctness and maintainability. They also examine the ability of \textit{ChatGPT} to repair bugs and code style issues by leveraging feedback information.} Different from their work, we conduct a systematical assessment with deep analysis for \textit{ChatGPT}-based code generation \revision{across five languages} in terms of correctness, \revision{complexity}, and security, including the multi-round process. \revision{Our research significantly extends the current understanding of \textit{ChatGPT}-based code generation. We not only evaluate the correctness of the generated code but also provide a deep dive into the underlying causes of incorrectness in \textit{ChatGPT}-generated code. We also deeply assess the \revision{complexity} and security of the generated code. Furthermore, we deeply investigate the impact of the multi-round fixing process on these aspects, providing a more realistic evaluation of \textit{ChatGPT}'s capabilities in iterative code generation scenarios. This comprehensive study underscores the practical implications of AI-generated code in real-world software development.}

\vspace{-1em}
\section{Conclusion}\label{sec:conclusion}

In this paper, we present a systematic assessment of \textit{ChatGPT}-based code generation. We comprehensively evaluate code snippets generated by \textit{ChatGPT} from three aspects of correctness, \revision{complexity}, and security, including the multi-round fixing process. Our experimental results demonstrate that (1) \ChatGPT is better at generating functionally correct code for Bef. problems in different languages than Aft. problems (the average \textit{Accepted} rate of the former exceeds the latter by $48.14\%$), but \textit{ChatGPT}'s ability to directly fix erroneous code to achieve correct functionality is relatively weak; (2) the distribution of cyclomatic and cognitive complexity levels for code snippets in different languages varies. Additionally, the multi-round fixing process with \ChatGPT generally preserves or increases the complexity levels of code snippets; (3) in \minorrevision{algorithm} scenarios \minorrevision{with languages of C, C++, and Java, and CWE scenarios with languages of C and Python3}, the code generated by \ChatGPT has relevant vulnerabilities. Fortunately, the multi-round fixing process for vulnerable code snippets demonstrates promising results, with a high percentage ($100\%$ and $89.4\%$) of vulnerabilities successfully addressed\revision{; and (4) code generation may be affected by \textit{ChatGPT}'s non-determinism factor, resulting in variations of code snippets in functional correctness, complexity, and security}. Overall, our findings uncover potential issues and limitations that arise in the \textit{ChatGPT}-based code generation and pave the way for improving AI and LLM-based code generation techniques.

\ifCLASSOPTIONcaptionsoff
  \newpage
\fi



%

\bibliographystyle{IEEEtran}
\normalem
\IEEEtriggeratref{73}
\bibliography{ref/reference}


%

\end{document}